\documentstyle[12pt]{article}

\countdef\pageno=0\pageno=57

\newcommand{\sr}{\stackrel{\rightarrow}{r}}
\newcommand{\ra}{\rightarrow}
\newcommand{\dgr}{\dagger}
\newcommand{\lgl}{\langle}
\newcommand{\rgl}{\rangle}
\newcommand{\bt}{\beta}
\newcommand{\gm}{\gamma}
\newcommand{\dlt}{\delta}
\newcommand{\be}{\begin{equation}}
\newcommand{\ee}{\end{equation}}

   \font\tenmsb=msbm10 scaled\magstep 1
   \font\sevenmsb=msbm7 scaled \magstep 1
   \font\faivemsb=msbm5 scaled \magstep 1
\newfam\msbfam
      \textfont\msbfam=\tenmsb
      \scriptfont\msbfam=\sevenmsb
      \scriptscriptfont\msbfam=\faivemsb
\def\Bbb#1{{\fam\msbfam #1}}
\font\tengothic=eufm10 scaled\magstep 1
\font\sevengothic=eufm7 scaled\magstep 1
\newfam\gothicfam
      \textfont\gothicfam=\tengothic
      \scriptfont\gothicfam=\sevengothic

\begin{document}

\section{Green's Functions}

\begin{sloppypar}

The method of Green's functions is one of the most powerful techniques in
quantum statistical physics [52--57] and quantum chemistry [58].

If {\it L} is a linear operator, and $f(x)$ is a given function, 
then a solution of $Lu=f$ can be expressed in the form 
$u(x)=\int G(x,\xi)f(\xi) d\xi$, where $G$ satisfies the equation 
$LG(x,\xi)=\delta(x-\xi)$. The function $G(1,2)$ which serves as the
kernel of the above integral operator, is called a {\it Green's Function}.
This, because it was George Green ($1793-1841$) who introduced these
functions into the study of linear equations. Certain Green's functions
are also called {\it propagators} because as follows from the above
integral we can think of $G$ as transfering an {\it L--modulated} effect
of a "charge" $f(\xi)$ at $\xi$ to the point $x$.

In recent years, the use of Green's functions has become widespread in
theoretical physics and especially in the study of the $N$--body
problem in quantum mechanics. As we shall see below they occur in many
different forms. A basic mode of classifying them is as {\it one}--particle, 
{\it two}--particle, $\ldots$, and $n$--particle propagators. A reduced 
density operator can be defined as an initial value of a corresponding 
type of a Green's function and so the latter are also restricted by the
necessity of being $N$--{\it representable}. This fact was often neglected
by early Green's function enthusiasts. For a statistical system a hierarchy 
of equations, involving $n-$ and $(n+1)$--particle Green's function from
$n=0$ to $n=N$, must be satisfied. Only thus can $N$--representability of
the Green's functions be {\it guaranteed}. For large $N$ this is manifestly 
impossible, so in practice the hierarchy of equations is usually truncated
by an {\it ansatz} at the level of the first, second or third order
Green's functions. But this always leaves some doubt as to the extent of
possible errors introduced by the {\it ansatz.} 

An even more difficult obstacle in discussing the $N$--body problem is
that it cannot be solved exactly if $N>1$! So to make any progress, some
form of approximation must be adopted. For 2 or 3 particles, the approach
which Hylleraas used to explain the spectrum of atomic helium can
sometimes give very accurate results. However, for larger $N$ one normally
starts from a tractable initial approximation attempting to improve this
by successive perturbations. During recent decades physicists have
developed increasingly sophisticated methods for obtaining approximations 
to  Green's functions, and for the application of perturbation techniques
to them.

It has been our aim in this chapter to provide the reader with a
comprehensive overview of the extensive resulting literature about
Green's functions. We have not followed any one previous treatment. Rather,
we have sought to gather together what we regard as the most important and
most generally applicable aspects of the Green's function approach to the
quantum mechanical $N$--body problem. It may be that, for some readers,
our condensed treament will bring to mind the well--known canard: "in
order to do theoretical physics all you need is a knowledge of perturbation
theory and a mastery of the Greek alphabet" and they will feel submerged
under formalism! If so, we apologize, and only plead that we were unable,
otherwise, to present in reasonable length what we consider to be the
state--of--the--art for Green's functions.

\subsection{Principal Definitions.}

We shall adopt the widespread convention of using Planck's constant divided
by $2\pi$ and Boltzmann's constant as units; so $h/2\pi=1,\; k_B=1$.

In order to have a compact notation we shall often denote the particle
coordinates $(x_i,t_i)$ by the index $i$, so that a
function of $n$ particles may be written as 
$$
f(12\ldots n) := f(x_1,t_1;x_2,t_2;\ldots;x_n,t_n), 
$$
and a differential for $n$ particles becomes 
$$
d(12\ldots n) := dx_1dt_1dx_2dt_2...dx_ndt_n. 
$$

Thus, for the field operator we shall often use 
$$
\psi \left( 1\right) :=\psi \left( x_1,t_1\right) . 
$$
The distinction between Bose and Fermi statistics is basic to quantum
mechanics and is encoded in the field operator. We have adopted the
convention that when the symbols $\pm$ or $\mp$ appear in our formulas 
{\it the upper sign refers to bosons}.

It is a basic metaphysical assumption of most of us that {\it cause}
precedes {\it effect} in time. In consequence a so--called {\it
chronological} or {\it time--ordering operator}, $\hat T$, acting on
field operators plays an important role in the theory of Green's 
functions. 
\begin{eqnarray}
\hat T \psi(1) \psi^\dagger(2) := \left\{ 
\begin{array}{cc}
\psi(1)\psi^\dagger(2) ,     & t_{12}> 0 \\ 
\pm \psi^\dagger(2) \psi(1), & t_{12} < 0 
\end{array} \right.  \; ,
\label{2.1}
\end{eqnarray}
where 
\begin{equation}
t_{12}:= t_1 - t_2 = -t_{21}.  
\label{2.2}
\end{equation}

Similarly, when $\hat{T}$ acts on the products $\psi(1)\psi(2)$ and 
$\psi^\dagger(1)\psi^\dagger(2)$, it arranges the order of the field
operators so that time increases from {\it right} to {\it left}. If a
permutation of the operators is neccessary, the product is multiplied by
$+1$ or $-1$ according as Bose or Fermi statistics is relevant.

Invoking the unit-step function: 
$$
\Theta(t):=\left\{ 
\begin{array}{cc}
0, &  t < 0 \\ 
1, &  t > 0
\end{array}
\right. , 
$$
the time--ordered product (2.1)  can be written as 
$$
\hat{T}\psi(1) \psi^\dagger(2) = \Theta \left(
t_{12}\right) \psi \left( 1\right) \psi ^{\dagger}\left( 2\right) \pm \Theta
\left( -t_{12}\right) \psi ^{\dagger}\left( 2\right) \psi \left( 1\right) . 
$$
The time--ordered products of any number of field operators can be defined
in a similar manner.

We now define five different types of Green's functions commonly used in
many--body theory:

\begin{itemize}
\item  {\em causal Green's function} {\em or propagator} 
\begin{equation}
G\left( 12\right) :=-{\em i}<\hat{T}\psi \left( 1\right) \psi ^{\dagger} 
\left( 2\right) >;  \label{2.3}
\end{equation}

\item  {\em retarded Green's function} 
\begin{equation}
G_{ret}(12):=-i\Theta \left( t_{12}\right) <\psi \left( 1\right)\psi^{\dagger}
\left( 2\right) >;  \label{2.4} \end{equation}

\item  {\em advanced Green's function} 
\begin{equation}
G_{adv}\left( 12\right) :=-i\Theta \left( t_{21}\right) <\psi ^{\dagger} 
\left( 2\right) \psi \left( 1\right) >;  \label{2.5}
\end{equation}

\item  {\em commutator and anticommutator retarded Green's function } 
\begin{equation}
G_{ret}^{\mp _{}}:=-i\Theta \left( t_{12}\right) <\left[ \psi \left(
1\right) ,\psi ^{\dagger}\left( 2\right) \right] _{\mp }>;  \label{2.6}
\end{equation}

\item  {\em commutator and anticommutator advanced Green's function } 
\begin{equation}
G_{adv}^{\mp }\left( 12\right) :=-i\Theta \left( t_{21}\right) <\left[ \psi
\left( 1\right) ,\psi ^{\dagger}\left( 2\right) \right] _{\mp }>.  
\label{2.7} 
\end{equation}
\end{itemize}

Note that the average $<\cdot >$ may , depending on the context, be with
respect to a pure state or with respect to an ensemble which may be
temperature dependent and which may be in equilibrium or not.

The properties of these different Green's functions are similar to one
another and there are fairly obvious relations among them so that in most
situations it is as convenient to work with one of them as another.
However, in our opinion the {\it causal Green's function} has a slight
edge for the following reasons:

\begin{enumerate}
\item  
As a linear combination of the retarded and advanced Green's functions,
the causal Green's function carries more information than either of those
separately: 
\begin{equation}
G \left( 12\right) =G_{ret}\left( 12\right) \pm G_{adv}\left( 12\right) ;  
\label{2.8}
\end{equation}

\item  
Dealing with the causal Green's function is rather simpler than with
the commutator or anticommutator Green's functions;

\item  
The causal Green's function, as we have defined it is accepted and
used, almost universally, in both quantum statistical mechanics and
quantum field theory.
\end{enumerate}

Notice, further, that in the usual set--up in Hilbert space, operators act
on the left of the symbols denoting states of the system. Thus when the
operator is a product we interpret its action to mean that the factor on 
the right acts first. The chronological operator, $\hat{T}$, was defined
with this in mind ensuring that the factor which acts first on the wave
vector is at the earlier time. This corresponds to our intuitive feeling
about causality, justifying the term "causal" applied to (2.3)

It turns out that $G\left( 12\right) $ is closely associated with the
time-dependent 1-matrix. In fact, 
\begin{equation}
\rho _1\left( x_1,x_2,t\right) =\pm i\lim_{t_i\rightarrow t}G\left(
12\right) ,  
\label{2.9}
\end{equation}
where the limit is taken subject to $t_2>t_1.$

Recall that here and elsewhere when $\pm$ or $\mp$ occur indicating
Fermi or Bose stastistics, the {\it upper} sign refers to Bose statistics.

The diagonal elements of the 1--matrix give the value of particle density
and are therefore positive, that is 
$$
\rho _1\left( x,x,t\right) =\rho \left( x,t\right) . 
$$
From this we obtain the following condition on $G(12)$,
\begin{equation}
\pm i \lim_{t_i\rightarrow t}\lim_{x_2\rightarrow x_1}
G\left( 12\right) \quad \geq 0,  
\label{2.10}
\end{equation}
where the limit is taken subject to the restriction $t_2>t_1$.

All of the Green's functions defined above are one--particle functions. In
an analogous manner we may define a two--particle causal Green's function,
or {\it two--particle propagator}: 
\begin{equation}
G_2\left( 1234\right) :=-<\hat{T}\psi \left( 1\right) \psi \left( 2\right)
\psi ^{\dagger}\left( 3\right) \psi ^{\dagger}\left( 4\right) >.  
\label{2.11} 
\end{equation}

The definition (2.1) for the chronological operator is generalized
to an arbitrary finite product of field operators as follows: the factors
are permuted so that time increases from right to left and the appropriate
sign is determined by decomposing the required permutation into a succession
of transpositions of neighbouring factors. It then follows that (2.11)
has the following symmetry properties, 
$$
G_2\left( 1234\right) =
\pm G_2\left( 2134\right) =\pm G_2\left( 1243\right) . 
$$
We again find that there is a simple relation between the 2--particle
propagator and the time--dependent 2--matrix, namely 
\begin{equation}
\rho _2\left( x_1,x_2,x_3,x_4,t\right) =\mp \lim_{t_i\rightarrow t}
G_2\left( 1234\right) ,  
\label{2.12}
\end{equation}
where the limit is taken subject to the restriction $t_4>t_3>t_1>t_2$.

Similarly, we define the {\it $n$--particle propagator} as 
\begin{equation}
G_n\left( 12\ldots 2n\right) :=
\left( -i\right) ^n<\hat{T}\psi \left( 1\right)
\ldots \psi \left( n\right) \psi ^{\dagger}\left( n+1\right) 
\ldots \psi^{\dagger}  \left( 2n\right) >.  
\label{2.13} 
\end{equation}
Again, we can recover the $n$--matrix from (2.13). It turns out that 
\begin{equation}
\rho _n\left( x_1,\ldots,x_{2n},t\right) = ( \pm 1)^{(3n-1)n/2} i^n
\lim_{t_i\rightarrow t} G_n\left(1\ldots 2n\right) ,  
\label{2.14}
\end{equation}
where the limit is subject to the restriction 
$$
t_{2n} > t_{2n-1} > \ldots > t_{n+1}> t_1 > t_2 >\ldots >t_n . 
$$

In all of the Green's functions defined so far there are the same number of
creation and annihilation operators. This is appropriate for situations in
which the number of particles is conserved for in such cases the expected
value for operators involving unequal numbers of creation an annihilation
operators will be zero. However, if the number of particles is not
conserved then such averages may be different from zero. Indeed, this can
happen for systems with a Bose condensate and for superconductors. In
these cases averages of operators such as $\psi(1)\psi(2)$,
$\psi^\dagger(1)\psi^\dagger(2)$, $\psi^\dagger(1)\psi(2)\psi(3)$ 
may differ from zero. Such averages, which can differ from zero only if 
the number of particles is not conserved are often called {\it anomalous 
averages} and the Green's functions defined by means of such averages are 
called {\it anomalous Green's functions}. The most commonly used are
certainly 
$$
F\left(12\right) := - i<\hat{T}\psi \left( 1\right) \psi(2) >,  
$$
\begin{equation}
\label{2.15} 
F^{+}\left( 12\right) := i<\hat{T}\psi^\dagger(2) \psi^\dagger(1) >.  
\end{equation}

In analogy with the definition of Green's functions by means of the field
operators, it is possible to employ any other operators to define 
Green's functions as statistical averages of products of operators. All
observable quantities can be expressed by means of integrals involving
Green's functions. For example it follows from (2.9) that the density of
particles is 
\begin{equation}
\rho(x,t) = \pm \lim_{x_1\rightarrow x}\lim_{(21)}G(12) .  
\label{2.16}
\end{equation}

Here we have employed an abbreviated notation for the limit to mean 
\begin{equation}
\lim_{(21)}\Leftrightarrow x_2\rightarrow x_1; \qquad
t_i\rightarrow t, \qquad  (t_2 > t_1).  
\label{2.17}
\end{equation}

Further, the density of particle current can be expressed as 
\begin{equation}
{\bf j}(x,t) =\pm \frac{1}{2m}\lim_{x_1\rightarrow x}
\lim_{(21)}\left( \nabla _1 - \nabla _2\right ) G(12)
\label{2.18}
\end{equation}
where the limit is interpreted by means of (2.17).

Any 1--particle operator, $\hat{A}_1(t)$, can be expressed by means of the
field operators and an operator-valued function $A_1(x,t)$ as 
$$
\hat{A}_1(t) = \int \psi^\dagger( x,t) A_1(x,t)\psi(x,t) dx, 
$$
and its {\it average} or {\it expected value} can be written as 
\begin{equation}
<\hat{A}_1>=\pm i\int \lim_{(21)} A_1(x_1,t) G( 12) dx_1\; .  
\label{2.19}
\end{equation}

Similarly, the average, or expected value, of a 2--particle operator 
$$
\hat{A}_2(t) = \frac{1}{2}\int \psi^\dagger(x_1,t) 
\psi^\dagger(x_2,t) A_2(x_1,x_2,t) \psi(x_2,t)\psi(x_1,t) dx_1dx_2 
$$
is 
\begin{equation}
<\hat{A}_2> = - \frac{1}{2}
\int \lim_{(4321)} A_2(x_1,x_2,t) G_2(1234) dx_1dx_2,  
\label{2.20}
\end{equation}
where in this limit: 
\begin{equation}
x_4\rightarrow x_1,\qquad x_3\rightarrow x_2;\qquad t_i\rightarrow t,
\qquad (t_4>t_3>t_2>t_1).  
\label{2.21}
\end{equation}

In order to abbreviate notation, we shall use $x^n$, with $n$ as a
{\it superscript}, to denote the $n$ variables $(x_1,x_2,\ldots,x_n)$ and 
thus $dx^n:=dx_1dx_2...dx_n$. Then an $n$--particle operator can be written 
as 
$$
\hat{A}_n(t) = \frac{1}{n!}\int \psi^\dagger(x_1,t) 
\ldots
\psi^\dagger(x_n,t) A_n(x^n,t) \psi(x_n,t) \ldots \psi(x_1,t) dx^n. 
$$
The average of this operator is given by the formula
\begin{equation}
<\hat{A}_n> = \frac{(\pm i)^n}{n!}\int \lim_{(2n\ldots 1)}
A_n(x^n,t) G_n(1\ldots 2n) dx^n
\label{2.22}
\end{equation}
which involves the $n$--particle propagator (2.13) and the generalization
of the limiting procedures (2.17) and (2.21) defined by the rule 
\begin{eqnarray}
\lim_{(2n\ldots 1)}:= \left\{ 
\begin{array}{cc}
x_{2n-i} \rightarrow x_{i+1}, &  0 \leq i\leq n-1  \\ 
t_j \rightarrow t,            &  t_{j+1} > t_j, \; \forall j .
\end{array}
\right.  
\label{2.23}
\end{eqnarray}

Applying this to the number--of--particles operator, $\hat{N}$, we find
that the average number of particles is 
\begin{equation}
N = <\hat{N}> = \pm i \int \lim_{(21)} G(12) dx_1.
\label{2.24}
\end{equation}

Averaging the Hamiltonian, $H$, we find 
$$
<H> = \pm i\int \lim_{(21)}\left [ K(\nabla _1)
+ U(x_1,t) - \mu \right ] G(12) dx_1 -
$$
\begin{equation}
-\frac{1}{2}\int \lim_{(4321)} V(x_1,x_2) G_2(1234) dx_1 dx_2.  
\label{2.25}
\end{equation}

Using this we obtain the internal energy as 
\begin{equation}
E = <H> + \mu N.  
\label{2.26}
\end{equation}
These examples should convince the reader that an {\it exact }knowledge of
the Green's functions for an $N$--body system contains complete 
information about its state whether or not it is in a pure or statistical
state. However, we also know that for $N>1$ it is too much to hope for 
{\it exactness}. So our problem is to develop methods which can give us
reasonable approximations.

\subsection{Spectral Representations}

The Green's functions defined in the previous section are appropriate
whether or not the system being studied is in equilibrium. When it is in
equilibrium, as suggested by Lehmann, taking the Fourier transform with
respect to the time variable leads to a form for the Green's functions which
is referred to as the {\it spectral representation} and with which it is
much easier to work. The properties of this representation give rise
to certain advantages of the Green's function method. In particular,
singularities in the spectral representation of the first order Green's
function occur at energy levels of quasi--particles.

For the equilibrium states we can introduce the {\it equilibrium
correlation functions: } 
$$
C_{+}( x_1,x_2,t):= <\psi(1)\psi^\dagger(2)>, 
$$
\begin{equation}
\label{2.27} 
C_{-}(x_1,x_2,t):= <\psi^\dagger(2)\psi(1)>,  
\end{equation}
in which 
\begin{equation}
t=t_{12}=t_1-t_2.  
\label{2.28}
\end{equation}

The right hand side of the equations in (2.27) apparently depends on $t_1$
and $t_2$ but, in fact, only on the difference $t$.  For recall that when
the Hamiltonian, $H$ , does not depend explicitly on $t$, i.e. $\partial
H/\partial t=0$, the evolution operator is given by 
$$
U(t) = e^{-iHt}; \qquad [ U(t) ,\hat{\rho} ] =0, 
$$
where $\hat{\rho}$ is the $N$--particle von Neumann density matrix of the
system. By definition, 
$$
<\psi( 1) \psi^\dagger( 2) >= 
Tr\left[ \hat{\rho}\psi(1)\psi^\dagger(2) \right] =
<\psi(x_1,t)\psi^\dagger(x_2,0) >
$$
since $\psi(x,t)=U^+(t)\psi(x,0)U(t)$

The functions (2.27) enjoy the following properties with respect to
complex conjugation. 
\begin{equation}
C_{+}^{*}(x_1,x_2,t) =C_{+}(x_2,x_1,t) \qquad 
C_{-}^{*}(x_1,x_2,t) =C_{-}(x_{2,}x_1,t).  
\label{2.29}
\end{equation}

Recall that $\beta$ denotes the inverse temperature. An important relation
between the correlation functions (2.27) is the {\it equilibrium
condition}: 
\begin{equation}
C_{-}(x_1,x_2,t) =C_{+}\left( x_1,x_2,t-i\beta \right) ,
\label{2.30}
\end{equation}
which is often referred to as the {\it Kubo--Martin--Schwinger conditon}. 
Its proof follows: 
$$
C_{-}(x_1,x_2,t) = 
Tr\left[ \hat{\rho}\psi^\dagger(x_2,0)\psi(x_1,t) \right] =
Tr\left[ \psi(x_1,t) \hat{\rho}\psi^\dagger(x_2,0) \right]=
$$ 
$$
= Tr\left[\hat{\rho}e^{\beta H}\psi(x_1,t) 
e^{-\beta H}\psi^\dagger(x_2,0) \right]  
= Tr[\hat{\rho}\psi(x_1,t-i\beta)\psi^\dagger( x_2,0) ] 
= C_{+}(x_1,x_2,t-i\beta)
$$
since, at thermal equilibrium, $\hat{\rho}=(\exp (-\beta H))/Tr(
\exp(-\beta H))$.

The Fourier transforms, $K_{\pm}$, of functions (2.27) are given
by 
$$
C_+( x_1,x_2,t) = \int_{-\infty }^{+\infty }
K_+(x_1,x_2,\omega) e^{-i\omega t}\frac{d\omega }{2\pi }\; , 
$$
$$
K_+(x_1,x_2,\omega) = \int_{-\infty }^{+\infty }
C_+(x_1,x_2,t) e^{i\omega t}dt\; ,, 
$$
$$
C_-(x_1,x_2,t) = \int_{-\infty}^{+\infty}
K_-(x_1,x_2,\omega) e^{-i\omega t}\frac{d\omega}{2\pi}\; , 
$$
$$
K_-( x_1,x_2,\omega) = \int_{-\infty}^{+\infty}
C_-(x_1,x_2,t) e^{i\omega t}dt \; .
$$
Then, taking the Fourier transform of (2.30) we obtain 
\begin{equation}
K_{-}\left( x_1,x_2,\omega \right) =e^{-\beta \omega }K_{+}\left(
x_1,x_2,\omega \right) .  
\label{2.31}
\end{equation}

The correlation functions (2.27) at $t=0$ satisfy the equation 
\begin{equation}
C_{+}\left( x_1,x_2,0\right) \mp C_{-}\left( x_1,x_2,0\right) =\delta \left(
x_1-x_2\right) ,  
\label{2.32}
\end{equation}
which follows from the commutation relations for the field operators. The
Fourier transform of  (2.32) reads 
\begin{equation}
\int_{-\infty }^{+\infty }K_{+}\left( x_1,x_2,\omega \right) \left( 1\mp
e^{-\beta \omega }\right) \frac{d\omega }{2\pi }=\delta \left(
x_1-x_2\right) ,  \label{5.33}
\end{equation}
when we take (2.31) into account.

By means of the correlation functions (2.27) it is possible to rewrite the
Green's functions of (2.4) and (2.5) as
\begin{equation}
G_{ret}(12) = -i\Theta(t)C_{+}(x_1,x_2,t) , \qquad
G_{adv}(12) = -i\Theta(-t) C_{-}(x_1,x_2,t) ,
\label{2.34}
\end{equation}
where the notation (2.28) is used. The Fourier transforms, $G_{+}$ and
$G_{-},$ of these Green's functions are defined by the relations: 
$$
G_{ret}(12) = \frac{1}{2\pi}\int_{-\infty}^{+\infty}
G_+(x_1,x_2,\omega) e^{-i\omega t}d\omega , $$
\begin{equation}
G_{adv}(12) = \frac{1}{2\pi}\int_{-\infty}^{+\infty}
G_-(x_1,x_2,\omega) e^{-i\omega t}d\omega .
\label{2.35}
\end{equation}

The unit step function, $\Theta$, can be expressed as a Fourier integral as
follows: 
\begin{equation}
\Theta \left( \pm t\right) =\pm i\int_{-\infty }^{+\infty }\frac{e^{-i\omega
t}}{\omega \pm i0}\frac{d\omega }{2\pi },  
\label{2.36}
\end{equation}
where the convention
\begin{equation}
\int_{-\infty }^{+\infty }f(\omega \pm i0)d\omega :=\lim_{\varepsilon
\rightarrow +0}\int_{-\infty }^{+\infty }f\left( \omega \pm i\epsilon
\right) d\omega 
\label{2.37}
\end{equation}
is used.

Recall the well--known formulas for the $\delta$--function: 
$$
\delta( t) =\int_{-\infty}^{+\infty}e^{-i\omega t}\frac{d\omega}{2\pi} 
\qquad
\int_{-\infty}^{+\infty}e^{i\omega t}dt=2\pi \delta(\omega) .
$$
To obtain expressions for $G_{+}$ and $G_{-}$, we invert equations (2.35)
making use of (2.34) and the convolution theorem for Fourier integrals.
This leads us to the {\it spectral representation} of the retarded and of
the advanced Green's functions
$$
G_{+}(x_1,x_2,\omega) = -\frac{1}{2\pi}
\int_{-\infty}^{+\infty}
\frac{K_+(x_1,x_2,\omega')}{\omega' -\omega - i0}d\omega' ,
$$
\begin{equation}
G_{-}(x_1,x_2,\omega) = \frac{1}{2\pi}\int_{-\infty}^{+\infty}
\frac{K_-(x_1,x_2,\omega')}{\omega' - \omega + i0}d\omega' .
\label{2.38}
\end{equation}

The spectral representation of the causal Green's function is an immediate
consequence of (2.38) and the definition of the Fourier transform of (2.3)
as 
\begin{equation}
G(x_1,x_2,\omega)= \int_{-\infty}^{+\infty}G(12) e^{i\omega t}dt, \qquad
G(12) = \int_{-\infty}^{+\infty}G(x_1,x_2,\omega)
e^{-i\omega t}\frac{d\omega}{2\pi}.
\label{2.39}
\end{equation}

According to (2.8) we have 
\begin{equation}
G\left( x_1,x_2,\omega \right) =G_{+}\left( x_1,x_2,\omega \right) \pm
G_{-}\left( x_{1,}x_2,\omega \right) .  
\label{2.40}
\end{equation}
Combining (2.38) and (2.40), we obtain the {\it spectral representation} 
\begin{equation}
G(x_1,x_2,\omega) = -\int_{-\infty}^{+\infty}K_+(x_1,x_2,\omega') 
\left( \frac{1}{\omega' -\omega - i0} \mp 
\frac{e^{-i\beta \omega'}}{\omega' - \omega + i0} \right) 
\frac{d\omega'}{2\pi },  
\label{2.41}
\end{equation}
of the causal Green's function.

For systems in equilibrium, the averages of operators are expressed by 
means of integrals involving $G(x_1,x_2,\omega)$ using the
formula (2.41). For example, the density of particles (2.16) is
\begin{equation}
\rho(x,t) =\pm i\lim_{t\rightarrow +0}
\int_{-\infty}^{+\infty}e^{i\omega t}G(x,x,\omega)\frac{d\omega}{2\pi }.
\label{2.42}
\end{equation}

The Green's function at coincident times is defined as 
\begin{equation}
G(12)\mid_{t=0}:= \lim_{t\rightarrow -0}G(12) .
\label{2.43}
\end{equation}
Substituting (2.41) into (2.42), we obtain 
$$
\rho \left( x,t\right) =\int_{-\infty }^{+\infty }K_{+}\left( x,x,\omega
\right) e^{-\beta \omega }\frac{d\omega }{2\pi }. 
$$

Instead of dealing with the two Fourier transforms $K_{+}$ and $K_{-}$ 
which are related by (2.31), we introduce the {\it spectral function} 
\begin{equation}
J(x_1,x_2,\omega) := \frac{K_+(x_1,x_2,\omega)}{1\pm n(\omega)}=
\frac{K_-(x_1,x_2,\omega)}{n(\omega)}\; .  
\label{2.44}
\end{equation}

Here, 
\begin{equation}
n(\omega) :=\frac 1{e^{\beta\omega }\mp 1}  
\label{2.45}
\end{equation}
is called the {\it Bose function} or the {\it Fermi function} according as
one takes the minus or plus sign. Recall that this is consistent with our
general convention that the {\it upper} of the two signs refers to Bose
statistics, the {\it lower} to Fermi statistics.

Using the equality $1\pm n(\omega)=n(\omega)e^{\beta\omega}$ and the 
spectral function (2.44) the spectral representation of the causal Green's
function, or propagator, can be expressed in the form 
\begin{equation}
G(x_1,x_2,\omega) = -\int_{-\infty}^{+\infty}J(x_1,x_2,\omega') 
\left[ \frac{1\pm n(\omega')}{\omega' - \omega -i0} \mp 
\frac{n(\omega')}{\omega' - \omega  + i0}\right] 
\frac{d\omega'}{2\pi}.  
\label{2.46}
\end{equation}

Expressions involving poles can be reorganized by invoking the symbolic
identities, introduced by Dirac, in which the symbol $P$ indicates
that the {\it principal} value is to be taken. 
$$
\frac{1}{\omega\pm i0} := P\frac{1}{\omega} \mp i\pi \delta(\omega) , 
$$
\begin{equation}
\label{2.47}
\frac{1}{f(\omega \pm i0)} := P\frac{1}{f(\omega)}\mp 
i\pi\delta(f(\omega)) .  
\end{equation}

Evaluating (2.46) by means of (2.47), we find that 
\begin{equation}
G(x_1,x_2,\omega) = 
P\int_{-\infty}^{+\infty}\frac{J(x_1,x_2,\omega')}{\omega -\omega'}
\frac{d\omega'}{2\pi} - 
\frac{i}{2}\left[ 1\pm 2n(\omega)\right] J(x_1,x_2,\omega) .  
\label{2.48}
\end{equation}

If $J$ has no poles on the real axis then again using (2.47)
we can show that (2.48) implies that 
\begin{equation}
J\left( x_1,x_2,\omega \right) =i\left[ G\left( x_{1,}x_2,\omega +i0\right)
-G\left( x_1,x_2,\omega -i0\right) \right] .  
\label{2.49}
\end{equation}

\subsection{Dispersion Relations}

As follows from their spectral representations (2.38)
the Fourier transforms of the retarded and advanced Green's functions, 
$G_{+}$ and $G_{-}$, have the following asymptotic behaviour, when 
$\omega \longrightarrow \pm \infty $
$$
G_{+}\left( x_{1,}x_2,\omega \right) \cong \frac 1\omega C_{+}(x_1,x_2,0), 
$$
and 
$$
G_{-}\left( x_1,x_2,\omega \right) \cong -\frac 1\omega C_{-}\left(
x_1,x_2,0\right) . 
$$
Thus 
$$
\lim_{\omega \rightarrow \pm \infty }G_{+}\left( x_1,x_2,\omega \right)
=\lim_{\omega \rightarrow \pm \infty }G_{-}(x_1,x_2,\omega )=0. 
$$

It follows from (2.34) that the retarded Green's function differs from
zero only if $t>0$, and the advanced Green's function only if $t<0.$
From this and the expression (2.35) for the Fourier transforms, we
may conclude that $G_{+}$ can be continued analytically into the upper
complex $\omega$ half--plane and $G_{-}$ into the lower half--plane.
For $\omega$ in the domain of analyticity of $G_{+}$ or $G_{-}$
one has the Cauchy integrals,
$$
G_{+}(x_1,x_2,\omega) = \frac{1}{2\pi i}\int_{-\infty}^{+\infty}
\frac{G_+(x_1,x_2,\omega')}{\omega' -\omega - i0} d\omega',
$$
\begin{equation}
\label{2.50} 
G_-(x_1,x_2,\omega) = - \frac{1}{2\pi i}\int_{-\infty}^{+\infty}
\frac{G_-(x_1,x_2,\omega')}{\omega' - \omega + i0}d\omega' .
\end{equation}
Using the identities (2.47) and (2.50) we obtain the {\it Hilbert 
transforms} 
$$
{\rm Im} G_\pm(x_1,x_2,\omega) = \mp \frac{P}{\pi} \int_{-\infty}^{+\infty}
\frac{{\rm Re}G_{\pm}(x_1,x_2,\omega')}{\omega' - \omega}d\omega' ,
$$
\begin{equation}
\label{2.51}
{\rm Re}G_\pm(x_1,x_2,\omega) = \pm \frac{P}{\pi}\int_{-\infty}^{+\infty}
=\frac{{\rm Im}G_\pm(x_1,x_2,\omega')}{\omega' - \omega}d\omega' .
\end{equation}
In analogy with similar equations in Optics, (2.51) are called the {\it
dispersion relations} for the retarded and advanced Green's functions.

The Fourier transform of the causal Green's function has the asymptotic
form 
$$
G(x_1,x_2,\omega) \cong \frac 1\omega \delta(x_1-x_2) ,\qquad 
(\omega \longrightarrow \pm \infty ) . 
$$

The propagator is a linear combination (2.40) of $G_+$ and $G_-$, but
since these have non--intersecting domains of analyticity for complex
$\omega$ there are no simple dispersion relations for the Fourier
transform of the causal Green's function. Even so, there are dispersion
relations for its diagonal element 
\begin{equation}
G\left( x,\omega \right) :=G\left( x,x,\omega \right) .  
\label{2.52}
\end{equation}
To derive these we first note that it follows from (2.10) that 
\begin{equation}
C_{-}\left( x,x,0\right) \geq 0.  
\label{2.53}
\end{equation}
From this and (2.31)
we obtain the inequality 
$$
\int_{-\infty}^{+\infty}K_+(x,x,\omega) e^{-\beta\omega}d\omega \geq 0, 
$$
valid for all $\beta\in\Bbb{R}_+$. Hence, 
\begin{equation}
K( x,\omega) := K_{+}( x,x,\omega) \geq 0.
\label{2.54}
\end{equation}

From the spectral representation (2.41), by means of (2.47), we obtain 
$$
{\rm Re}G(x,\omega) = \frac{P}{\pi}\int_{-\infty}^{+\infty}
\frac{{\rm Im}G(x,\omega')}{\omega' - \omega}
\left( \frac{e^{\beta\omega'}\mp 1}{e^{\beta\omega'}\pm 1}\right) 
d\omega' ,  
$$
\begin{equation}
\label{2.55}
{\rm Im}G(x,\omega) = - \frac{1}{2} K(x,\omega)
\left ( 1\pm e^{-\beta\omega} \right ) . 
\end{equation}
These are the {\it dispersion relations} for the causal Green's function
or the propagator. Rewriting (2.55) separately for the Bose and Fermi
cases, we obtain 
$$
{\rm Re}G(x,\omega) =\frac{P}{\pi} \int_{-\infty}^{+\infty}
\frac{{\rm Im}G(x,\omega')}{\omega' - \omega }
\tanh \left ( \frac{\beta\omega'}{2} \right )
d\omega' , \qquad (Bose) 
$$
$$
{\rm Re}G(x,\omega) = \frac{P}{\pi}\int_{-\infty}^{+\infty}
\frac{{\rm Im}G(x,\omega')}{\omega' - \omega }
\coth \left ( \frac{\beta\omega'}{2} \right )
d\omega' .\qquad (Fermi) 
$$

The equality (2.55) gives
$$
K(x,\omega) = -\frac{2{\rm Im}G(x,\omega)}{1\pm e^{-\beta \omega }}. 
$$
Using this, we may write the density of particles for a system in
equilibrium as 
\begin{equation}
\rho( x,t) = - \frac{1}{\pi}\int_{-\infty}^{+\infty}
\frac{{\rm Im}G(x,\omega)}{e^{\beta\omega }\pm 1}d\omega .
\label{2.56}
\end{equation}
Recall that for equilibrium this density does not vary with time.

Another form of the dispersion relations for the causal Green's function
can be expressed by means of the diagonal element of the spectral function
(2.44) which we denote by 
\begin{equation}
J\left( x,\omega \right) :=J\left( x,x,\omega \right) ,  
\label{2.57}
\end{equation}
which, as follows from (2.54), is a real valued function. Then, from the
spectral representation (2.46), we obtain the dispersion relations 
$$
{\rm Re}G(x,\omega) = - P\int_{-\infty}^{+\infty}
\frac{J(x,\omega')}{\omega' - \omega}\frac{d\omega'}{2\pi}, 
$$
\begin{equation}
\label{2.58}
{\rm Im}G(x,\omega) = - \frac{1}{2} [ 1\pm 2n(\omega)] J(x,\omega) , 
\end{equation}
where the equality 
$$
1\pm 2n(\omega) = \frac{e^{\beta\omega }\pm 1}{e^{\beta\omega }\mp 1} 
$$
has been used.

If we make use of (2.57), the expression (2.56) for the particle density
takes the simpler form 
\begin{equation}
\rho (x,t)=\int_{-\infty}^{+\infty}J(x,\omega) n(\omega) 
\frac{d\omega}{2\pi}\; .  
\label{2.59}
\end{equation}
Other observables can similarly be expressed by means of integrals 
involving the spectral function (2.57).

For equilibrium systems there is still one more very important type of
Green's function -- the {\it temperature Green's functions}, $G_{tem}$ ,
the definition of which is analogous to that for the causal  Green's
function except that {\it imaginary} time variables are employed.
Set 
$$
t_j=-i\tau _j,\qquad \left( j=1,2,...\right) 
$$
with $\tau _j$ in the interval $0\leq\tau _j\leq\beta$. The definition
of the {\it one -particle temperature Green's function} then takes the
form 
\begin{equation}
G_{tem}(12):=-i\left[ \Theta(\tau) C_{+}( x_1,x_2,-i\tau) \pm 
\Theta( -\tau) C_{-}( x_1,x_2,-i\tau) \right] ,  
\label{2.60}
\end{equation}
where $\tau:=\tau _1-\tau _2=i(t_1-t_2)$.

The equilibrium condition (2.30) for the temperture Green's function
then becomes: 
\begin{equation}
G_{tem}(12) \mid _{\tau =0}=\pm G_{tem}(12)_{\tau=\beta }.  
\label{2.61}
\end{equation}
Because of this periodicity condition, $G_{tem}$ may be expressed as a
Fourier series. Spectral representations and dispersion relations for the
temperature Green's function can be derived which are strictly analogous to
those for the real--time propagator.

Which type of Green's function one wishes to use is largely a matter of
taste. However, in our view, the real--time causal Green's functions are
the most convenient and so, normally, we prefer to deal with them.

\subsection{Evolution Equations}

The previous two sections were devoted to the study of Green's functions 
for systems of identical particles in equilibrium. We now consider
arbitrary statistical systems which may or may not be in equilibrium and
obtain the equations of motion for the various Green's functions. These
equations are practically the same for Advanced, Retarded and Causal
Green's functions. For the sake of concreteness we shall work mostly with
Causal Green's functions that is with {\it propagators}.

Assume that the system is specified by the Hamiltonian: 
$$
H(t) = \int \psi^\dagger(x,t) 
\left [ K(\nabla) + U(x,t) -\mu \right ] \psi(x,t) dx +
$$
$$
+ \frac{1}{2} \int\psi^\dagger(x,t) \psi^\dagger(x',t) V(x,x') 
\psi(x',t)\psi(x,t) dxdx' ,
$$
in which $K(\nabla)$ is a kinetic--energy operator. Introduce the {\it
effective chemical potential}
\begin{equation}
\mu(1) := \mu - U(x_1,t_1) ,  
\label{2.62}
\end{equation}
which incorporates the external fields in $U(1)$, and define the {\it 
interaction potential}
\begin{equation}
\Phi(12) := \Phi(x_1,x_2) \delta(t_1-t_2) ,  
\label{2.63}
\end{equation}
where $\Phi(x_1,x_2):=\frac{1}{2}[ V(x_1,x_2)+V(x_2,x_1)]$.

In what follows we shall occasionally employ propagators with coincident
variables. In order to avoid ambiguity, we generalize the definition of the
chronological operator in (2.1) to the case of two field operators with
the same argument and adopt the following convention:
\begin{equation}
\hat{T}\psi(1)\psi^\dagger(1) := \pm \psi^\dagger(1)\psi(1)  \qquad
\hat{T}\psi^\dagger(1)\psi(1) := \psi^\dagger(1)\psi(1) .
\label{2.64} 
\end{equation}
A product of two field operators in which the creation operator is to the
left of the annihilator has been called the {\it normal form}. Thus, the
chronological operator by acting on the product of $\psi(1)$ and
$\psi^\dagger(1)$, reduces it to normal form.  This language occurred
first in connection with the well--known theorem of Wick .

The equation of motion for the one--particle propagators follows from the
Heisenberg evolution equation for the field operator and the equality
$$
\frac d{dt}\Theta \left( \pm t\right) =\pm \delta \left( t\right) .
$$
This implies
\begin{equation}
\left [ i\frac{\partial}{\partial t_1} - K(\nabla_1) +\mu(1) \right ]
G(12) \mp  i\int \Phi(13) G_2(1332)d(3) = \delta(12) ,  
\label{2.65}
\end{equation}
where 
$$
\delta(12) =\delta(x_1-x_2) \delta(t_1-t_2) . 
$$

A quantity which plays an important role in the theory of Green's 
functions is the {\it self--energy}, $\Sigma(12)$, which is defined
implicitly by
the equation 
\begin{equation}
\int \Sigma(13) G(32) d(3) =\pm \int \Phi(13) G_2(1332) d(3) .  
\label{2.66}
\end{equation}
Then (2.65) can be transformed to the form
\begin{equation}
\left [ i\frac{\partial}{\partial t} - K(\nabla_1) + \mu(1)\right ]
G(12) -\int \Sigma(13) G(32) d(3) = \delta(12) .  
\label{2.67}
\end{equation}
Defining the {\it inverse propagator} 
\begin{equation}
G^{-1}(12) :=\left [ i\frac{\partial}{\partial t} - K(\nabla_1) +
\mu(1) \right ] \delta(12) - \Sigma(12) ,  
\label{2.68}
\end{equation}
we may write (2.67) as
\begin{equation}
\int G^{-1}(13) G(32) d(3) = \delta (12) .  
\label{2.69}
\end{equation}
It then follows from ( 2.66) that 
\begin{equation}
\Sigma(12) =\pm i\int \Phi(13) G_2(1334) G^{-1}(42) d(34) .
\label{2.70}
\end{equation}

Suppose that $G_0(12)$ is any propagator for which there is an inverse,
$G_0^{-1}(12)$, satisfying the equation 
\begin{equation}
\int G_0^{-1}(13) G_0(32) d(3) = \delta(12) ,  
\label{2.71}
\end{equation}
then (2.69) can be transformed into the integral equation 
\begin{equation}
G(12) = G_0(12) +\int B(13) G(32) d(3) ,  
\label{2.72}
\end{equation}
where the kernel 
\begin{equation}
B(12) := \int G_0(13)[ G_0^{-1}(32) - G^{-1}(32) ]d(3) .  
\label{2.73}
\end{equation}
The equation (2.72) is called the {\it Dyson equation} which is thus a
transform, via (2.69), of the equation of motion (2.65).

If we choose 
\begin{equation}
G_0^{-1}(12) :=\left[ i\frac{\partial}{\partial t_1} - K(\nabla_1) + 
\mu(1) \right ] \delta(12) -\Sigma _0(12) ,  
\label{2.74}
\end{equation}
then the kernel (2.73) becomes 
$$
B(12) = \int G_0(13) [ \Sigma(32) -\Sigma_0(32) ] d(3) . 
$$

Since (2.70) and therefore $B(12)$ involve $G_2$, Dyson's equation 
does not by itself determine the one--particle Green's function $G(12)$.
The equation of motion for $G_2$ involves $G_3$ and so on up to $G_N$ .
There is thus a hierarchy of equations for Green's functions as there is
for Reduced Density Matrices. In order to short--circuit this chain of
equations one makes an {\it ansatz} for the Green's function of some
chosen order thus introducing approximations.

Perhaps the most common such an {\it ansatz} is to assume that 
\begin{equation}
G_2(1234) =G_2^{HF} (1234) := G(14) G(23) \pm G(13) G(24),  
\label{2.75}
\end{equation}
which is called the {\it Hartree--Fock approximation}. Making this ansatz
will lead us to equations determing the 1--propagator.

The {\it Hartree--Fock ansatz} enables us to obtain the following
expression for the {\it self--energy} (5.70),
\begin{equation}
\Sigma_{HF}(12) = \pm i\delta(12) \int \Phi(3) G (33) d(3) +
i\Phi(12) G(12) ,  
\label{2.76}
\end{equation}
where 
\begin{equation}
G(11) := \lim_{x_2\rightarrow x_1}\lim_{t_{12}\rightarrow -0}G(12) ,  
\label{2.77}
\end{equation}
in accord with the convention (2.64). Introducing the {\it Hartree 
potential} 
\begin{equation}
V_H(1) :=\int \Phi(12) \rho(2) d(2) ,  
\label{2.78}
\end{equation}
in which 
$$
\rho \left( 1\right) =\pm iG\left( 11\right) =\rho \left( x_1,t_1\right) 
$$
is the density of particles, we may reduce (2.76) to
$$
\Sigma _{HF}\left( 12\right) =V_H\left( 1\right) \delta \left( 12\right)
+i\Phi \left( 12\right) G\left( 12\right) . 
$$

There is a more general {\it ansatz} which appeared in connection with 
the theory of superconductivity and is now usually referred to as the {\it
Hartree--Fock--Bogolubov approximation}. This implies that the
two--particle propagator decouples as 
\begin{equation}
G_2^{HFB}(1234) = G(14) G(23) \pm G(13) G(24) \mp F(12) F^{+}(34),  
\label{2.79}
\end{equation}
where $F$ and $F^{+}$ are the {\it anomalous Green's functions} defined in
(2.15). This approximation is used not only for superconductors but also
often for discussing Bose systems with condensate. Under assumption (2.79) 
the self--energy (2.70) takes the form 
\begin{equation}
\Sigma_{HFB}(12) = \Sigma_{HF}(12) -
i\int \Phi(13)F(13) F^{+}(34) G^{-1}(42) d(34) .  
\label{2.80}
\end{equation}
In addition, we will need equations of motion for the anomalous Green's
functions which would enable us to obtain an approximation to the
one--particle Green's function.

In these two examples we made an assumption about the nature of the
second--order propagator which enables us to obtain an approximation 
for the first--order propagator. In order to approximate the two--particle
propagator one must decouple it from the third and higher order propagators 
by an appropriate ansatz. We now derive the {\it Bethe--Salpeter equation}
and the {\it Brueckner approximation} which are two widely used methods
of moving beyond the 1--propagator and the Hartree--Fock or
Hartree--Fock--Bogolubov approximations.

The equation of motion for $G_2$, which involves $G_3$, is as follows: 
$$
\left [ i\frac{\partial}{\partial t_1} - K(\nabla_1) + \mu(1) \right ]
G_2\left( 1234\right) \mp i\int \Phi(15)G_3(125534) d(5) =
$$
\begin{equation}
= \delta (14) G (23) \pm \delta(13)G(24) .  
\label{2.81}
\end{equation}
Using (2.68) the first term in this equation can be rewritten as
$$
\left [ i\frac{\partial}{\partial t_1} - K(\nabla_1) + \mu(1)\right ]
G_2(1234)= $$
$$
= \int \left [ G^{-1}(15) + \Sigma(15) \right ]G_2(5234) d(5) . 
$$
Operating on (2.81) by $G$, leads to the integral equation 
$$
G_2(1234) = G(14) G(23)\pm G(13)G(24) +
i\int G(15) \Phi(56) G_3(256634) d(56) -
$$ 
\begin{equation}
-\int G (15) \Sigma(56) G_2(6234) d(56) .  
\label{2.82}
\end{equation}

Unhappily this involves $G_3$, otherwise it would be an equation 
determining $G_2$. The standard procedure is to "decouple" by an {\it
ansatz} for $G_3$ in terms of $G=G_1$ and $G_2$. In analogy with (2.75),
we assume that 
\begin{equation}
G_3(123456) = G(16) G_2(2345) \pm G(15) G_2(2346) + G(14)G_2(2356) ,
\label{2.83}
\end{equation}
which is an expression with appropriate symmetry properties. Substituting
(2.83) in (2.82) leads to the integral equation 
$$
G_2(1234) = G_2^0(1234) + i\int G(15)\Phi(56) G(26) G_2(5634) d(56) \pm
$$
\begin{equation}
\pm \int G(15) \Sigma(56) \left[ G_2^0(2634) -G_2(2634) \right] d(56) .  
\label{2.84} 
\end{equation}
In this we used the definition (2.66) for the self--energy and the
notation 
\begin{equation}
G_2^0(1234) := G(14) G(23) \pm G(13) G(24) .  
\label{2,85}
\end{equation}
Since the difference between $G_2$ and $G_2^0$ arises from interaction
between particles, the last term in (2.84) includes the interaction
potential $\Phi$ to the second degree and higher. Dropping the third term
on the right--hand side of (2.84) leads to the {\it Bethe--Salpeter
equation} 
\begin{equation}
G_2\left( 1234\right) =G_2^0\left( 1234\right) +i\int G(15)\Phi \left(
56\right) G\left( 26\right) G_2\left( 5634\right) d\left( 56\right) ,
\label{2.86}
\end{equation}
which includes interaction in first order.

One could try to solve this equation directly or adopt an approach due to
Brueckner by introducing the so--called $T$--matrix which is defined
implicitly by the relation 
\begin{equation}
\Phi \left( 12\right) G_2\left( 1234\right) =\int T\left( 1256\right)
G_2^0\left( 5634\right) d\left( 56\right) .  
\label{2.87}
\end{equation}
Then from (2.86) and (2.87), using the property 
$$
\int G_2^0\left( 1256\right) G^{-1}\left( 53\right) G^{-1}\left( 64\right)
d\left( 56\right) =\delta \left( 14\right) \delta \left( 23\right) \pm
\delta \left( 13\right) \delta \left( 24\right) , 
$$
it is possible to show that 
\begin{equation}
T\left( 1234\right) =\Phi \left( 12\right) \delta \left( 13\right) \delta
\left( 24\right) +i\Phi \left( 12\right) \int G\left( 15\right) G\left(
26\right) T\left( 5634\right) d\left( 56\right) .  
\label{2.88}
\end{equation}
The approach based on (2.88) is called the {\it Brueckner approximation}.
If (2.88) can be solved for $T $, then (2.87) can be used to obtain $G_2$.

In concluding this section dealing with the equations of motion, we shall
prove an important -- even somewhat startling -- result which echos the
famous theorem of Kohn--Hohenberg. From (2.65)
assuming that $t_1\neq t_2$, we have 
$$
\int \Phi \left( 13\right) G_2\left( 1332\right) d\left( 3\right) =\mp
\left[ i\frac \partial {\partial t_1}-K\left(\nabla_1\right) +\mu
\left( 1\right) \right] G\left( 12\right) . 
$$
If, as in the standard case, the interaction betwen particles is symmetric
so $V(x_1,x_2)=\Phi(x_1,x_2)$, then the average of the Hamiltonian (2.25)
 can be cast in the form 
\begin{equation}
<H>=\pm \frac i2\int \lim_{\left( 21\right) }\left[ i\frac \partial
{\partial t_1}+K\left(\nabla_1\right) -\mu \left( 1\right) \right]
G\left( 12\right) dx_1.  
\label{2.89}
\end{equation}
Further, the internal energy (2.26) becomes 
\begin{equation}
E=\pm \frac i2\int \lim_{\left( 21\right) }\left[ i\frac \partial {\partial
t_1}+K\left( {\bf \nabla }_1\right) +U\left( x_1,t_1\right) +\mu \right]
G\left( 12\right) dx_1.  
\label{2.90}
\end{equation}
For equilibrium, when $U(x,t)=U(x)$, we may invoke the Fourier transform
in (2.90) to obtain 
\begin{equation}
E=\pm \frac i2\int dx\int_{-\infty}^{+\infty}
\left[ \omega +K(\nabla) + U(x) + \mu \right] e^{+i\omega 0}G(x,\omega)
\frac{d\omega}{2\pi},  
\label{2.91}
\end{equation}
where the notation (2.52) is used. Finally, using the
spectral representation (2.46) we obtain from (2.91)
\begin{equation}
E = \frac{1}{2}\int dx\int_{-\infty}^{+\infty}
\left [ \omega + K(\nabla) + U(x) + \mu \right ] J(x,\omega)n(\omega) 
\frac{d\omega}{2\pi},  
\label{2.92}
\end{equation}
where $J(x,\omega)$ is the diagonal element (2.57) of the spectral
function (2.44).

This means that in order to calculate the internal energy of a system in
equilibrium we need to know only the 1--propagator, or the diagonal of the
corresponding spectral function. Note that the expressions (2.91) and 
(2.92) are exact. Of course, it will only be for moderately simple models
that we shall be able to obtain the exact 1--propagator.

\subsection{Wave Functions}

In this section we shall introduce a family of 1--particle functions which
can be thought of as Fourier transforms of time--dependent natural
orbitals. Since, in quantum mechanics, time and energy are conjugate
observables the {\it effective wave functions} which we shall define will
be associated with 1--particle energies. The effective Hamiltonian of
which these wave functions are eigen-functions contains the self-energy
operator and therefore, in analogy with Hartree--Fock orbitals, they may
be thought of as describing "dressed" quasi--particles which encapsulate a
major portion of the correlation between the bare particles. As far as we
are aware the precise relation between natural orbitals and the effective
wave functions has never been studied so we do not assert that the effective
wave functions in the limit specified in (2.9) become the natural orbitals
of the Hartree--Fock approximation to the state of the system. We do not
rule this out as a possibility but will be surprised if it proves to be
the case since, while the Hartree--Fock approximation and the method of
effective Hamiltonian of this section are both attempts to take into
account a portion of the interaction energy, they are probably not
strictly equivalent. We believe that this is an issue which deserves
further study. 

For a system in equilibrium we reformulate the equations of motion in terms
of their Fourier transforms. Thus the inverse propagator can be expanded
as 
\begin{equation}
G^{-1}(12) = \int_{-\infty}^{+\infty}G^{-1}(x_1,x_2,\omega) 
e^{-i\omega t}\frac{d\omega}{2\pi}\; ,  
\label{2.93}
\end{equation}
in which as in (2.39) we have used the notation $t=t_{12}$. Then the
evolution equation (2.69) takes the form 
\begin{equation}
\int G^{-1}\left( x_1,x_3,\omega \right) G\left( x_3,x_2,\omega \right)
dx_3=\delta \left( x_1-x_2\right) .  
\label{2.94}
\end{equation}

Similarly for the self--energy, 
\begin{equation}
\Sigma \left( 12\right) =\int_{-\infty }^{+\infty }\Sigma \left(
x_1,x_2,\omega \right) \frac{d\omega }{2\pi }\; .  
\label{2.95}
\end{equation}
Introducing the notation 
\begin{equation}
H(x,x',\omega) := [ K(\nabla) + U(x) ] \delta( x-x') +
\Sigma( x,x',\omega) ,  
\label{2.96}
\end{equation}
for the Fourier transform of the inverse propagator (2.68) we find 
\begin{equation}
G^{-1}(x,x',\omega) = ( \omega +\mu) \delta( x-x') - H(x,x',\omega) .
\label{2.97}
\end{equation}

With respect to a complete orthonormal set of one--particle functions, 
$\{\vartheta_n(x)\}$, we have expansions 
$$
G(x,x',\omega) =
\sum_{m,n}G_{mn}(\omega) \vartheta_m(x) \vartheta_n^{*}(x') \; ,
$$
\begin{equation}
\label{2.98} 
G^{-1}(x,x',\omega) =
\sum_{m,n}G_{mn}^{-1}(\omega) \vartheta_m(x) \vartheta_n^{*}(x')\;  , 
\end{equation}
and hence, by (2.97),
\begin{equation}
G_{mn}^{-1}\left( \omega \right) =\left( \omega +\mu \right) \delta
_{mn}-H_{mn}\left( \omega \right) ,  
\label{2.99}
\end{equation}
where 
\begin{equation}
H_{mn}(\omega) := 
\int \vartheta_m^{*}(x) H(x,x',\omega) \vartheta_n(x') dxdx'.  
\label{2.100}
\end{equation}
The equation of motion (2.94) becomes 
\begin{equation}
\sum_pG_{mp}^{-1}(\omega) G_{pn}(\omega) =\delta_{mn}.  
\label{2.101}
\end{equation}
Similarly, we can express the spectral function (2.49) in the form 
\begin{equation}
J(x,x',\omega) =\sum_{m,n}J_{mn}(\omega)\vartheta_m(x) \vartheta_n^{*}(x') ,
\label{2.102}
\end{equation}
with 
$$
J_{mn}\left( \omega \right) =i\left[ G_{mn}\left( \omega +i0\right)
-G_{mn}\left( \omega -i0\right) \right] . 
$$
In terms of these we may write the corresponding spectral representations
and dispersion relations.

Expressions such as the above are most useful if $G_{mn}(\omega)$ is a 
diagonal matrix. We define an {\it effective Hamiltonian}, $H(x,\omega)$, 
as an integral operator such that 
\begin{equation}
H(x,\omega) \varphi(x,\omega) :=\int H(x,x',\omega) \varphi(x',\omega)dx',  
\label{2.103}
\end{equation}
with kernel (2.96). From (2.94) and (2.97) the equation of motion then 
takes the form 
\begin{equation}
[\omega + \mu - H(x,\omega) ] G( x,x',\omega) = \delta( x-x') .  
\label{2.104}
\end{equation}
If we substitute (2.98) in this and assume that $\vartheta _n$ can be 
chosen so that $G_{mn}(\omega)=G_m(\omega)\delta_{mn}$, we find 
$$
\sum_nG_n(\omega) [ \omega + \mu - H(x,\omega) ] \vartheta_n(x)
\vartheta_n^{*}(x')  = \delta ( x - x') = 
\sum_n\vartheta_n(x) \vartheta_n^{*}(x') .
$$
In the last step we have used the assumption that $\{\vartheta_n\}$ is a
complete orthonormal set. This equality will be possible only if 
$$
G_n\left( \omega \right) \left[ \omega +\mu -H\left( x,\omega \right)
\right] \vartheta _n\left( x\right) =\vartheta _n\left( x\right) . 
$$

In particular it is clear that the $\vartheta_n$ are eigenfunctions of 
$H(x,\omega)$ with eigenvalues depending only on $\omega$ and the constant
$\mu$. Set 
\begin{equation}
H\left( x,\omega \right) \vartheta _n\left( x\right) =H_n\left( \omega
\right) \vartheta _n\left( x\right) .  
\label{2.105}
\end{equation}
Equation  (2.105) is called the {\it effective wave equation} and the
functions $\vartheta_n$, {\it effective wave functions}. Some authors also
call (2.105) the {\it effective Schr\"{o}dinger equation}. Comparing
(2.104) and (2.105), we see that the poles of the propagator are defined
by the eigenvalues of the effective Hamiltonian. The eigenfunctions, 
$\vartheta_n(x)$, describe quantum states, indexed by $n$, analogous to
the familiar wave functions for the 1--particle Schr\"{o}dinger equation.
However, the effective Schr\"{o}dinger equation is an {\it integral}
rather than a differential equation and contains the additional parameter
$\omega$.

Because of (2.105), the matrix elements (2.100) take the form 
\begin{equation}
H_{mn}\left( \omega \right) =H_n\left( \omega \right) \delta _{mn}.
\label{2.106}
\end{equation}
The Fourier coefficients of the inverse propagator  (2.99)  then become 
\begin{equation}
G_{mn}^{-1}\left( \omega \right) =\left[ \omega +\mu -H_n\left( \omega
\right) \right] \delta _{mn}.  
\label{2.107}
\end{equation}
Substituting (2.107) into the equation of motion (2.101), we obtain 
\begin{equation}
G_{mn}(\omega) = G_n(\omega) \delta _{mn}, 
\qquad
G_n(\omega) = \frac{1}{\omega - H_n(\omega) +\mu}\; .
\label{2.108} 
\end{equation}
Equation (2.108) defines $G_n(\omega)$ as a function of $\omega $ on the
whole of the complex plane apart from the real axis. For real values of
$\omega$, $G_n$ is defined by means of the spectral representation (2.46).
This, together with the spectral function 
$J_{mn}(\omega)=J_n(\omega)\delta_{mn}$, yields 
\begin{equation}
G_n(\omega) = \int_{-\infty}^{+\infty}J_n(\omega) 
\left [ \frac{1\pm n(\omega')}{\omega - \omega' + i0} \mp 
\frac{n(\omega')}{\omega -\omega' - i0} \right ] \frac{d\omega'}{2\pi}.
\label{2.109}
\end{equation}
When the spectral function, $J_n(\omega)$, has no poles on the
real axis, (2.109) implies that 
\begin{equation}
J_n\left( \omega \right) =i[G_n\left( \omega +i0\right) -G_n(\omega -i0)].
\label{2.110}
\end{equation}
Indeed, in the expression in the square bracket of (2.109), the terms 
involving $n(\omega')$ cancel and leave an integrand 
$J_n(\omega)/(\omega -\omega')$  to which Cauchy's formula applies.

In order to calculate $J_n(\omega)$, we need to know how (2.106) depends 
on $\omega$, so in view of (2.96)  we separate the self--energy into two
parts, which are respectively independent and dependent on $\omega$, by
defining $\Lambda(x,x',\omega)$ to satisfy 
\begin{equation}
\Sigma(x,x',\omega) = \Sigma(x,x',0) + \Lambda(x,x',\omega) .  
\label{2.111}
\end{equation}
Then , in view of (2.103) and (2.105),
\begin{equation}
H_n\left( \omega \right) =H_n\left( 0\right) +\Lambda _n\left( \omega
\right) ,  
\label{2.112}
\end{equation}
where 
$$
H_n(0) = \int\vartheta_n^{*}(x) \left[ K(\nabla) + U(x) \right]
\vartheta_n(x) dx + 
\int \vartheta_n^{*}(x) \Sigma(x,x',0) \vartheta_n(x') dxdx' ,
$$
and 
$$
\Lambda_n(\omega) = \int \vartheta_n^{*}(x)
\Lambda(x,x',\omega) \vartheta_n(x') dxdx'. 
$$
The second term in (2.112) can be related to the propagator $G_n(\omega)$
by means of the equation of motion and has analytical properties similar
to those of $G_n(\omega)$. Just as $J_n(\omega)$ was defined in connection
with the spectral representation of $G_n(\omega)$, so we may introduce
a spectral function, $\Gamma_n(\omega)$ such that for complex
values of $\omega$, with Im$(\omega) \neq 0$, 
\begin{equation}
\Lambda_n(\omega) = \int_{-\infty}^{+\infty}
\frac{\Gamma_n(\omega')}{\omega -\omega'} \frac{d\omega'}{2\pi}\; .  
\label{2.113}
\end{equation}
For real $\omega$, the spectral representation is 
\begin{equation}
\Lambda_n(\omega) = \int_{-\infty}^{+\infty}\Gamma_n(\omega') 
\left [ \frac{1\pm n(\omega')}{\omega -\omega'+i0} \mp 
\frac{n(\omega')}{\omega -\omega'-i0} \right ] 
\frac{d\omega'}{2\pi}\; .
\label{2.114}
\end{equation}
In analogy with (2.110), it can be proved that 
\begin{equation}
\Gamma _n\left( \omega \right) =i\left[ \Lambda _n\left( \omega +i0\right)
-\Lambda _n\left( \omega -i0\right) \right] .  
\label{2.115}
\end{equation}

Thus, when we take (2.112) and (2.113) into account, the propagator 
(2.108) assumes, for complex $\omega$ off the real axis, the value 
\begin{equation}
G_n(\omega) = \left\{ \omega - H_n(0) + \mu - \int_{-\infty}^{+\infty}
\frac{\Gamma_n(\omega')}{\omega -\omega'} 
\frac{d\omega'}{2\pi} \right\} ^{-1}.
\label{2.116}
\end{equation}
In preparation for applications, we define an {\em effective energy }as 
\begin{equation}
E_n(\omega) := H_n(0) + P\int_{-\infty}^{+\infty}
\frac{\Gamma_n(\omega')}{\omega -\omega'} \frac{d\omega'}{2\pi}.  
\label{2.117}
\end{equation}
Then substituting (2.116) in (2.110) we obtain the following important and
very useful expression for the spectral function 
\begin{equation}
J_n(\omega) =
\frac{\Gamma_n(\omega)}{[\omega -E_n(\omega) +\mu]^2+\Gamma_n^2(\omega) /4}.  
\label{2.118}
\end{equation}
The expression 
\begin{equation}
{\rm Re}\Lambda_n(\omega) = P\int_{-\infty}^{+\infty}
\frac{\Gamma_n(\omega')}{\omega -\omega'}
\frac{d\omega'}{2\pi}  
\label{2.119}
\end{equation}
is called the {\it energy shift}, and 
\begin{equation}
\Gamma_n(\omega) = -
\frac{2{\rm Im}\Lambda_n(\omega)}{1\pm 2n(\omega)}  
\label{2.120}
\end{equation}
is called the {\it decay rate} for the state $\vartheta_n(x)$.

Neglecting the decay is equivalent to retaining only the real part of the
poles of the propagator. In this case, (2.110) together with (2.47) imply
that 
\begin{equation}
J_n\left( \omega \right) =2\pi \delta \left( \omega -E_n\left( \omega
\right) +\mu \right) .  
\label{2.121}
\end{equation}

When a smooth function $f(\omega)$ has only simple zeros, $\omega_i$, that 
is $f'(\omega_i)\neq 0$ for all $i$, 
$$
\delta(f(\omega)) =
\sum_i\frac{\delta(\omega -\omega_i)}{|f'(\omega_i)|}\; . 
$$

Applying this formula to (2.121) yields 
\begin{equation}
J_n(\omega) = \sum_i
\frac{2\pi\delta(\omega -\omega_{ni})}{| 1-E_n'(\omega_{ni}) |}\; ,
\label{2.122}
\end{equation}
where $\omega_{ni}$ are defined by the equation 
$$
\omega_{ni} = E_n(\omega_{ni}) - \mu ,\qquad (i=1,2,\ldots) 
$$
and 
$$
E_n'(\omega) = \frac{d}{d\omega} E_n(\omega) ;
\qquad E_n'(\omega_{ni}) \neq 1. 
$$

If we use the spectral function (2.122), the propagator (2.109) takes the
form 
\begin{equation}
G_n(\omega) = \sum_i \frac{1}{|1-E_n'(\omega_{ni})|}
\left [ \frac{1\pm n(\omega_{ni})}{\omega-\omega _{ni}+i0} \mp 
\frac{n(\omega_{ni})}{\omega -\omega_{ni}-i0} \right ] ,  
\label{2.123}
\end{equation}
which is simpler and easier to use than when the spectral function (2.118)
is inserted in (2.109). But one should remember that this last form is
appropriate only if decay processes may validly be neglected.

\subsection{Scattering Matrix}

In his very first paper about Quantum Mechanics, Heisenberg presented a
theory in terms of infinite matrices which, as far as possible, involved
only observable properties of the physical situation under examination.
Later, he introduced his {\it scattering matrix} appropriate for a
situation in which one or more particles came into a "black box" and
emerged in a different direction or as different species. He defined a
matrix which connected the intial to the final state of the system. In
this section, we use propagators to take into account the nature of the
"black box" and to calculate the scattering matrix. Our $S_\infty$,
defined in (2.139), is a particular case of Heisenberg's concept. The
ideas of this section play an important role in providing a practical
method of calculating Green's functions.

We envisage a system which is evolving under the control of a Hamiltonian 
$H$ and ask how it behaves if the Hamiltonian is modified by the addition
of a term, $\Delta H$, which may or may not be "small". Both $H$ and
$\Delta H$ are functionals of the field operators, which we indicate by 
\begin{equation}
H=H\left\{ \psi \right\} ,\qquad \Delta H=\Delta H\left\{ \psi \right\} .
\label{2.124}
\end{equation}
The new Hamiltonian is 
\begin{equation}
H_\Delta :=\tilde{H}+\Delta \tilde{H},  
\label{2.125}
\end{equation}
where 
\begin{equation}
\tilde{H}=H\left\{ \psi _\Delta \right\} ,\qquad \Delta \tilde{H}=\Delta
H\left\{ \psi _\Delta \right\}  
\label{2.126}
\end{equation}
are the same functionals as in (2.124) but functionals of new field
operators $\psi _\Delta $ which satisfy the Heisenberg equation 
\begin{equation}
i\frac \partial {\partial t}\psi _\Delta \left( x,t\right) =\left[ \psi
_\Delta ( x,t), H_\Delta(t) \right] .  
\label{2.127}
\end{equation}
Thus, the time dependence of the new field operators is given by 
\begin{equation}
\psi _\Delta \left( x,t\right) =U_\Delta ^{+}\left( t\right) \psi _\Delta
\left( x,0\right) U_\Delta \left( t\right) ,  
\label{2.128}
\end{equation}
where the evolution operator satisfies the equation 
$$
i\frac d{dt}U_\Delta \left( t\right) =H_\Delta U_\Delta \left( t\right) , 
$$
with $H_\Delta$ and $H_\Delta(t)$ related, analogously to (2.128), by 
\begin{equation}
H_\Delta \left( t\right) =U_\Delta ^{+}\left( t\right) H_\Delta U\left(
t\right) .  
\label{2.129}
\end{equation}
Since the statistics of particles is independent of the Hamiltonian, the 
new field operators satisfy the same commutation relations as the old
ones. In other words, 
$$
\left[ \psi_\Delta(x,t) ,\psi_\Delta(x',t) \right ] _{\mp}=0, 
$$
and 
$$
\left [ \psi_\Delta(x,t) ,\psi_\Delta^\dagger(x',t) \right ]_{\mp}
= \delta(x-x') . 
$$

Since $\tilde{H}$ will be assumed to be self--adjoint and we want the
creation operators $\psi_\Delta^\dagger(x,t)$ to satisfy the same
equation (2.127) as $\psi_\Delta(x,t)$, we shall also assume that 
$\Delta\tilde{H}$ is self--adjoint.
Thus 
\begin{equation}
\Delta \tilde{H}^{+}=\Delta \tilde{H}.  
\label{2.130}
\end{equation}
This imposes a restriction on $\Delta\tilde{H}$ so that if, for example, 
\begin{equation}
\Delta \tilde{H}:=
-\int \psi_\Delta^\dagger(x) \Delta \mu(x,x',t) \psi_\Delta(x') dxdx',
\label{2.131}
\end{equation}
in which $\psi_\Delta(x):=\psi_\Delta(x,0)$ and $\Delta\mu$ is a
complex--valued function, then (2.130) implies that 
\begin{equation}
\Delta \mu ^{*} (x,x',t) = \Delta \mu (x,x',t) .  
\label{2.132}
\end{equation}

Corresponding to the new field operators, new propagators can be defined
just as the old ones were in (2.3), (2.11), and (2.13). Thus, for the
one--
and two--particle propagators we have, respectively 
$$
\tilde{G}(12) := - i <\hat{T}\psi_\Delta(1)
\psi_\Delta^\dagger(2) >,
$$
\begin{equation}
\label{2.133} 
\tilde{G}_2(1234) : = - <\hat{T}\psi_\Delta(1) \psi_\Delta(2) 
\psi_\Delta^\dagger(3) \psi_\Delta^\dagger(4) >. 
\end{equation}
The statistical averaging in (2.133) does not presuppose that the system
is in equilbrium. It is taken with respect to the system density matrix,
$\hat{\rho}(t)$, prescribed by $\hat{\rho}(0)$ at initial time zero.

As compared to the equations of motion for the old propagators, those for
the new ones, (2.133), contain an additional term which results from the
commutation of $\psi_\Delta$ with 
$$
\Delta \tilde{H}\left( t\right ) =U_\Delta^{+}\left( t\right ) \Delta 
\tilde{H }U_\Delta \left( t\right ) . 
$$
For the one--particle propagator $\tilde{G}$, the equation of motion takes
the form 
\begin{equation}
i\frac \partial {\partial t_1}\tilde{G}\left( 12\right) =\delta \left(
12\right) -i<\hat{T}\left[ \psi _\Delta \left( 1\right) ,H_\Delta \left(
t_1\right) \right] \psi _\Delta ^{\dagger}\left( 2\right) >.  
\label{2.134}
\end{equation}
If we take the additive term in the form (2.131) we find 
$$
\left[ \psi _\Delta \left( x,t\right) ,\Delta \tilde{H}\left( t\right)
\right] =-\int \Delta \mu ( x,x',t) \psi_\Delta(x',t) dx', 
$$
and obtain 
$$
\left[ i\frac \partial {\partial t_1}-K\left( \nabla _1\right) +\mu
\left( 1\right) \right] \tilde{G}\left( 12\right) 
 + \int \Delta \mu \left( x_1,x_3,t_1\right) \delta \left( t_1-t_3\right) 
\tilde{G}\left( 32\right) d\left( 3\right) \mp
$$
\begin{equation}
\mp i\int \Phi \left( 13\right) \tilde{G}_2\left( 1332\right) d\left(
3\right) = \delta \left( 12\right) . 
\label{2.135}
\end{equation}

Notice that when $\Delta\mu\rightarrow 0$, so $\Delta\tilde{H}$ tends to
zero, 
$$
\psi _\Delta \left( x,t\right) \rightarrow \psi \left( x,t\right) ,
\qquad H_\Delta \rightarrow \tilde{H}\rightarrow H, 
$$
while (2.135) becomes (2.65). Since we have not assumed that
$\Delta\tilde{H}$ is small, the equation (2.135) for the propagator is
valid for any $\Delta\tilde{H}$ of the form (2.131).

We now introduce the {\em scattering matrix} 
\begin{equation}
S\left( t_1,t_2\right) :=\hat{T}\exp \left\{ -i\int_{t_2}^{t_1}\Delta
H\left( t\right) dt\right\} ,  
\label{2.136}
\end{equation}
in which 
$$
\Delta H\left( t\right) =U^{+}\left( t\right) \Delta HU\left( t\right) , 
$$
and the term
\begin{equation}
\Delta H:=-\int \psi^{\dagger}\left( x\right) 
\Delta \mu \left( x,x',t\right) \psi \left( x'\right) dxdx'
\label{2.137}
\end{equation}
has the same form as $\Delta\tilde{H}$ in (2.131) but contains the
original field operators $\psi$ and $\psi^\dagger$.

The scattering matrix, $S(t_1,t_2)$, has the following
properties: 
$$
S(t,t)=1,\qquad \lim_{\Delta H\rightarrow 0}S\left( t_1,t_2\right) = 1 ,
$$
\begin{equation}
S(t_1,t_2)S\left( t_2,t_3\right) = S\left( t_1,t_3\right) ,  
\label{2.138}
\end{equation}
$$
S^{+}\left( t_1,t_2\right) = S^{-1}\left( t_1,t_2\right) = 
S\left( t_2,t_1\right) . 
$$
The limits of integration, $t_1$ and $t_2$ in (2.136), can be any real
numbers. In particular we may define the balanced double limit 
\begin{equation}
S_\infty :=\lim_{t\rightarrow +\infty }S\left( t,-t\right) =S\left( +\infty
,-\infty \right) ,  
\label{2.139}
\end{equation}
giving 
\begin{equation}
S_\infty :=\hat{T}\exp \left\{ -i\int_{-\infty }^{+\infty }\Delta H\left(
t\right) dt\right\} .  
\label{2.140}
\end{equation}
By considering the time--ordered product of $S_\infty$ with the field
operator $\psi(1)$, we infer 
$$
\hat{T}S_\infty \psi \left( 1\right) =S\left( +\infty ,t_1\right) \psi
\left( 1\right) S\left( t_1,-\infty \right) . 
$$

In a similar manner, we can obtain the time--ordered product of $S_\infty$
with several field operators. For example, 
$$
\hat{T}S_\infty\psi(1)\psi^\dagger(2) =
\Theta(t_{12}) S(+\infty,t_1) \psi(1)
S(t_1,t_2) \psi^\dagger(2) S(t_2,-\infty) \pm
$$
$$
\pm \Theta(t_{21}) S(+\infty,t_2) \psi^\dagger(2) S(t_2,t_1) 
\psi(1) S(t_1,-\infty) .
$$

With the help of the scattering matrix we can express the new propagators 
in  (2.133) by means of the old field operators: 
$$
\tilde{G}\left( 12\right) = -i\frac{<\hat{T}S_\infty \psi \left(
1\right)
\psi ^{\dagger}\left( 2\right) >}{<S_\infty >}, 
$$
\begin{equation}
\tilde{G}_2\left( 1234\right) = -\frac{<\hat{T}S_\infty 
\psi \left( 1\right) \psi (2)\psi ^{\dagger}\left( 3\right) 
\psi^{\dagger}\left( 4\right) >}{ <S_\infty >}.  
\label{2.141}
\end{equation}
Similarly, the propagator, $\tilde{G}_n$, of order $n$, can be expressed
by means of the field operators as 
\begin{equation}
\tilde{G}_n\left( 1\ldots 2n\right) =
\left( -i\right ) ^n\frac{<\hat{T}S_\infty\psi \left( 1\right ) \ldots
\psi \left( n\right ) \psi^{\dagger}\left( n+1\right )
\ldots \psi^{\dagger}\left( 2n\right ) >}{<S_\infty >} \; .
\label{2.142}
\end{equation}

When $\psi$ in (2.13) is replaced by $\psi_\Delta$, we obtain (2.142). The
proof is obtained by considering their equations of motion. We illustrate
this in the case of the first order propagator in (2.141). Differentiate
$\tilde{G}$ with respect to $t_1$ and take into account the relations 
$$
i\frac \partial {\partial t_1}S\left( +\infty,t_1\right) = -S\left(
+\infty ,t_1\right) \Delta H\left( t_1\right) , 
$$
$$
i\frac \partial {\partial t_1}S(t_1,t_2) = \Delta H\left( t_1\right)
S\left( t_1,t_2\right) , 
$$
$$
i\frac \partial {\partial t_1}S\left( t_2,t_1\right) = 
-S\left( t_2,t_1\right) \Delta H\left( t_1\right) , 
$$
$$
i\frac \partial {\partial t_1}S\left( t_1,-\infty \right) =
\Delta H\left( t_1\right) S\left( t_1,-\infty \right) .
$$
From (2.141) we then find 
$$
i\frac \partial {\partial t_1}\tilde{G}\left( 12\right) =\delta \left(
12\right) -i\frac{<\hat{T}S_\infty \left[ \psi \left( 1\right) ,H\left(
t_1\right) +\Delta H\left( t_1\right) \right] \psi^{\dagger}\left( 2\right) 
>}{ <S_\infty >}\; . 
$$
This is seen to reduce to (2.135), when we recall the definition of $H(t)$
and note that (2.137) implies that 
$$
\left[ \psi \left( x,t\right) ,\Delta H\left( t\right) \right] =
- \int \Delta \mu(x,x',t) \psi(x',t) dx'. 
$$

It is important to realize that the representation (2.141)
for the propagators (2.133) is valid for arbitary shifts $\Delta H$ of the
Hamiltonian (2.125) whether or not the system is in equilibrium. It is
frequently assumed that $H$ is the Hamiltonian for non--interacting
particles while $\Delta H$ describes their interaction. In this case, 
(2.141) is referred to as the {\it interaction representation} for the
Green's functions.

Consider a particular case for which $\Delta H$ is so small that we may
restrict ourselves to terms which are linear in 
\begin{equation}
\Delta G\left( 12\right) :=\tilde{G}\left( 12\right) -G\left( 12\right) .
\label{2.143}
\end{equation}
For such an approximation, linear in $\Delta H$, the scattering
matrix\thinspace (2.140) is 
\begin{equation}
S_\infty \simeq 
1-i\hat{T}\int_{-\infty }^{+\infty }\Delta H\left( t\right) dt.  
\label{2.144}
\end{equation}
From (2.141), it follows that 
$$
\tilde{G}\left( 12\right) \simeq G\left( 12\right) + iG\left( 12\right)
\int_{-\infty }^{+\infty }<\hat{T}\Delta H\left( t\right) > dt -
$$
$$
-\int_{-\infty }^{+\infty } < \hat{T}\Delta H\left( t\right) \psi \left(
1\right) \psi ^{\dagger}\left( 2\right) >dt.
$$
If $\Delta H$ has the form (2.137), then 
$$
< \hat{T}\Delta H\left( t\right) >=\mp \int \Delta \mu \left(
x_3,x_4,t\right) \delta \left( t_3-t\right) \delta \left( t_4-t\right)
G\left( 43\right) d\left( 34\right) , 
$$
$$
<\hat{T}\Delta H\left( t\right) \psi \left( 1\right)
\psi^{\dagger}\left( 2\right) >= 
\pm \int \Delta \mu \left( x_3,x_4,t\right) \delta \left( t_3-t\right)
\delta \left( t_4-t\right) G_2\left( 1432\right) d\left( 34\right) .
$$
Therefore the difference (2.143) becomes 
$$
\Delta G\left( 12\right) = 
\pm \int \left[ G\left( 12\right) G\left( 43\right) -G_2\left( 1432\right)
\right] \Delta \mu \left( x_3,x_4,t_3\right) \delta \left( t_3-t_4\right)
d\left( 34\right) .
$$
This last expression, with the notation 
\begin{equation}
\Delta \mu \left( 12\right) :=\Delta \mu \left( x_1,x_2,t_1\right) \delta
\left( t_1-t_2\right) ,  
\label{2.145}
\end{equation}
takes the form 
\begin{equation}
\Delta G\left( 12\right) \simeq 
\pm \int \left[ G\left( 12\right) G\left(43\right) -
G_2\left( 1432\right) \right] \Delta \mu \left( 34\right) d\left(34\right) .  
\label{2.146}
\end{equation}

Define the variational derivative 
\begin{equation}
\frac{\delta G\left( 12\right) }{\delta \mu \left( 34\right) }:=
\lim_{\Delta\mu \rightarrow 0}
\frac{\Delta G\left( 12\right) }{\Delta \mu \left(34\right) }.  
\label{2.147}
\end{equation}
Then (2.146) yields the {\it variational representation} 
\begin{equation}
G_2\left( 1234\right) =G\left( 14\right) G\left( 23\right) \mp \frac{\delta
G\left( 14\right) }{\delta \mu \left( 32\right) },  
\label{2.148}
\end{equation}
for the two--particle propagator.

If $\Delta\mu$ is diagonal in $x$, so that 
$$
\Delta \mu \left( x,x^{\prime },t\right) =\Delta \mu \left( x,t\right)
\delta \left( x-x^{\prime }\right) , 
$$
then (2.146) gives 
\begin{equation}
\Delta G\left( 12\right) \simeq \pm \int \left[ G(12)G\left( 33\right)
-G_2\left( 1332\right) \right] \Delta \mu \left( 3\right) d\left( 3\right) ,
\label{2.149}
\end{equation}
where 
$$
\Delta \mu \left( 1\right) =\Delta \mu \left( x_1,t_1\right) . 
$$
Defining the variational derivative 
\begin{equation}
\frac{\delta G\left( 12\right) }{\delta \mu \left( 3\right) }:=
\lim_{\Delta\mu \rightarrow 0}
\frac{\Delta G\left( 12\right) }{\Delta \mu \left(3\right) },  
\label{2.150}
\end{equation}
we obtain the variational representation 
\begin{equation}
G_2\left( 1332\right) = G\left( 12\right) G\left( 33\right) \mp
\frac{\delta G\left( 12\right) }{\delta \mu \left( 3\right) }
\label{2.151}
\end{equation}
for the two--particle propagator. Note that this has precisely the form
which we need in the evolution equation (2.65). 

\subsection{Perturbation Theory}

Using the definition of the scattering matrix it is straightforward to
develop perturbation theory for Green's functions.

Suppose that the {\it unperturbed Hamiltonian} in (2.124) is $H\equiv H_0$ 
with 
\begin{equation}
H_0=\int \psi^{\dagger}\left( x\right) \left[ K\left( \nabla \right) -\mu 
\left( x,t\right) \right] \psi \left( x\right) dx,  
\label{2.152}
\end{equation}
where $\mu\left( x,t\right)$ may include the chemical potential, external
fields and some self--consistent mean--fields. It might, for example,
include the Hartree potential (2.78). We assume a perturbation,
corresponding to a two particles interaction, in the form 
\begin{equation}
\Delta H=\frac 12\int \psi^{\dagger}\left( x\right) \psi^{\dagger}
\left( x'\right) V\left( x,x'\right) 
\psi \left( x' \right) 
\psi\left( x\right) dxdx'.  
\label{2.153}
\end{equation}

Denote the {\it unperturbed Green's function} by $G_0$. Employing the
commutator 
$$
\left[ \psi \left( 1\right) ,H_0\left( t_1\right) \right] =\left\{ K\left(
\nabla _1\right) -\mu \left( 1\right) \right\} \psi \left( 1\right) , 
$$
we obtain the equation of motion for the {\it unperturbed} propagator,
$G_0$:
\begin{equation}
\left[ i\frac \partial {\partial t_1}-K\left( \nabla _1\right) +\mu \left(
1\right) \right] G_0\left( 12\right) =\delta \left( 12\right) .
\label{2.154}
\end{equation}
This equation can be solved by the method of expansion in wave functons,
discussed in Section 2.4.

We proceed as follows in order to obtain perturbative corrections to $G_0$.
Expand the scattering matrix (2.140) in the series 
\begin{equation}
S_\infty = 1 + \sum_{k=1}^\infty 
\frac{\left( -i\right) ^k}{k!}\int_{-\infty}^{+\infty }
\hat{T}\prod_{p=1}^k\Delta H\left( t_p\right) dt_p.
\label{2.155}
\end{equation}
Truncating this series at order $k$ and substituting it into the 
interaction representation (2.141) gives rise to the $k$--th order
approximation, $G^{(k)}$, to the propagator $\tilde{G}$. This
approximation includes powers of $\Delta H$ up to the $k$--th.
Clearly, the zero--th order approximation is the unperturbed propagator 
$$
G^{\left( 0\right) }\left( 12\right) :=G_0\left( 12\right) . 
$$
The $k$--th order approximation, $G_n^{(k}$, for the $n$--particle 
propagator, $\tilde{G}_n$, can be obtained in the same way.

Taking account of the equality 
$$
\int_{-\infty }^{+\infty }<\tilde{T}\Delta H\left( t\right) >dt=-\int
G_2^{\left( 0\right) }\left( 1221\right) \Phi \left( 12\right) d\left(
12\right) , 
$$
the first--order approximation for the one--particle propagator becomes 
$$
G^{\left( 1\right) }\left( 12\right) = G_0\left( 12\right) +
$$
\begin{equation}
+ i\int \left[ G_3^{\left( 0\right) }\left( 134432\right) -G_0\left(
12\right) G_2^{\left( 0\right) }\left( 3443\right) \right] \Phi \left(
34\right) d\left( 34\right) .  
\label{2.156}
\end{equation}
Here, the $n$--particle propagators, $G_n^{( 0)}$, satisfy
equations of motion involving the unperturbed Hamiltonian, $H_0$.

The equation of motion for $G_2^{(0)}$ is as follows: 
\begin{equation}
\left[ i\frac \partial {\partial t_1}-K\left( \nabla _1\right) +\mu \left(
1\right) \right] G_2^{\left( 0\right) }\left( 1234\right) =\delta \left(
14\right) G_0\left( 23\right) \pm \delta \left( 13\right) G_0\left(
24\right) .  
\label{2.157}
\end{equation}
Since the solution of a first--order differential equation is determined
by the intial values, by comparing this equation with (2.154) we may 
conclude that 
\begin{equation}
G_2^{\left( 0\right) }\left( 1234\right) =G_0\left( 14\right) G_0\left(
23\right) \pm G_0\left( 13\right) G_0\left( 24\right) .  
\label{2.158}
\end{equation}
In a similar manner, from the equation of motion for $G_3^{(0)}$
together with (2.154) and (2.157), we conclude that 
$$
G_3^{\left( 0\right) }\left( 123456\right) = G_0\left( 16\right)
G_2^{\left( 0\right) }\left( 2345\right)   \pm
$$
\begin{equation}
\pm G_0\left( 15\right) G_2^{\left( 0\right) }\left( 2346\right)
+G_0\left( 14\right) G_2^{\left( 0\right) }\left( 2356\right) .
\label{2.159}
\end{equation}
From (2.158) it follows that $G_3^{(0)}$ can be expressed as a sum of six
terms involving only the first--order propagator $G_0$. We symbolize this
by 
$$
G_3^{\left( 0\right) }=\sum_P\left( \pm 1\right) ^PG_0G_0G_0,
$$
where $P$ runs over the group of permutations of $\{ 4,5,6\}$. By 
$(\pm 1)^P$ we mean $+1$ for bosons and, for fermions, the parity, 
$+ 1$ or $-1$, of the permutation $P$. Each $G_0$ must contain a variable
from the first three and from the second three variables in $G_3^{(0)}$.
The terms which occur can be obtained from $G_0(16)G_0(25)G_0(34)$, in
which $\{ 1,2,3\}$ are in natural order and $\{ 4,5,6\}$ are in reversed
order, by the six possible permutations of $\{4,5,6\}$.

Similarly, any $n$--particle unperturbed propagator can be presented as a
sum 
\begin{equation}
G_n^{\left( 0\right) }=\sum_P\left( \pm 1\right)^PG_0G_0\ldots G_0.
\label{2.160}
\end{equation}
The sum will contain $n!$ terms obtained by a rule which is the obvious
generalization of the preceding example. In quantum field theory, this
result is usually referred to as {\it Wick's theorem}.

Using 
(2.158) and (1.159) in (2.156) leads us to 
\begin{equation}
G^{\left( 1\right) }\left( 12\right) =G_0\left( 12\right) \pm i\int
G_0\left( 13\right) \Phi \left( 34\right) G_2^{\left( 0\right) }\left(
3442\right) d\left( 34\right) ,  
\label{2.161}
\end{equation}
for the first--order approximation to $G(12)$. The second--order
approximation is given by 
$$
G^{(2)}(12) = G^{(1)}(12) + \frac{1}{2} \int_{-\infty}^{+\infty}
< \hat{T}\left \{ \left [ G_0(12)
+ i \psi(1) \psi^\dagger(2) \right ] \times \right.
$$
$$
\times \left. \left [ \Delta H(t_3) - 
2 < \Delta H(t_3) > \right ] \Delta H(t_4) \right \}>dt_3dt_4, 
$$
in which we need to substitute (2.153) and make use of (2.160).

In a similar manner we can obtain any $k$--th order approximation for an 
{\it n}-particle propagator $G_n^{(k)}$.

Another approach to perturbation theory uses the Dyson equation (2.72).
Setting 
\begin{equation}
\Sigma _0\left( 12\right) =0,  
\label{2.162}
\end{equation}
we find that the inverse propagator (2.74) is 
\begin{equation}
G_0^{-1}\left( 12\right) =\left[ i\frac \partial {\partial t_1}-K\left(
\nabla _1\right) +\mu \left( 1\right) \right] \delta \left( 12\right) ,
\label{2.163}
\end{equation}
and the kernel (2.73) becomes 
$$
B\left( 12\right) =
\int G_0\left( 13\right) \Sigma \left( 32\right) d\left( 3\right) . 
$$

The Dyson equation, (2.72), takes the form 
\begin{equation}
G\left( 12\right) =G_0\left( 12\right) +\int G_0\left( 13\right) \Sigma
\left( 34\right) G\left( 42\right) d\left( 34\right) .  
\label{2.164}
\end{equation}
Successive approximations to $G$ are obtained by iterating (2.164) 
according to the scheme 
\begin{equation}
G^{\left( k\right) }\rightarrow \Sigma ^{\left( k+1\right) }\rightarrow
G^{\left( k+1\right) }.  
\label{2.165}
\end{equation}
The first--order approximation gives 
\begin{equation}
G^{\left( 1\right) }\left( 12\right) =G_0\left( 12\right) +\int G_0\left(
13\right) \Sigma ^{\left( 1\right) }\left( 34\right) G_0\left( 42\right)
d\left( 34\right) ,  
\label{2.166}
\end{equation}
in which 
$$
\Sigma ^{\left( 1\right) }\left( 12\right) =\pm i\int \Phi \left(
13\right) G_2^{\left( 0\right) }\left( 1334\right) G_0^{-1}\left( 42\right)
d\left( 34\right) .
$$
With $G_2^{(0)}$ given by (2.158), we have 
\begin{equation}
\Sigma ^{\left( 1\right) }\left( 12\right) =i\Phi \left( 12\right)
G_0\left( 12\right) \pm i\delta \left( 12\right) \int \Phi \left(
13\right) G_0\left( 33\right) d\left( 3\right) .  
\label{2.167}
\end{equation}
Therefore (2.166) coincides with (2.161).

Note that (2.167) has the same apparent structure as the Hartree--Fock
self--energy (2.76) but is quite  different since $\Sigma^{(1)}$ involves
$G_0$, while $\Sigma _{HF}$ involves $G$.

By (2.66) the Dyson equation (2.164) can be transformed into 
\begin{equation}
G\left( 12\right) =G_0\left( 12\right) \pm i\int G_0\left( 13\right) \Phi
\left( 34\right) G_2\left( 3442\right) d\left( 34\right) .  
\label{2.168}
\end{equation}

There is an equivalent approach to obtaining successive approximations to 
$G$ in which we adopt the recursive scheme 
\begin{equation}
G^{\left( k\right) }\rightarrow 
G_2^{\left( k\right) }\rightarrow G^{\left( k+1\right) }.  
\label{2.169}
\end{equation}
Substituting $G_2^{(0)}$ from (2.158) in the right--hand side of equation
(2.168), again gives us (2.161). In order to find $G^{(2)},$ we need 
$G_2^{(1)}$. This can be obtained from either 
(2.141) or (2.65), giving
$$
G_2^{\left( 1\right ) }\left( 1234\right)  =
G_2^{\left( 0\right )}\left( 1234\right )  +
$$
$$
+ i\int_{-\infty}^{+\infty} < 
\hat{T}\left[ G_2^{\left( 0\right )}\left(1234\right ) +
\psi \left( 1\right ) \psi \left( 2\right) \psi^{\dagger}\left(3\right ) 
\psi^{\dagger}\left( 4\right ) \right ] \Delta H\left( t\right ) >dt. 
$$
With the help of (2.153), this yields 
$$
G_2^{\left( 1\right) }\left( 1234\right) =
G_2^{\left( 0\right) }\left( 1234\right)   +
$$
\begin{equation}
\label{2.170} 
+ i\int \left[ G_4^{\left( 0\right) }\left( 12566534\right) -G_2^{\left(
0\right) }\left( 1234\right) G_2^{\left( 0\right) }\left( 5665\right)
\right] \Phi \left( 56\right) d\left( 56\right) . 
\end{equation}
The equation of motion for $G_4^{(0)}$, together with the Wick's
theorem (2.160), yields 
$$
G_4^{\left( 0\right) }\left( 12345678\right) = G_0\left( 18\right)
G_3^{\left( 0\right) }\left( 234567\right) \pm G_0\left( 17\right)
G_3^{\left( 0\right) }\left( 234568\right) + 
$$
$$
+ G_0\left( 16\right) G_3^{\left( 0\right) }\left( 234578\right) \pm
G_0\left( 15\right) G_3^{\left( 0\right) }\left( 234678\right) .
$$
Invoking (2.159), we obtain 
$$
G_4^{\left( 0\right) }\left( 12345678\right) = 
$$
$$
= G_2^{\left( 0\right) }\left( 1278\right) G_2^{\left( 0\right) }\left(
3456\right) \pm G_2^{\left( 0\right) }\left( 1378\right) G_2^{\left(
0\right) }\left( 2456\right) +
$$
$$
+ G_2^{\left( 0\right) }\left( 1478\right) G_2^{\left( 0\right) }\left(
2356\right) +G_2^{\left( 0\right) }\left( 1256\right) 
G_2^{\left( 0\right)}\left( 3478\right)  \pm 
$$
$$
\pm G_2^{\left( 0\right) }\left( 1356\right)G_2^{\left( 0\right) }\left(
2478\right) +G_2^{\left( 0\right) }\left( 1456\right) G_2^{\left( 0\right)
}\left( 2378\right) ,
$$
where $G_2^{(0)}$ is given by (2.158).

The two approaches to perturbation theory, are equivalent to one another,
and applicable whether or not the system is in equilibrium, as long as 
$\Delta H$ can be considered small. Since Green's functions may be
interpreted as generalized functions, convergence of the perturbation
sequence is to be understood as convergence of functionals, -- for example,
as the convergence of the expected values for some local observables.

It may happen that terms in the perturbation expansion diverge when they
involve integrals of products of several propagators. We therefore
illustrate, by means of an example, two methods, which we call {\it 
implicit} and {\it explicit}, of avoiding such divergences [59].

Consider the Fourier transform of (2.163) for a system in equilibrium, 
\begin{equation}
G_0^{-1}\left( x,x',\omega \right) =\left[ \omega +\mu -H\left(
x\right) \right] \delta \left( x-x'\right) ,  
\label{2.171}
\end{equation}
where 
\begin{equation}
H\left( x\right) =K\left( \nabla \right) +U\left( x\right)   
\label{2.172}
\end{equation}
is hermitian. The propagator corresponding to (2.171) can be expanded in 
the form 
\begin{equation}
G_0\left( x,x',\omega \right) =
\sum_nG_n(\omega) \psi_n( x) \psi_n^{*}( x')   
\label{2.173}
\end{equation}
with respect to one--particle wave functions satisfying the equation 
\begin{equation}
H\left( x\right) \psi _n\left( x\right) =E_n\psi _n\left( x\right) ,
\label{2.174}
\end{equation}
with expansion coefficients 
\begin{equation}
G_n\left( \omega \right) =\frac 1{\omega -E_n+\mu }  
\label{2.175}
\end{equation}
defined on the complex $\omega$--plane. As was explained in Section 2.4,
on the real $\omega$--axis 
\begin{equation}
G_n\left( \omega \right) =
\frac{1\pm n\left( E_n\right) }{\omega -E_n+i0} \mp 
\frac{n\left( E_n\right) }{\omega -E_n-i0}.  
\label{2.176}
\end{equation}
The eigenvalues $E_n$ are real since we assumed that $H\left(x\right)$ is
hermitian.

Starting from the zero order approximation (2.173) for the
propagator, in successive approximations we meet a variety of integrals 
with integrands containing products of factors such as (2.176). For
example: 
$$
I_{mn} = \pm i\int_{-\infty }^{+\infty }e^{+i\omega 0}G_m\left( \omega
\right) G_n\left( \omega \right) \frac{d\omega }{2\pi },  
$$
$$
I_{mn}' = \pm i\int_{-\infty }^{+\infty }e^{+i\omega 0}\omega
G_m\left( \omega \right) G_n\left( \omega \right) \frac{d\omega }{2\pi }, 
$$
$$
I_{m\ell n} = \pm i\int_{-\infty }^{+\infty }e^{+i\omega 0} 
G_m(\omega) G_\ell(\omega) G_n(\omega) \frac{d\omega }{2\pi},  
$$
\begin{equation}
\label{2.177} 
I_{m\ell n}' = \pm i\int_{-\infty }^{+\infty}e^{+i\omega 0}
\omega G_m\left( \omega \right) G_\ell \left( \omega \right) 
G_n\left(\omega \right) \frac{d\omega }{2\pi }.  
\end{equation}
The integrals in (2.177) are symmetric with respect to
permutation of their indices.

When two or more of the indices in (2.177) coincide, a direct integration
of $G_n^2(\omega)$, or in general of $G_n^k(\omega)$, would lead to a
divergence. This would be due to the fact that the propagator $G_n(\omega)$ 
must be treated as a {\it generalized function} or {\it distribution} and
products of distributions with coincident singualarities are not uniquely
determined. They require additional definition.

There are two approaches, so--called {\it implicit} and {\it explicit},
which avoid the divergences [59].

$(i)$ For the {\it implicit} approach, instead of defining $G_n^k(\omega)$ 
directly we replace it by the limit of a sequence. We first calculate the
integrals in (2.177) for distinct indices, giving 
$$
I_{mn} = A_{mn} + A_{nm} \; ,
$$
$$
I_{mn}^{\prime }  = E_mA_{mn} + E_nA_{nm}\; ,
$$
$$
I_{m\ell n}  = B_{m\ell n} + B_{\ell nm} + B_{nm\ell }\; ,
$$
$$ 
I_{m\ell n}' = E_mB_{m\ell n} + E_\ell B_{\ell nm} + E_nB_{nm\ell },
$$
where 
$$
A_{mn} = \frac{n\left( E_m\right) }{E_m-E_n}, 
$$
and 
$$
B_{m\ell n}=\frac{n\left( E_m\right) \left[ 1\pm n\left( E_\ell \right)
\right] \left[ 1\pm n\left( E_n\right) \right] \pm \left[ 1\pm n\left(
E_m\right) \right] n\left( E_\ell \right) n\left( E_n\right) }{\left(
E_m-E_\ell \right) \left( E_m-E_n\right) }. 
$$
We then define the integrals (2.177), for coincident indices, by the
following limiting procedure: 
$$
I_{nn} := \lim_{m\rightarrow n}I_{mn}=-\beta n\left( E_n\right) \left[
1\pm n\left( E_n\right) \right] , 
$$
$$
I_{nn}' := \lim_{m\rightarrow n}I_{mn}^{\prime }=n\left(
E_n\right) +E_nI_{nn},  
$$
\begin{equation}
I_{n\ell n} :=\lim_{m\rightarrow n}I_{m\ell n}=
\frac{I_{nn}-I_{n\ell }}{E_n-E_\ell },  
\label{2.178} 
\end{equation}
$$
I_{n\ell n}' := \lim_{m\rightarrow n}I_{m\ell n}' =
\frac{ E_nI_{nn}-E_\ell I_{n\ell }}{E_n-E_\ell }, 
$$
$$
I_{nnn} := \lim_{\ell \rightarrow n}I_{n\ell n}=\frac{\beta ^2}2n\left(
E_n\right) \left[ 1\pm n\left( E_n\right) \right] \left[ 1\pm 2n\left(
E_n\right) \right] ,  
$$
$$
I_{nnn}' : =\lim_{\ell \rightarrow n}I_{n\ell n}' = I_{nn}+E_nI_{nnn}.  
$$
That is, we let the indices coincide only after integrating but not in
the integrands.

$(ii)$ For the {\it explicit} approach, we first define the 
$k$--th power of $G_n$ as 
\begin{equation}
G_n^k\left( \omega \right) :=\frac{1\pm n\left( \omega \right) }{\left(
\omega -E_n+i0\right) ^k}\mp \frac{n\left( \omega \right) }{\left( \omega
-E_n-i0\right) ^k}.  
\label{2.179}
\end{equation}
We then substitute the expression (2.179) directly into the integrands. If
under the integral, it is multiplied by a function, $f(\omega)$ say, which 
is differentiable of order $k$ or more and vanishes sufficiently rapidly
at infinity, then integration by parts gives 
$$
\pm i\int_{-\infty }^{+\infty }e^{+i\omega 0}f\left( \omega \right)
G_n^k\left( \omega \right) \frac{d\omega }{2\pi }=
\frac 1{\left( k-1\right)!}\frac{d^{k-1}}{dE_n^{k-1}}
\left[ f\left( E_n\right) n\left( E_n\right) \right] . 
$$
Using the relations 
$$
\frac d{d\omega }n\left( \omega \right) =-\beta n\left( \omega \right)
\left[ 1\pm n\left( \omega \right) \right] , 
$$
and 
$$
\frac{d^2}{d\omega ^2}n\left( \omega \right) =-\beta \left[ 1\pm 2n\left(
\omega \right) \right] \frac d{d\omega }n\left( \omega \right) ,
$$
it is straightforward to show that these two approaches to defining the
integrals (2.177) with coincident indices are equivalent.

Thus, by using either perturbation method, we obtain a sequence, 
$\{G^{(k)}\}$, of approximations for a propagator. With these we may then
find a sequence, $\{<{\hat A>}_k\}$, of expected values for an element,
${\hat A}$, of an algebra of observables. For any particular problem, it
will be the convergence properties of the sequences $<{\hat A}>_k$ that
determine convergence of our procedure.

\subsection{Excited States}

It frequently happens that we are more interested in the spectrum of the
quantum states of an equilibrium system than in its statistical states. As
we proved in Section 2.4, the spectrum is given by the poles of the
propagators in the $\omega$--representation and can be found from the
effective wave equation (2.105).

The {\it effective Hamiltonian} 
\begin{equation}
H\left( x,\omega \right) =H\left( x\right) +M\left( x,\omega \right)
\label{2.180}
\end{equation}
consists of two terms. The first is the Schr\"{o}dinger operator 
(2.172) and the second, called the {\it mass operator, }is an
integral operator the kernel of which is the self--energy, $\Sigma$: 
\begin{equation}
M\left( x,\omega \right) \varphi \left( x,\omega \right) :=
\int \Sigma \left( x,x',\omega \right) 
\varphi \left( x',\omega \right) dx' .  
\label{2.181}
\end{equation}
Thus the effective wave equation is 
\begin{equation}
\left[ H\left( x\right) + 
M\left( x,\omega \right) \right] \varphi_n\left( x,\omega \right) =
H_n\left( \omega \right) \varphi _n\left(x,\omega \right) .  
\label{2.182}
\end{equation}
This equation differs from the usual Schr\"{o}dinger equation in three
respects:

\begin{enumerate}
\item[1.]  It is an integro--differential equation;

\item[2.]  Since, in general, the effective Hamiltonian, (2.180), is not 
hermitian, the eigenvalues, $H_n(\omega)$, are not necessarily real;

\item[3.]  The self--energy is normally defined only by means of the
perturbation theory described in the previous section.
\end{enumerate}

Therefore in order to solve (2.182) it will be necessary to use
perturbation theory, but $(3)$ implies that the familiar method of solving
the Schr\"{o}dinger equation is not  applicable since this assumes that
the Hamiltonian is known {\it before} recursion for successive
approximations commences. We therefore attempt a solution by setting 
\begin{equation}
M\left( x,\omega \right) =
\sum_{k=0}^\infty \varepsilon ^kM^{\left( k\right)}\left( x,\omega \right)  
\label{2.183}
\end{equation}
in which $\varepsilon$ is a formal expansion parameter introduced to keep
track of the successive orders of approximation. At the end of the
calculation we let $\varepsilon \rightarrow 1$, hoping that the series
converges! Each $M^{(k)}$ is an integral operator such that 
$$
M^{\left( k\right) }\left( x,\omega \right) \varphi \left( x,\omega \right)
:=\int M^{\left( k\right) }\left( x,x',\omega \right) \varphi
\left( x',\omega \right) dx', 
$$
with the kernel, $M^{(k)}(x,x',\omega)$, defined by the corresponding
approximation for the self--energy. In the zero approximation, 
$\Sigma_0(12)=0$, we have 
\begin{equation}
M^{\left( 0\right) }\left( x,x',\omega \right) =
\Sigma _0\left( x,x',\omega \right) =0.  
\label{2.184}
\end{equation}

The first approximation $M^{(1)}$ is given by 
\begin{equation}
M^{\left( 1\right) }\left( x_1,x_2,\omega \right) =
\int_{-\infty }^{+\infty}\Sigma ^{\left( 1\right) }
\left( 12\right) e^{i\omega t}dt,  
\label{2.185}
\end{equation}
with self--energy (2.167) and $t\equiv t_{12}$. Employing the expansion
(2.173) for the zero--order propagator $G_0$, we find 
$$
i\int_{-\infty}^{+\infty}e^{+i\omega 0}G_0\left( x,x',\omega\right) 
\frac{d\omega}{2\pi}=
\pm \rho _1^{\left( 0\right) }\left(x,x'\right) , 
$$
where 
\begin{equation}
\rho _1^{\left( 0\right) }\left( x,x'\right) :=\sum_mn\left(
E_m\right) \psi _m\left( x\right) \psi _m^{*}\left( x'\right)
\label{2.186}
\end{equation}
is the $1$--matrix in zero approximation and $E_k$ and $\psi_k$ are
defined by (2.174). Thus (2.185) reduces to the expression 
\begin{equation}
M^{(1)}(x,x',\omega) = \pm \Phi(x,x')\rho_1^{(0)}( x,x')
+ \delta( x-x') \int \Phi(x,x'') \rho ^{(0)}(x'') dx'',  
\label{2.187}
\end{equation}
in which 
\begin{equation}
\rho ^{\left( 0\right) }\left( x\right) :=
\rho _1^{\left( 0\right) }\left(x,x\right) =
\sum_mn\left( E_m\right) \left| \psi _m\left( x\right) \right| ^2
\label{2.188}
\end{equation}
is the particle density. Note that (2.187) does not depend on $\omega$.

However, in higher approximations, that is for $k\geq 2$, the kernel 
\begin{equation}
M^{\left( k\right) }\left( x_1,x_2,\omega \right) =
\int_{-\infty }^{+\infty}\left[ \Sigma ^{\left( k\right) }
\left( 12\right) -\Sigma ^{\left(k-1\right) }\left( 12\right) \right] 
e^{i\omega t}dt  
\label{2.189}
\end{equation}
will, in general, depend on $\omega$.

The series (2.183) for the mass operator suggests that we should seek a
solution of the eigenvalue problem (2.182) in the form 
$$
H_n\left( \omega \right) = E_n+\sum_{k=1}^\infty \varepsilon
^kH_n^{\left( k\right) }\left( \omega \right) ,  
$$
\begin{equation}
\label{2.190}
\varphi _n\left( x,\omega \right) = \psi _n\left( x\right)
+\sum_{k=1}^\infty \varepsilon ^k\varphi _n^{\left( k\right) }\left(
x,\omega \right) .  
\end{equation}

The functions $\varphi_n^{(k)}$ can be expanded in the basis $\{\psi_n\}$, 
so that 
\begin{equation}
\varphi_n^{\left( k\right) }\left( x,\omega \right) =
\sum_mc_{mn}^{\left(k\right) }\left(\omega \right) \psi_m\left( x\right) ,
\label{2.191}
\end{equation}
and, since we assume that the $\psi_m$ are orthonormal, 
\begin{equation}
c_{mn}^{\left( k\right) }\left( \omega \right) =
\left( \psi _m,\varphi_n^{\left( k\right) }\right) ,\qquad k\geq 1.  
\label{2.192}
\end{equation}
There is an additional condition on these coefficients expressing the fact
that we want $\varphi_n(x,\omega)$ in 
(2.190) to be normalized, 
\begin{equation}
\left( \varphi _n,\varphi _n\right) =1.  
\label{2.193}
\end{equation}

Substitute the expansions (2.183) and (2.190) into (2.182) and equate like
powers of $\varepsilon$. In zero order, this gives (2.174) and in first 
order: 
\begin{equation}
\left [ H(x) - E_n \right ] \varphi_n^{(1)}(x,\omega) =
\left [ H_n^{(1)}(\omega) - M^{(1)}(x,\omega) \right ] \psi_n(x) .
\label{2.194}
\end{equation}
This can be considered as a nonhomogeneous equation defining 
$\varphi_n^{(1)}$. Using (2.191) we may transform (2.194) to 
\begin{equation}
c_{mn}^{\left( 1\right) }\left( \omega \right) \left( E_m-E_n\right) =
\delta_{mn}H_n^{\left( 1\right) }\left( \omega \right) -
M_{mn}^{\left( 1\right) }\left( \omega \right) ,  
\label{2.195}
\end{equation}
in which 
\begin{equation}
M_{mn}^{(k)}(\omega) :=
\int \psi_m^{*}(x) M^{(k)}(x,x',\omega) \psi_n( x') dxdx'.  
\label{2.196}
\end{equation}
The coefficients $c_{nn}^{(1)}(\omega)$, when $m=n$, are not specified by
(2.195) but can be obtained from the normalization condition  (2.193). If
we substitute in this condition the expansion (2.191) and equate equal
powers of $\varepsilon$, we find 
\begin{equation}
c_{nn}^{\left( 1\right) }\left( \omega \right) =\left( \psi _n,\varphi
_n^{\left( 1\right) }\right) =0.  
\label{2.197}
\end{equation}
It follows that the first order corrections to the eigenvalues and
eigenfunctions are 
$$
H_n^{\left( 1\right) }\left( \omega \right) = M_{nn}^{\left( 1\right)
}\left( \omega \right) ,  
$$
\begin{equation}
\label{2.198} 
\varphi_n^{\left( 1\right) }\left( \omega \right) = 
\sum_mc_{mn}^{\left(1\right) }\left( \omega \right) \psi _m\left( x\right) , 
\end{equation}
where 
\begin{equation}
c_{mn}^{\left( 1\right) }\left( \omega \right) =
-\left( 1-\delta_{mn}\right) 
\frac{M_{mn}^{\left( 1\right) }\left( \omega \right) }{E_m-E_n}.
\label{2.199}
\end{equation}

In the second order, we find 
$$
\left [ H(x) - E_n \right ] \varphi_n^{(2)}(x,\omega) =
\left [ H_n^{(1)}(\omega)  - M^{(1)}(x,\omega) \right ] 
\varphi_n^{(1)}(x,\omega) +
$$
\begin{equation}  
+ \left[ H_n^{(2)}(\omega) - M^{(2)}(x,\omega) \right ] \psi_n(x) .  
\label{2.200} 
\end{equation} 
Together with (2.191) this leads to 
$$ c_{mn}^{(2)}(\omega) \left ( E_m-E_n \right) = 
\delta _{mn}H_n^{\left( 2\right)}\left( \omega \right) +
c_{mn}^{\left( 1\right) }\left( \omega \right)
H_n^{\left( 1\right) }\left( \omega \right) - $$ 
\begin{equation} 
-\sum_\ell M_{m\ell }^{\left( 1\right) }\left (\omega\right ) c_{\ell
n}^{\left( 1\right)  }\left( \omega \right) -M_{mn}^{\left( 2\right)
}\left( \omega \right) .  
\label{2.201} 
\end{equation} 
From this we obtain
$$
H_n^{\left( 2\right) }\left( \omega \right) = -\sum_{m\neq n}
\frac{M_{mn}^{(1)}(\omega) M_{nm}^{(1)}(\omega)}{E_m-E_n} +
M_{nn}^{(2)}(\omega) ,
$$ 
\begin{equation}
\label{2.202} 
c_{mn}^{(2)}(\omega) = \sum_{\ell \neq n}
\frac{M_{m\ell}^{(1)}(\omega)M_{\ell n}^{(1)}(\omega)}
{( E_m-E_n)( E_\ell-E_n)} -
\frac{M_{mn}^{(1)}(\omega) M_{nn}^{(1)}(\omega)}{( E_m-E_n)^2} - 
\frac{M_{mn}^{(2)}(\omega)}{E_m-E_n}\; ,
\end{equation}
where in this last equation $m\neq n$. As before, $c_{nn}^{(2)}(\omega)$, 
with coincident indices, cannot be obtained from (2.201), but can be
deduced from the normalization requirement (2.193) which, to within a
phase factor, implies 
\begin{equation}
c_{nn}^{\left( 2\right) }\left( \omega \right) =
-\frac{1}{2}\sum_{m\neq n}
\left| \frac{M_{mn}^{(1)}(\omega)}{E_m-E_n} \right| ^2.  
\label{2.203}
\end{equation}

Another method of defining the second-order correction, 
$H_n^{(2)}(\omega)$, is to multiply  (2.200) by $\psi_n^{*}(x)$ and
integrate the resulting equation with respect to $x$. Noting that $H(x)$
is hermitian and using (2.107) leads to 
\begin{equation}
H_n^{\left( 2\right) }\left( \omega \right) =\left( \psi _n,M^{\left(
1\right) }\varphi _n^{\left( 1\right) }\right) +M_{nn}^{\left( 2\right)
}\left( \omega \right) ,  
\label{2.204}
\end{equation}
where $\varphi_n^{(1)}$ is defined as the solution of the nonhomogeneous
equation (2.194). Thus the correction, $H_n^{(2)}$, is expressed by means 
of quantities with the common label $n$. At first sight, (2.204) looks
simpler than (2.202) which involves a summation. However, this second
approach requires the solution of the differential equation (2.194) in
order to evaluate the scalar product on the right hand side of (2.204).
Which is the optimal approach will depend on the particular problem under
consideration.

The procedures sketched above may be continued to any desired order of
approximation. The approach based on the summation formulas, as in 
(2.198) and (2.202), is a version, in this context, of the standard {\it
Rayleigh--Schr\"{o}dinger perturbation theory}. The second approach, using
(2.204) and solving a differential equation is analogous to the {\it
Dalgarno-Lewis perturbation theory} [60,61] (see also [62,63]).

However as already mentioned, there are a number of differences between the
problem of calculating the spectrum of a one--particle operator and that
of obtaining the spectrum of excited states for a many--particle system.
The most obvious is that the effective wave equation  (2.182) contains the
mass operator which is not given {\it a priori} but is itself defined
recursively as the successive terms in (2.183) are obtained. Each
$M^{(k)}$ is expressed in terms of the propagators, $G_k^{(0)}$ or the
reduced density matrices, $\rho_k^{(0)}$. For example, $M^{(1)}$ contains
the density matrix $\rho_1^{(0)}$ of (2.186), while $M^{(2)}$ contains the
density matrix 
\begin{equation}
\rho _2^{\left( 0\right) }\left( x_1,x_2,x_3,x_4\right) =\mp
\lim_{t_i\rightarrow t}G_2^{\left( 0\right) }\left( 1234\right) ,
\label{2.205}
\end{equation}
where the limit is taken such that 
$$
t_4>t_3>t_1>t_2.
$$
For the one--particle propagator, with Fourier transform (2.98), we find 
$$
G_0\left( 12\right) = -i\sum_n\left\{ \Theta \left( t_{12}\right)
\left[
1\pm n\left( E_n\right) \right] \pm \Theta \left( -t_{12}\right)
n(E_n\right\} \times
$$
\begin{equation}
\times \psi _n\left( x_1\right) \psi _n^{*}\left( x_2\right) \exp
\left( -iE_nt_{12}\right) .  
\label{2.206}
\end{equation}
Substituting (2.206) into (2.158) we obtain the two--particle propagator
$G_2^{(0)}$. In view of (2.205), this defines the 2--matrix 
$$
\rho _2^{\left( 0\right) }\left( x_1,x_2,x_3,x_4\right) = 
\sum_{m,n}n(E_m)n\left( E_n\right) \psi _m\left( x_1\right) 
\psi_n\left( x_2\right)  \times
$$
\begin{equation}
\times \left[ \psi_n^{*}( x_3) \psi_m^{*}( x_4)
\pm \psi_m^{*}( x_3) \psi_n^{*}( x_4) \right] .
\label{2.207}
\end{equation}
Defining the {\it two--particle wave function} 
$$
\psi _{mn}\left( x_1,x_2\right) :=\frac 1{\sqrt{2}}\left[ \psi _m\left(
x_1\right) \psi _n\left( x_2\right) \pm \psi _n\left( x_1\right) \psi
_m\left( x_2\right) \right] ,
$$
and the {\it pairon density of states} 
$$
B_{mn}:=n\left( E_m\right) n\left( E_n\right) ,
$$
the zero--order 2--matrix takes the form 
\begin{equation}
\rho _2^{\left( 0\right) }\left( x_1,x_2,x_3,x_4\right) =
\sum_{m,n}B_{mn}\psi _{mn}\left( x_1,x_2\right)
\psi_{nm}^{*}\left( x_3,x_4\right) .  
\label{2.208}
\end{equation}

Another difference between the two eigenvalue problems arises from the fact
the Hamiltonian (2.180) is not necessarily hermitian. Not only can this
give rise to complex eigenvalues signalling the presence of decay, but it
also precludes the use of the virial or hypervirial theorems which are
widely employed in quantum mechanics. These theorems may be used in
situations, discussed in Section 2.4, in which decay is negligible and the
effective Hamiltonian can be approximated sufficiently accurately by
a hermitian operator $H=K+U$.

If $K=-p^2/2m$ and ${\bf p}=i\nabla$, then if $H=K+U$ is hermitian with
eigenfucntions $\psi_n,$ then the {\it virial theorem} asserts that 
$$
2\left( \psi _n,K\psi _n\right) =\left( \psi _n,\left ( {\bf r\cdot }\nabla 
U\right ) \psi _n\right) . 
$$
The proof of this is based on 
$$
\lbrack {\bf r\cdot p,}H]=2iK-i{\bf r\cdot }\nabla U, 
$$
and on the equality 
$$
\left( \psi _n,\left[ {\bf r\cdot p,}H\right] \psi _n\right) =0. 
$$

The {\it hypervirial theorem} is a generalization of the virial theorem:
If $H$ is a hermitian operator on Hilbert space with eigenfunctions, 
$\psi_n,$ and if $A$ is a well--defined linear operator on the same space,
then $(\psi_n,[A,H]\psi_n)=0$. The proof uses the fact that $H $ is 
hermitian.

These theorems can sometimes be used to greatly simplify calculations in
perturbation theory for the propagators but only if the effective
Hamiltonian is strictly, or approximately, hermitian.

Perhaps the most important peculiarity of the many--body problem is that
the eigenvalues of the effective equation (2.182) do not give us the
spectrum of the excited states directly. This spectrum is obtained from
the poles of the propagator (2.108) by solving the equation 
\begin{equation}
\omega _n=H_n\left( \omega _n\right) -\mu  
\label{2.209}
\end{equation}
for $\omega_n$. This defines the {\it single--particle spectrum}. 
Similarly, one can find the poles of the two--particle propagator which
define the {\it spectrum of the collective excitations}. Expanding the
two--particle propagator, in analogy with (2.98), one can derive a
two--particle effective wave equation. One may then proceed to obtain
$n$--particle effective wave equations of successive orders
$n=1,2,3,\ldots$, and attempt to solve them by a procedure similar to that
described for (2.198). Note that the spectra for the effective 
Hamiltonians may depend on thermodynamic variables such as temperature.

\subsection{Screened Potential}

We remarked in Section 2.7, that the unperturbed Hamiltonian (2.152) may 
include potential fields. It is a widely used trick to include in such
potentials an averaged form of the interactions among the particles. In
this way, the convergence properties of the perturbation procedure can
often be significantly improved.

An obvious candidate for a potential to be included in the unperturbed
Hamiltonian for this purpose is the Hartree potential (2.78) which has the
form 
\begin{equation}
V_H\left( 1\right) =
\int \Phi \left( 12\right) \rho \left( 2\right) d\left(2\right) =
\pm i\int \Phi \left( 12\right) G\left( 22\right) d\left( 2\right),  
\label{2.210}
\end{equation}
where $\Phi(12)$, introduced in (2.63), is often referred to as the bare 
interaction potential. The unperturbed inverse propagator can then be
written as 
\begin{equation}
G_0^{-1}\left( 12\right) =\left[ i\frac \partial {\partial t_1}-K\left(
\nabla _1\right) -V\left( 1\right) \right] \delta \left( 12\right) ,
\label{2.211}
\end{equation}
in which we have introduced the {\em effective potential} 
\begin{equation}
V\left( 1\right) :=V_H\left( 1\right) -\mu \left( 1\right) .  
\label{2.212}
\end{equation}
Thus, already in the zero approximation we partially take account of
interaction among the particles by means of the self--consistent field
(2.210). It is therefore appropriate to call this the {\it Hartree} or the
{\it mean--field} approximation.

The total inverse propagator then takes the form 
\begin{equation}
G^{-1}\left( 12\right) =G_0^{-1}(12)-\Sigma ^{\prime }\left( 12\right) ,
\label{2.213}
\end{equation}
in which the supplemental term, 
\begin{equation}
\Sigma' \left( 12\right) := 
\Sigma \left( 12\right) -V_H\left(1\right) \delta \left( 12\right) ,  
\label{2.214}
\end{equation}
is called the {\it exchange--correlation self--energy} since it takes into
account exchange and correlation effects not included in the Hartree
approximation. We may then proceed further by following the perturbation
scheme outlined in Section 2.7.

However, there is an alternative procedure which makes use of the
variational derivatives discussed in Section 2.6. In following this route
an important role is played by the {\it screened potential } 
\begin{equation}
W\left( 12\right) :=
\int\varepsilon^{-1}\left( 13\right)\Phi \left(32\right) d\left( 3\right) ,  
\label{2.215}
\end{equation}
in which 
\begin{equation}
\varepsilon ^{-1}\left( 12\right) =
-\frac{\delta V\left( 1\right) }{\delta \mu \left( 2\right) }  
\label{2.216}
\end{equation}
is the so--called {\em inverse dielectric function}. The screened 
potential (2.215) is sometimes also called the {\em shielded potential}; 
and the function $\varepsilon^{-1}(12)$ in (2.216), the {\it dielectric 
response function}. Further, the solution, $\varepsilon(12)$, when it
exists, of the integral equation 
$$
\int \varepsilon ^{-1}\left( 13\right) \varepsilon \left( 32\right) 
d\left(3\right) =\delta \left( 12\right) , 
$$
is called the {\it dielectric function. }This is an appropriate name since
it is a generalization of the familiar dielectric constant of
electro--magnetic theory.

To obtain the inverse dielectric function we evaluate 
$$
\frac{\delta V\left( 1\right) }{\delta \mu \left( 2\right) }=
-\delta \left(12\right) \pm i\int \Phi \left( 13\right) 
\frac{\delta G\left( 33\right) }{\delta \mu \left( 2\right) }
d\left( 3\right) , 
$$
and 
$$
\frac{\delta G\left( 12\right) }{\delta \mu \left( 3\right) }=
\int \frac{\delta G\left( 12\right) }{\delta V\left( 4\right) }
\frac{\delta V\left(4\right) }{\delta \mu \left( 3\right) }
d\left( 4\right) . 
$$
Introducing the {\it polarization function} 
\begin{equation}
\Pi \left( 12\right) :=
\frac{\delta \rho \left( 1\right) }{\delta V\left( 2\right) }=
\pm i\frac{\delta G\left( 11\right) }{\delta V\left( 2\right) },
\label{2.217}
\end{equation}
we finally have an integral equation 
\begin{equation}
\varepsilon^{-1}\left( 12\right) =
\delta \left( 12\right) +\int \Phi\left( 13\right) 
\Pi \left( 34\right) \varepsilon ^{-1}\left( 42\right)
d\left( 34\right)  
\label{2.218}
\end{equation}
satisfied by the inverse dielectric function.

This equation must be supplemented by an equation for the polarization
function (2.217). Varying the equation of motion (2.69) with respect to
the effective potential (2.212), we are led to 
$$
\int \frac{\delta G^{-1}\left( 13\right) }{\delta V\left( 4\right) }
G(32)d\left( 3\right) + \int G^{-1}\left( 13\right) 
\frac{\delta G\left(32\right) }{\delta V\left( 4\right) }
d\left( 3\right) =0, 
$$
which implies that 
$$
\frac{\delta G\left( 12\right) }{\delta V\left( 3\right) }=
-\int G\left(14\right) 
\frac{\delta G^{-1}\left( 45\right) }{\delta V\left( 3\right) }
G\left( 52\right) d\left( 45\right) . 
$$
In order to calculate the polarization function we shall need the {\it
vertex function}, 
\begin{equation}
\Gamma \left( 123\right) =
-\frac{\delta G^{-1}\left( 12\right) }{\delta V\left( 3\right) },  
\label{2.219}
\end{equation}
which derives its name from the graphical representation of Green's
functions by Feynman diagrams. We find 
\begin{equation}
\Pi \left( 12\right) =
\pm i\int G\left( 13\right) G\left( 41\right) \Gamma\left( 342\right) 
d\left( 34\right) .  
\label{2.220}
\end{equation}
In order to make use of this formula we need an equation for the vertex
function. A variation of (2.213) leads to 
$$
\frac{\delta G^{-1}\left( 12\right) }{\delta V\left( 3\right) }=
-\delta\left( 12\right) \delta \left( 13\right) -
\frac{\delta \Sigma '\left( 12\right) }{\delta V\left( 3\right) }, 
$$
and 
$$
\frac{\delta \Sigma'\left( 12\right)}{\delta V\left( 3\right)} =
\int \frac{\delta \Sigma'\left( 12\right) }{\delta G\left(45\right) }
\frac{\delta G\left( 45\right) }{\delta V\left( 3\right) }
d\left( 45\right) . 
$$
Using these in (2.219) gives 
$$
\Gamma \left( 123\right) =
\delta \left( 12\right) \delta \left( 13\right) + \int 
\frac{\delta \Sigma'\left( 12\right) }{\delta G\left(45\right) }
\frac{\delta G\left( 45\right) }{\delta V\left( 3\right) }d\left(
45\right) . 
$$
Since, as was shown above, 
$$
\frac{\delta G\left( 12\right) }{\delta V\left( 3\right) }=\int G\left(
14\right) \Gamma \left( 453\right) G\left( 52\right) d\left( 45\right) , 
$$
we finally obtain the integral equation 
\begin{equation}
\Gamma \left( 123\right) =
\delta \left( 12\right) \delta \left( 13\right) + \int 
\frac{\delta \Sigma'\left( 12\right) }{\delta G\left( 45\right) }
G\left( 46\right) G\left( 75\right) \Gamma \left( 673\right)
d\left( 4567\right)  
\label{2.221}
\end{equation}
for the vertex function.

We still need an equation for the exchange--correlation self--energy
(2.214). Using the expression (2.70) for the self--energy and the
variational representation (2.150) for the two--particle propagator, we
have 
\begin{equation}
\Sigma'\left( 12\right) = - i\int \Phi \left( 13\right) 
\frac{\delta G\left( 14\right)}{\delta \mu \left( 3\right)}
G^{-1}\left( 42\right) d\left( 34\right) .  
\label{2.222}
\end{equation}
By varying (2.69) with respect to $\mu(1)$, we find 
$$
\int \frac{\delta G\left( 13\right)}{\delta \mu \left( 4\right)}
G^{-1}\left( 32\right) d\left( 3\right) + \int G\left( 13\right) 
\frac{\delta G^{-1}\left( 32\right)}{\delta \mu \left( 4\right)}
d\left( 3\right) =0. 
$$
Therefore (2.222) can be rewritten as 
\begin{equation}
\Sigma'\left( 12\right) =
i \int \Phi \left( 13\right) G\left(14\right) 
\frac{\delta G^{-1}\left( 42\right) }{\delta \mu \left( 3\right)}
d\left( 34\right) .  
\label{2.223}
\end{equation}
Using the relation 
$$
\frac{\delta G^{-1}\left( 12\right) }{\delta \mu \left( 3\right) }=
\int \frac{\delta G^{-1}\left( 12\right) }{\delta V\left( 4\right) }
\frac{\delta V\left( 4\right) }{\delta \mu \left( 3\right) }
d\left( 4\right) , 
$$
and the definitions (2.216) and (2.219), the equation (2.223) takes the 
form 
\begin{equation}
\Sigma'\left( 12\right) =
i\int G\left( 13\right) \Gamma \left(324\right) W\left( 41\right) 
d\left( 34\right) .  
\label{2.224}
\end{equation}

We have thus obtained a closed system of five equations (2.215), (2.218), 
(2.220), (2.221), and (2.224) for $W,\varepsilon^{-1},\Pi,\Gamma$, and
$\Sigma'$, respectively. The
equations (2.215) and  (2.218) may be combined to yield the equation 
\begin{equation}
W\left( 12\right) =
\Phi\left ( 12\right ) + \int \Phi \left( 13\right) 
\Pi \left( 34\right) W\left( 42\right) d\left( 34\right)  
\label{2.225}
\end{equation}
for the screened potential.

To provide a conspectus of this set of equations we write them in symboolic
form as follows: 
$$
W = \varepsilon ^{-1}\Phi ,  
$$
$$
\varepsilon^{-1} = 1+\Phi \Pi \varepsilon ^{-1}, 
$$
\begin{equation}
\Pi = \pm iGG\Gamma ,  
\label{2.226} 
\end{equation}
$$
\Gamma = 1+\frac{\delta \Sigma' }{\delta G}GG\Gamma , 
$$
$$
\Sigma' =  iG\Gamma W.  
$$

This system of equations may be solved by an iteration procedure, starting
from the Hartree approximation corresponding to 
\begin{equation}
\Sigma _0'\left( 12\right) =0.  
\label{2.227}
\end{equation}
Substituting this into the right--hand side of (2.226) gives 
\begin{equation}
\Gamma _0\left( 123\right) =
\delta \left( 12\right) \delta \left( 13\right)
\label{2.228}
\end{equation}
and hence 
\begin{equation}
\Pi _0\left( 12\right) =\pm iG\left( 12\right) G\left( 21\right) .
\label{2.229}
\end{equation}
Let $\varepsilon^{-1}(12)=\delta(12)$, then 
\begin{equation}
W_0\left( 12\right) =\Phi \left( 12\right) .  
\label{2.230}
\end{equation}
This implies that screening is absent in the zero approximation and the
second equation of (2.226) gives the inverse dielectric function 
\begin{equation}
\varepsilon _{HF}^{-1}\left( 12\right) =\delta \left( 12\right) +\int \Phi
\left( 13\right) \Pi _0\left( 32\right) d\left( 3\right)  
\label{2.231}
\end{equation}
in the Hartree--Fock approximation in which the exchange--correlation
self-energy is 
\begin{equation}
\Sigma _{HF}^{\prime }\left( 12\right) =iG\left( 12\right) \Phi \left(
12\right) .  
\label{2.232}
\end{equation}
Since we began with Hartree approximation, the propagators in (2.231) and
(2.232) are also to be taken in this approximation.

Substituting (2.229) into the second equation (2.226) gives the equation 
\begin{equation}
\varepsilon _{RP}^{-1}\left( 12\right) =\delta \left( 12\right) +\int \Phi
\left( 13\right) \Pi _0\left( 34\right) \varepsilon _{RP}^{-1}\left(
42\right) d\left( 34\right) ,  
\label{2.233}
\end{equation}
the solution of which is the inverse dielectric function in the {\it
random--phase approximation}. The corresponding screened potential is 
\begin{equation}
W_{RP}\left( 12\right) =\int \varepsilon _{RP}^{-1}\left( 13\right) \Phi
\left( 32\right) d\left( 3\right) .  
\label{2.234}
\end{equation}
Inserting this into the fifth equation (2.226) defines the
exchange--correlation self--energy 
\begin{equation}
\Sigma _{SP}^{\prime }\left( 12\right) =iG(12)W_{RP}\left( 12\right)
\label{2.235}
\end{equation}
in the {\it screened--potential approximation. }

The iteration procedure can be continued employing either (2.232) or 
(2.235). However, successive approximations soon become overly
complicated. In order to avoid excessive complications, it is sometimes
more convenient to try to include in the zero approxiamtion more
information about the properites of the system. For example, instead of
starting with the Hartree approximation (2.227), we might begin with the
{\it local--density approximation} 
\begin{equation}
\Sigma _{LD}'\left( 12\right) =V_{LD}\left( 1\right) \delta \left(
12\right) ,  
\label{2.236}
\end{equation}
in which the {\it local--density potential } 
\begin{equation}
V_{LD}\left( 1\right) =V_{LD}\left[ \rho \left( 1\right) \right]
\label{2.237}
\end{equation}
is a {\it functional} of the local density $\rho(1)$ such that (2.237) 
accounts for part of the exchange and correlation effects. Because of
this, (2.237) is also called the {\it local exchange--correlation
potential}.

Variation of (2.236) with respect to the propagator $G$ gives 
$$
\frac{\delta \Sigma _{LD}'\left( 12\right) }{\delta G\left(
34\right) }=\pm i\delta \left( 12\right) K_{LD}\left( 23\right) \delta
\left( 34\right) ,
$$
where 
\begin{equation}
K_{LD}\left( 12\right) :=\frac{\delta V_{LD}\left( 1\right) }{\delta \rho
\left( 2\right) }.  
\label{2.238}
\end{equation}
It then follows by means of (2.226) that we can express the vertex 
function in the form 
\begin{equation}
\Gamma \left( 123\right) =\delta \left( 12\right) \Gamma _{LD}\left(
23\right) ,  
\label{2.239}
\end{equation}
where the new function, $\Gamma _{LD}(12)$, called the {\it local--density
vertex,} satisfies the equation 
\begin{equation}
\Gamma _{LD}\left( 12\right) =\delta \left( 12\right) +\int K_{LD}\left(
13\right) \Pi _0\left( 34\right) \Gamma _{LD}\left( 42\right) d\left(
34\right) .  
\label{2.240}
\end{equation}
Substituting for $\Gamma(123)$ in the third equation (2.226) leads to the
following formula for the polarization function 
\begin{equation}
\Pi _{LD}\left( 12\right) =\int \Pi _0\left( 13\right) \Gamma _{LD}\left(
32\right) d\left( 3\right)   
\label{2.241}
\end{equation}
in the local--density approximation. In this approximation, the inverse
dielectric function is given by a solution of the integral equation 
\begin{equation}
\varepsilon_{LD}^{-1}\left( 12\right) =
\delta \left( 12\right) +\int \Phi \left( 13\right) \Pi_0\left( 34\right)
\Gamma _{LD}(45)\varepsilon_{LD}^{-1}(52) d(345) .  
\label{2.242}
\end{equation}
Consequently, the screened potential in the local--density approximation
is 
\begin{equation}
W_{LD}\left( 12\right) =\int \varepsilon _{LD}^{-1}\left( 13\right) \Phi
\left( 32\right) d\left( 3\right) .  
\label{2.243}
\end{equation}
Substituting this in (2.226), we define the screened potential 
\begin{equation}
W_{LV}(12) :=\int \Gamma _{LD}(13) W_{LD}(32) d(3)   
\label{2.244}
\end{equation}
in the {\it local--vertex approximation. }It follows from 
(2.240) that the potential (2.244) satisfies the
equation 
$$
W_{LV}\left( 12\right) = \int \Gamma _{LD}\left( 13\right) \Phi \left(
32\right) d\left( 3\right)  +
$$
$$
+ \int \Gamma _{LD}\left( 13\right) \Phi \left( 34\right) \Pi _0\left(
45\right) W_{LV}\left( 52\right) d(345).
$$
With a solution of this equation we may define the local--vertex
approximation, 
\begin{equation}
\Sigma _{LV}'\left( 12\right) =iG\left( 12\right) W_{LV}\left(
21\right) ,  
\label{2.245}
\end{equation}
for the exchange--correlation self--energy [65--69].

Since we started from the local--density approximation (2.236), we should
take the propagators in (2.244) and (2.245) in the same approximation.

Analysing the structure of perturbation series in quantum electrodynamics, 
Feynman suggested a graphycal interpretation of perturbative terms in the 
form of diagrams. Such an interpretation permitted to understand better 
the physical meaning of perturbative expansions. It helped also to 
separate an infinite class of diagrams which could be summed, thus, 
renormalizing and improving perturbation theory. This resummation is 
usually done by selecting those diagrams which correspond to a set of 
terms forming a geometrical progression. Together with the penetration 
of the field--theory methods into the many--body problem, the diagram 
technique has become popular among the statistical--physics community [70].

We shall not consider here such diagrams because of the following:
A diagram is nothing but the illustration of an analytical expression. 
It is possible to accomplish calculations without diagrams, while, dealing
with diagrams, one always has to return finally to analytical formulas. In 
many cases the diagrammatic illustration helps to clarify the resummation 
procedure. However, the latter can always be done without invoking the 
former.

\subsection{Heterophase States}

Till now, speaking about a statistical system we have tacitly assumed that 
the same type of order occurs in the whole volume of the system. It may 
happen, however, that different types of order appear in different parts 
of the system. If these regions of different order are macroscopic then 
one says that the system is divided into domains. In this case, each domain 
plays the role of a separate system, so that the problem is actually reduced
to that of a system with one fixed order, with a slight complication coming 
from the boundary conditions describing interdomain layers.

A much more complex situation develops if the regions of different order 
are of mesoscopic sizes and, moreover, are randomly distributed in space.
Since each type of order is related to a thermodynamic phase, we may say 
then that the system is heterophase. The mesoscopic regions of different 
phases not only are randomly distributed in space but may also oscillate 
in time. Because of this, they are called {\it heterophase fluctuations}. 
To describe the system with such fluctuations is a problem far from 
trivial. In this section we present a general approach for treating 
statistical systems with heterophase fluctuations [19,71--80].

Consider several thermodynamic phases which we enumerate by the index 
$\nu=1,2,\ldots$. Recall that different thermodynamic phases correspond 
to qualitatively different statistical states. At the same time, to each 
phase one can put into correspondence a {\it space of typical states} 
which is a Hilbert space ${\cal H}_\nu$ of microscopic states 
$h_\nu\in{\cal H}_\nu$ such that one can define an {\it order operator} 
$\hat\eta$ whose mathematical expectation
$$ (h_\nu,\hat\eta h_\nu) =\eta(h_\nu) $$ 
gives an {\it order parameter} $\eta(h_\nu)$ which is qualitatively 
different from $\eta(h_\mu)$ if $\mu\neq\nu$.

The direct sum
\be
{\cal X} =\oplus_\nu {\cal H}_\nu
\label{7.216}
\ee
describes a system which can be in either one or another thermodynamic 
phase. In order to describe the states of coexisting phases we need to 
deal with the space
\be
{\cal Y} =\otimes_\nu{\cal H}_\nu
\label{7.217}
\ee
which is the tenzor product of ${\cal H}_\nu$. The spaces ${\cal X}$ and 
${\cal Y}$ are uniquely isomorphic but correspond to different physical 
situations.

Let a statistical system in the real--space region ${\bf V}\subset{\bf 
R}^3$ be divided into subregions ${\bf V}_\nu\subset{\bf V}$ filled by 
different thermodynamic phases. A family $\{ {\bf V}_\nu\}$ of subregions
${\bf V}_\nu$ forms a covering of the region ${\bf V}$ when
\be
{\bf V} =\cup_\nu {\bf V}_\nu, \qquad V=\sum_\nu V_\nu ,
\label{7.218}
\ee
where $$ V=mes{\bf V}, \qquad V_\nu = mes {\bf V}_\nu . $$

The space separation of subregions ${\bf V}_\nu$ can be characterized by 
the {\it manifold indicator functions}
\begin{eqnarray}
\xi_\nu(\sr) :=\left\{ \begin{array}{cc}
1, & \sr\in {\bf V}_\nu \\
0, & \sr\not\in {\bf V}_\nu .
\end{array}\right.
\label{7.219}
\end{eqnarray}
A set
\be
\xi:=\{\xi_\nu(\sr)|\; \sr\in{\bf V}, \nu=1,2,\ldots\} 
\label{7.220}
\ee
of the indicator functions (\ref{7.219}) uniquely defines a {\it phase 
configuration} in real space.

Operators of the density observables acting on ${\cal H}_\nu$ have the 
general form
$$ \hat A_\nu(\xi,\sr) 
=\sum_{k=0}^\infty\int\xi_\nu(\sr)\xi_\nu(\sr_1)\xi_\nu(\sr_2)\ldots
\xi_\nu(\sr_k)\times $$
\be
\times A_{\nu k}(\sr_,\sr_1,\sr_2,\ldots,\sr_k)d\sr_1d\sr_2\ldots d\sr_k ,
\label{7.221}
\ee
in which $A_{\nu k}(\ldots)$ is an operator distribution. Here and in 
what follows the integrals in real space are assumed to be over the whole 
region ${\bf V}$, if not specified. Operators of local observables, for a 
subregion ${\bf\Lambda}$, are given by
\be
A_\nu(\xi,{\bf\Lambda}) =\int_{\bf\Lambda}\hat A_\nu(\xi,\sr)d\sr .
\label{7.222}
\ee
The operator density (\ref{7.221}) can be considered as a limiting case of
(\ref{7.222}) in the sense of the limit
$$ \lim_{mes{\bf\Lambda}\ra 0}
\frac{A_\nu(\xi,{\bf\Lambda})}{mes{\bf\Lambda}} =\hat A_\nu(\xi,\sr) . $$

The local operators acting on the space (\ref{7.217}) are defined as
\be
\hat A(\xi,\sr)=\oplus_\nu\hat A_\nu(\xi,\sr) , \qquad 
A(\xi,{\bf\Lambda}) =\oplus_\nu A_\nu(\xi,{\bf\Lambda}) .
\label{7.223}
\ee
The family
\be
{\cal A}(\xi) :=\{\hat A(\xi,\sr),\; A(\xi,{\bf\Lambda})|\; 
\sr\in{\bf V}, \; {\bf\Lambda}\subset{\bf V}\} 
\label{7.224}
\ee
of all operators of local observables forms the algebra of local observables.

When a phase configuration characterized by the set (\ref{7.220}) is 
fixed, then we have the case of frozen phase separation. However, when the 
phase configuration is randomly distributed in space and, in addition, 
may fluctuate in time, we have to consider the set (\ref{7.220}) as a 
stochastic variable. Then observable quantities are given by the averages 
of operators from the algebra of local observables (\ref{7.224}), with 
the averaging procedure containing the trace over microscopic degrees of 
freedom and, also, an averaging over phase configurations. The latter is 
a functional integral over the manifold indicator functions 
(\ref{7.219}), which we shall denote by $\int{\cal D}\xi$. The statistical 
averaging, as usual, must include a statistical operator $\rho(\xi)$. Thus,
the observable quantities are given by the average
\be
\lgl{\cal A}\rgl = Tr\int\rho(\xi){\cal A}(\xi){\cal D}\xi .
\label{7.225}
\ee
Here and everywhere in what follows it is assumed that the operators of 
taking trace and of integrating over $\xi$ always commute.

As examples of the operators (\ref{7.223}), we may write the operator of 
the energy density
\be
\hat E(\xi,\sr) =\oplus_\nu \hat E_\nu(\xi,\sr) ,
\label{7.226}
\ee
where
\be
\hat E_\nu(\xi,\sr) =\hat K_\nu(\xi,\sr) +\hat V_\nu(\xi,\sr)
\label{7.227}
\ee
is the operator of the energy density for a $\nu$--phase, consisting of 
the kinetic--energy operator density.
\be
\hat K_\nu(\xi,\sr) =\xi_\nu(\sr)\psi_\nu^\dgr(\sr) \left [ -
\frac{\nabla^2}{2m} +U(\sr)\right ]\psi_\nu(\sr) 
\label{7.228}
\ee
and of the potential--energy operator density
$$
\hat V_\nu(\xi,\sr)=\frac{1}{2}\int\xi_\nu(\sr)\xi_\nu(\sr')
\psi_\nu^\dgr(\sr)\psi_\nu^\dgr(\sr')\times $$
\be
\times \Phi(\sr-\sr')\psi_\nu(\sr')\psi_\nu(\sr)d\sr' .
\label{7.229}
\ee

Another example is the operator of particle density
\be
\hat N(\xi,\sr) =\oplus_\nu\hat N_\nu(\xi,\sr)
\label{7.230}
\ee
with
\be
\hat N_\nu(\xi,\sr) =\xi_\nu(\sr)\psi_\nu^\dgr(\sr)\psi_\nu(\sr) .
\label{7.231}
\ee

To calculate the average (\ref{7.225}), we need to know the statistical 
operator $\rho(\xi)$. This can be defined by invoking the {\it principle 
of unbiased guess}, that is, by maximizing the information entropy. The 
entropy itself is defined by the expression
\be
S(\rho) = -Tr\int\rho(\xi)\ln\rho(\xi){\cal D}\xi .
\label{7.232}
\ee
As additional conditions for $\rho(\xi)$, we require that this 
statistical operator is normalized so that
\be
Tr\int\rho(\xi){\cal D}\xi = 1 ,
\label{7.233}
\ee
that the local energy density is given by
\be
E(\sr) =Tr\int\rho(\xi)\hat E(\xi,\sr){\cal D}\xi ,
\label{7.234}
\ee
and that the local particle density is
\be
N(\sr) =Tr\int\rho(\xi)\hat N(\xi,\sr){\cal D}\xi .
\label{7.235}
\ee
For the informational entropy we have
$$ S_{inf} = S(\rho) +\zeta\left [ Tr\int\rho(\xi){\cal D}\xi - 1\right ] 
+ $$ $$ + \int\bt(\sr)\left [ E(\sr) - Tr\int\rho(\xi)\hat E(\xi,\sr){\cal D}
\xi \right ]d\sr + $$
\be
+\int\gm(\sr)\left [ N(\sr) - Tr\int\rho(\xi)\hat N(\xi,\sr){\cal D}
\xi\right ]d\sr , \label{7.236}
\ee
where $\zeta,\; \bt(\sr)$, and $\gm(\sr)$ are the Lagrange multipliers. 
Conditions (\ref{7.233}), (\ref{7.234}), and (\ref{7.235}) are recovered
from the variational equations
$$ \frac{\partial S_{inf}}{\partial\zeta} = 0 , \qquad
\frac{\dlt S_{inf}}{\dlt\bt(\sr)} = 0 , \qquad
\frac{\dlt S_{inf}}{\dlt\gm(\sr)} = 0 . $$

The statistical operator $\rho(\xi)$ is defined as a maximizer of 
(\ref{7.236}), that is, from the equations
\be
\frac{\dlt S_{inf}}{\dlt\rho(\xi)} = 0 , \qquad
\frac{\dlt^2S_{inf}}{\dlt\rho^2(\xi)} < 0 .
\label{7.237}
\ee
Introducing the notation
$$ \zeta = 1 -\ln Z, \qquad
\gm(\sr) =-\bt(\sr)\mu(\sr) , $$
from the extremum condition of (\ref{7.237}) we obtain
\be
\rho(\xi) =\frac{1}{Z}\exp\left\{ -Q(\xi)\right \} ,
\label{7.238}
\ee
where
\be
Z= Tr\int\exp\{ -Q(\xi)\}{\cal D}\xi
\label{7.239}
\ee
is the partition function and
\be
Q(\xi)=\int\bt(\sr)\left [\hat E(\xi,\sr) -\mu(\sr)\hat N(\xi,\sr)
\right ] d\sr \label{7.240}
\ee
is the {\it quasi--Hamiltonian}.

Now we need to concretize the procedure of averaging over phase 
configurations, that is, we have to define explicitly the functional 
integration over the stochastic variable (\ref{7.220}). All conceivable 
variants of the phase configurations form a topological space
${\cal K}=\{\xi\}$. To integrate over the stochastic variable $\xi$, we 
have to define a functional measure $\int{\cal D}\xi$ on the space ${\cal K}$.

Note that averaging over space configurations contains two types of 
actions. The first one averages over all possible phase configurations 
under a fixed set
\be
x:= \{ x_\nu|\; \nu=1,2,\ldots\}
\label{7.241}
\ee
of geometric probabilities
\be
x_\nu :=\frac{1}{V}\int\xi_\nu(\sr)d\sr = \frac{V_\nu}{V}
\label{7.242}
\ee
with the evident properties
\be
\sum_\nu x_\nu = 1 , \qquad 0 \leq x_\nu \leq 1 .
\label{7.243}
\ee
The second action is the variation of each probability (\ref{7.242}) from zero
to unity taking account of their normalization (\ref{7.243}).

The resulting differential functional measure ${\cal D}\xi$ can be written as
\be
{\cal D}\xi = dx{\cal D}_x\xi
\label{7.244}
\ee
with
\be
dx =\dlt\left (\sum_\nu x_\nu -1\right )\prod_\nu dx_\nu ,
\label{7.245}
\ee
where $x_\nu\in[0,1]$.

To define ${\cal D}_x\xi$, let us divide the whole system into $n$ parts, so 
that each subregion ${\bf V}_\nu$ is covered by a family of 
$n_\nu$--subregions ${\bf V}_{\nu i}$ such that
$$ {\bf V}_\nu =\bigcup_{i=1}^{n_\nu}{\bf V}_{\nu i} , \qquad 
V_\nu  = \sum_{i=1}^{n_\nu}v_{\nu i} $$
and with
$$ n=\sum_\nu n_\nu , \qquad v_{\nu i} =mes{\bf V}_{\nu i} . $$
The shape of each sub--region ${\bf V}_{\nu i}$ can be arbitrary. For 
each ${\bf V}_{\nu i}$, let us fix a vector $\stackrel{\ra}{a}_{\nu i}\in
{\bf V}_{\nu i}$ defined as the center of the local coordinate system in 
${\bf V}_{\nu i}$. Then the manifold indicator function (\ref{7.219}) can 
be presented as
\be
\xi_\nu(\sr) =\sum_{i=1}^{n_\nu}\xi_{\nu i}(\sr-\stackrel{\ra}{a}_{\nu i}) ,
\label{7.246}
\ee
where
\begin{eqnarray}
\xi_{\nu i}(\sr -\stackrel{\ra}{a}_{\nu i}) =\left\{ \begin{array}{cc}
1, & \sr \in {\bf V}_{\nu i} \\
\\ \nonumber
0, & \sr\not\in {\bf V}_{\nu i} .
\end{array}\right.
\end{eqnarray}
It is clear that if the sub--subregions ${\bf V}_{\nu i}$ are small 
enough, then they can be used as blocks for producing any phase 
configuration. A displacement of the blocks can be accomplished by moving 
the block centers $\stackrel{\ra}{a}_{\nu i}$.

In this way, we come to the definition of the functional differential
\be
{\cal D}_x\xi := \prod_\nu\prod_{i=1}^{n_\nu}\frac{a_{\nu i}}{V} \qquad
(n\ra\infty) ,
\label{7.247}
\ee
in which $\stackrel{\ra}{a}_{\nu i}\in{\bf V}$ and the limit $n\ra\infty$
implies that 
\be
n\ra\infty: \; n_\nu\ra\infty , \quad v_{\nu i}\ra 0 , \quad x_\nu\ra
const .
\label{7.248}
\ee
The limit (\ref{7.248}) is to be taken after the corresponding functional 
integration. Thus, for a function of $\xi$ denoted by $F(\xi)$ we have
\be
\int F(\xi){\cal D}_x\xi = \lim_{n\ra\infty}\int F(\xi)\prod_\nu
\prod_{i=1}^{n_\nu}\frac{d\stackrel{\ra}{a}_{\nu i}}{V} .
\label{7.249}
\ee
For example, it is straightforward to check that
$$ \int{\cal D}_x\xi = 1 , \qquad \int\xi_\nu(\sr){\cal D}_x\xi = x_\nu . $$
More generally, the following statement holds.

\vspace{3mm}

{\bf Proposition 2.1.}

\vspace{2mm}

Consider a class of functionals of the form
$$ F(\xi) =\sum_{k=0}^\infty\sum_{\nu_1}\sum_{\nu_2}\ldots\sum_{\nu_k}
\int\xi_{\nu_1}(\sr_1)\xi_{\nu_2}(\sr_2)\ldots\xi_{\nu_k}(\sr_k)\times $$
\be
\times F_{\nu_1\nu_2\ldots\nu_k}(\sr_1,\sr_2,\ldots,\sr_k)d\sr_1 d\sr_2
\ldots d\sr_k .
\label{7.250}
\ee
Then the functional integration (\ref{7.249}) gives
\be
\int F(\xi){\cal D}_x\xi = F(x) ,
\label{7.251}
\ee
where $F(x)$ has the form of (\ref{7.250}) but with all $\xi_\nu(\sr)$ 
changed by $x_\nu$.

\vspace{3mm}

{\bf Proof} is based on a direct integration of products of the manifold
indicator functions (\ref{7.246}) over the block vectors 
$\stackrel{\ra}{a}_{\nu i}$ with the differential measure (\ref{7.247}). 
In this integration, the products of $\xi_{\nu i}$ with coinciding vectors
$\stackrel{\ra}{a}_{\nu i}$ yield Lesbegue zeros. For example,
$$ \sum_{i=1}^{n_\nu}\int\xi_{\nu i}(\sr_1 -\stackrel{\ra}{a}_{\nu i})
\xi_{\nu i}(\sr_2-\stackrel{\ra}{a}_{\nu i}){\cal D}_x\xi = $$
$$ = \lim_{n\ra\infty}\sum_{i=1}^{n_\nu}\int\xi_{\nu i}
(\sr_1-\stackrel{\ra}{a}_{\nu i})\xi_{\nu i}(\sr_2-\stackrel{\ra}{a}_{\nu i})
\frac{d\stackrel{\ra}{a}_{\nu i}}{V} = x_\nu\dlt_{r_1r_2} , $$
where $\dlt_{r_1r_2}$ is the Kronecker delta. Therefore, only the 
manifold indicator functions with different block vectors 
$\stackrel{\ra}{a}_{\nu i}$
contribute to (\ref{7.249}) which leads to (\ref{7.251}).

\vspace{3mm}

{\bf Proposition 2.2.}

\vspace{2mm}

The partition function (\ref{7.239}) with the quasi--Hamiltonian 
(\ref{7.240}) and the differential functional measure (\ref{7.244}) can 
be written as
\be
Z = Tr\int\exp\{-Q(x)\}dx ,
\label{7.252}
\ee
where $Q(x)$ has the form of (\ref{7.240}) but with all $\xi_\nu(\sr)$ 
changed by $x_\nu$, and $dx$ is defined in (\ref{7.245}).

\vspace{3mm}

{\bf Proof} involves the previous proposition after expanding the 
exponential in (\ref{7.239}) in powers of $Q(\xi)$ and obtaining
$$ \int\exp\{-Q(\xi)\}{\cal D}_x\xi =\exp\{-Q(x)\} . $$

The integration in (\ref{7.252}) with the differential measure 
(\ref{7.245}), if accomplished in the thermodynamic limit $N\ra\infty$, 
defines a natural thermodynamic potential for a heterophase system as 
follows.

\vspace{3mm}

{\bf Proposition 2.3.}

\vspace{2mm}

Let a function
\be
y(x) := -\lim_{N\ra\infty}\frac{1}{N}\ln Tr\exp\{-Q(x)\}
\label{7.253}
\ee of $x$ in the hypercube
$$ {\bf H} :=\{ x|\; \sum_\nu x_\nu = 1 , \; 0\leq x_\nu \leq 1\} $$
have an absolute minimum
\be
y(w) =abs\min_{x\in{\bf H}}y(x) ,
\label{7.254}
\ee
in which the set
\be
w:=\{ w_\nu|\; \nu=1,2\ldots\}
\label{7.255}
\ee
consists of elements with the property
\be
\sum_\nu w_\nu = 1 ,\qquad 0 \leq w_\nu \leq 1 .
\label{7.256}
\ee
Then, in the thermodynamic limit,
\be
y(w) =-\lim_{N\ra\infty}\frac{1}{N}\ln Z .
\label{7.257}
\ee

\vspace{3mm}

{\bf Proof} uses the asymptotic, as $N\ra\infty$, presentation
$$ Z\simeq \int\exp\{-Ny(x)\} dx $$ 
for the partition function (\ref{7.252}) with definition (\ref{7.253}). Then,
the Laplace method of integration immediately yields (\ref{7.257}).

Thus, we come to the conclusion that the role of a natural thermodynamic 
potential for a heterophase system is played by the function
\be
y(w) =-\lim_{N\ra\infty}\frac{1}{N}\ln Tr\exp\{-Q(w)\} .
\label{7.258}
\ee
The set (\ref{7.255}) consists of the geometric probabilities $w_\nu$ 
defining relative weights of each thermodynamic phase. Consequently, 
$w_\nu$ may be called the {\it phase probability}. To find these 
probabilities, we have to minimize the {\it heterophase thermodynamic 
potential} (\ref{7.258}). The minimization means that the probabilities 
$w_\nu$ are to be defined by the equations
\be
\frac{\partial}{\partial w_\nu} y(w) = 0 , \qquad
\frac{\partial^2}{\partial w_\nu^2}y(w) > 0 
\label{7.259}
\ee
under condition (\ref{7.256}). In addition, one has to analyse the 
thermodynamic potential $y(w)$ for $w$ on the boundary of the hupercube 
${\bf H}$, that is, when one or several $w_\nu$ are equal to one or zero. 
The first equation in (\ref{7.259}) can be called the {\it equation for 
phase probabilities}, and the inequality in (\ref{7.259}) is the {\it 
condition of heterophase stability}. 

Taking account of the form of the heterophase thermodynamic potential 
(\ref{7.258}), as the equation for phase probabilities we get
\be
\lim_{N\ra\infty}\frac{1}{N}\left\lgl
\frac{\partial}{\partial w_\nu} Q(w) \right \rgl = 0 .
\label{7.260}
\ee
And the condition of heterophase stability becomes
\be
\lim_{N\ra\infty}
\frac{1}{N}\left [\left\lgl\frac{\partial^2}{\partial w_\nu^2} Q(w)\right\rgl -
\left\lgl\frac{\partial}{\partial w_\nu}Q(w)\right\rgl^2\right ] > 0 .
\label{7.261}
\ee
The quasi--Hamiltonian $Q(w)$ is defined in (\ref{7.240}) which together 
with (\ref{7.226})--(\ref{7.231}) gives
\be
Q(w) =\oplus_\nu Q_\nu(w) ,
\label{7.262}
\ee
where
$$ Q_\nu(w) =w_\nu\int\bt(\sr)\psi_\nu^\dgr(\sr)\left [-\frac{\nabla^2}{2m}
+U(\sr) -\mu(\sr)\right ]\psi_\nu(\sr)d\sr + $$
\be
+\frac{1}{2}w_\nu^2\int\bt(\sr)\psi_\nu^\dgr(\sr)\psi_\nu^\dgr(\sr')
\Phi(\sr-\sr')\psi_\nu(\sr')\psi_\nu(\sr)d\sr d\sr' .
\label{7.263}
\ee

Finally, we need to find an effective presentation for the expectation 
values (\ref{7.225}) of the algebra of local observables (\ref{7.224}).

\vspace{3mm}

{\bf Proposition 2.4.}

\vspace{2mm}

Let the conditions of the Proposition 2.3 be valid, then the expectation 
values (\ref{7.225}) asymptotically, as $N\ra\infty$, take the form
\be
\lgl{\cal A}\rgl\simeq Tr\rho_{eff}(w){\cal A}(w) \qquad (N\ra\infty) ,
\label{7.264}
\ee
where ${\cal A}(w)$ is the algebra of local observables (\ref{7.224}) with 
all $\xi_\nu(\sr)$ changed by $w_\nu$, and the {\it effective statistical 
operator} is
\be
\rho_{eff}(w) :=\frac{1}{Z_{eff}}\exp\{-Q(w)\}
\label{7.265}
\ee
with the {\it effective partition function}
\be
Z_{eff} := Tr\exp\{-Q(w)\} .
\label{7.266}
\ee

\vspace{3mm}

{\bf Proof} starts from Proposition 2.1, according to which
$$ \int\rho(\xi){\cal A}(\xi){\cal D}_x\xi = \rho(x){\cal A}(x) , $$
where
$$ \rho(x) =\frac{1}{Z}\exp\{-Q(x)\} $$
with $Z$ from (\ref{7.252}). Then, introducing
\be
\bar{\cal A}(x) :=\frac{Tr\exp\{-Q(x)\}{\cal A}(x)}{\exp\{-Ny(x)\}} ,
\label{7.267}
\ee
we have
$$ Tr\rho(x){\cal A}(x) =\frac{1}{Z}\exp\{-Ny(x)\}\bar{\cal A}(x) . $$
Using the method of steepest descent, for $N\ra\infty$, we obtain
$$ \int\exp\{-Ny(x)\}\bar{\cal A}(x)dx \simeq \exp\{-Ny(w)\}\bar{\cal A}(w)
\prod_\nu\left (\frac{2\pi}{Ny_\nu''}\right )^{1/2} , $$
where the product over $\nu$ contains the number of factors by one less 
than the number of phases, since one of $w_\nu$ must be expressed through 
others because of the normalization (\ref{7.256}), and
$$ y_\nu'' \equiv \frac{\partial^2y(w)}{\partial w_\nu^2} > 0 . $$
Similarly, we find
$$ Z\simeq \exp\{-Ny(w)\}
\prod_\nu\left (\frac{2\pi}{Ny_\nu''}\right )^{1/2} . $$
It follows that
$$ \frac{1}{Z}\int\exp\{-Ny(x)\}\bar{\cal A}(x)dx \simeq \bar{\cal A}(w) , $$
as $N\ra\infty$, that is
$$ \lgl{\cal A}\rgl \simeq \bar{\cal A}(w) . $$
Remembering the definition (\ref{7.267}), we come to (\ref{7.264}) with 
(\ref{7.265}) and (\ref{7.266}).

When there are no external fields specially inducing nonuniformity, the 
local inverse temperature and chemical potential are to be uniform through
the system:
\be
\bt(\sr)=\bt , \qquad \mu(\sr) =\mu .
\label{7.268}
\ee
These equations have the standard form of equilibrium conditions. 
However, we should not forget that a heterophase system, as is explained 
above, is quasiequilibrium. Condition (\ref{7.268}) appears after 
averaging over phase configurations. Therefore this condition has the 
meaning of the {\it condition of equilibrium on average} or it may be 
called the {\it condition of heterophase equilibrium}.

Under condition (\ref{7.268}), the quasi--Hamiltonian (\ref{7.262}) becomes
\be
Q(w) =\bt H(w)
\label{7.269}
\ee
with the {\it renormalized} or {\it effective heterophase Hamiltonian}
\be
H(w) :=\oplus_\nu H_\nu(w)
\label{7.270}
\ee
consisting of the terms
$$ H_\nu(w) =w_\nu\int\psi_\nu^\dgr(\sr)\left [-\frac{\nabla^2}{2m} +
U(\sr) -\mu\right ]\psi_\nu(\sr) d\sr + $$
\be
+\frac{1}{2}w_\nu^2\int\psi_\nu^\dgr(\sr)\psi_\nu^\dgr(\sr')
\Phi(\sr-\sr')\psi_\nu(\sr')\psi_\nu(\sr) d\sr d\sr' .
\label{7.271}
\ee
Emphasize that (\ref{7.270}) represents not just one system with a given 
phase separation but an infinite number of systems with all possible 
phase configurations. Because of this, each term (\ref{7.271}) may be 
called the {\it phase--replica Hamiltonian}. Owing to the nonlinear 
renormalization in (\ref{7.271}), interphase effects are included into 
the description. The {\it interphase state} is defined as
\be
\lgl{\cal A}\rgl_{int} :=\lgl{\cal A}\rgl -\sum_\nu w_\nu\lgl{\cal 
A}\rgl_\nu , 
\label{7.272}
\ee
that is, as the difference between the heterophase state $\lgl{\cal A}\rgl$,
and the respectively weighted sum of pure states
$$ \lgl{\cal A}\rgl_\nu :=\lim_{w_\nu\ra 1}\lgl{\cal A}\rgl . $$

As an example, consider two coexisting phases. Then the phase--probability 
equation (\ref{7.260}), with the quasi--hamiltonian (\ref{7.269}) gives
\be
w_1 =\frac{2\Phi_2+K_2-K_1+\mu(R_1-R_2)}{2(\Phi_1+\Phi_2)} , \qquad
w_2 = 1-w_1 ,
\label{7.273}
\ee
where
$$ K_\nu =\frac{1}{N}\int\left \lgl\psi_\nu^\dgr(\sr)\left [ 
-\frac{\nabla^2}{2m} +U(\sr)\right ]\psi_\nu(\sr)\right \rgl d\sr , $$
$$ \Phi_\nu =\frac{1}{N}\int\left\lgl \psi_\nu^\dgr(\sr)\psi_\nu^\dgr(\sr')
\Phi(\sr-\sr')\psi_\nu(\sr')\psi_\nu(\sr)\right\rgl d\sr d\sr' , $$
$$ R_\nu =\frac{1}{N} \int\left\lgl \psi_\nu^\dgr(\sr)\psi_\nu(\sr)\right\rgl
d\sr  . $$
Thus, all phase probabilities are defined in a self--consistent way.

Each field operator $\psi_\nu(\sr)$ is defined on a weighted Hilbert space
${\cal H}_\nu$ and has the standard commutation or anticommutation
relations,
$$ 
\left [ \psi_\nu(\sr) , \psi_\nu(\sr') \right ]_\mp = 0 , \qquad
\left [ \psi_\nu(\sr) ,\psi_\nu^\dgr(\sr') \right ] =\delta(\sr -\sr') ,
$$
depending on the kind of statistics. Field operators defined on different
spaces, by definition, commute with each other:
\begin{equation}
\label{2.304}
\left [ \psi_\nu(\sr) , \psi_{\nu'}(\sr') \right ] = 0 =
\left [ \psi_\nu(\sr) ,\psi_{\nu'}^\dgr(\sr')\right ] \qquad
(\nu\neq \nu')
\end{equation}
Therefore, the evolution equation for
$$
\psi_\nu(\sr,t) = U^+_\nu(t)\psi_\nu(\sr) U_\nu(t) ; \qquad
U_\nu)t) = e^{-iH_\nu t} \; ,
$$ has the usual Heisenberg form
$$
i\frac{\partial}{\partial t}\psi_\nu(\sr,t) = \left [ 
\psi_\nu(\sr,t) , H_\nu\right ] \; .
$$
For each phase replica, indexed by $\nu=1,2,\ldots$, we may introduce a
propagator
\begin{equation}
\label{2.305}
G_\nu(\sr,t;\sr',t') := 
- i\lgl \hat T\psi_\nu(\sr,t)\psi_\nu^\dgr(\sr',t')\rgl \; .
\end{equation}
Similarly, we may introduce many--particle propagators. Then we can write
the evolution equation for (\ref{2.305}) and proceed in analogy with one
of the techniques described in the previous sections. Dealing with the
effective Hamiltonian (\ref{7.270}), we have the dynamical equations for
different phase replicas formally decoupled, although thermodynamically
all equations are related with each other through the geometric
probabilities (\ref{7.255}). The possibility of realizing this dynamical
decoupling is one of the main advantages of the approach formulated in the
theory of heterophase fluctuations [19,71--80].

\end{sloppypar}

\end{document}


\begin{center}

\vspace{3cm}

{\Large{\bf Statistical Green's Functions \\ [3cm]
V.I. Yukalov}}

\vspace{15cm}

{\Large 1998}

\end{center}

\newpage

\hspace{7cm} Dedicated to A.J. Coleman

\hspace{7cm} on the occasion of his 80-th birthday

\vspace{1cm}

{\Large{\bf Preface}}

\vspace{5mm}

\begin{sloppypar}

Green's functions are widely used in various branches of mathematics and 
theoretical physics. Despite its rather long history, the method of Green's 
functions continues to develop. This development concerns first of all the
application of this technique to statistical mechanics. The latter is, 
probably, the most complicated subject which the method of Green's functions 
has been applied to. Generally speaking, practically all parts of theoretical 
physics can be treated as subsections of statistical mechanics (this 
statement is explained more precisely in Appendix). It is just the complexity
of statistical mechanics that perpetually forces us to develop the method of
Green's functions.

There exists a volumnous literature on this method and its various 
applications. In no way my aim was to give a review of all that. The aim 
of this work is to describe the basic mathematical properties of Green's 
functions used in statistical mechanics, as well as the equations defining
these functions and the techniques of solving these equations. No 
particular applications are touched in order not to be distracted from 
the main aim of analysing the underlying mathematical structure. Such a 
concentrated analysis of the fundamentals of the method may, to my mind,
help to grasp its most important features and to better understand its
strong and weak points. This could also give hints of how to improve and
develop the approach for overcoming existing difficulties. One such a
development that has not yet been published in a review form is presented 
here in Part 3. This is {\it Correlated Iteration Theory} that differs from
all other known variants of perturbation theory for Green's functions by 
the combination of two factors: the systematic formulation of an algorithm
for obtaining subsequent approximations and the consistent consideration 
of interparticle correlations at each step of the procedure.

The formalism of the correlated iteration theory is based on general 
properties of Green's functions, which are expounded in Part 2. Without
studying these general properties it is impossible to proceed further. And,
in turn, to introduce Green's functions for statistical systems and to
study the properties of these functions, it is necessary to have those
main definitions that make the foundation of statistical mechanics. These 
basic notions are formulated in Part 1.

Statistical Green's functions are a natural generalization of correlation
functions and reduced density matrices. The relation between these objects
is repeatedly traced throughout the text.

\newpage

{\bf Acknowledgements}

\vspace{2mm}

Many facts about reduced density matrices as well as a lot of important 
things in many--body problem I have learnt from A.J. Coleman with whom I
had a pleasure of enjoying numerous discussions. I greatly benefited from 
a lucky opportunity of having a close collaboration with A.J. Coleman, to 
whom I am indebted so much.

I highly appreciate permanent support and help from E.P. Yukalova.

During four years I worked at the Department of Mathematics and Statistics of
Queen's University. I am very grateful to my colleagues from this Department
for their friendly support and advice.

\vspace{1cm}

Kingston--London, May 1998.

\newpage

{\parindent=0pt

{\Large{\bf Contents}}

\vspace{1cm}

{\bf 1.} {\large{\bf Basic Notions}}}

\vspace{5mm}

{\bf 1.1.} Pure States

\vspace{3mm}

{\bf 1.2.} Mixed States

\vspace{3mm}

{\bf 1.3.} Field Operators

\vspace{3mm}

{\bf 1.4.} Local Observables

\vspace{3mm}

{\bf 1.5.} Order Indices

\vspace{3mm}

{\bf 1.6.} Information Functionals

\vspace{3mm}

{\bf 1.7.} Equilibrium States

\vspace{3mm}

{\bf 1.8.} Quasiequilibrium States

\vspace{3mm}

{\bf 1.9.} Evolution Equations

\vspace{3mm}

{\bf 1.10.} Coherent States

\vspace{1cm}

{\parindent=0pt

{\bf 2.} {\large{\bf Green's Functions}}}

\vspace{5mm}

{\bf 2.1.} Principal Definitions

\vspace{3mm}

{\bf 2.2.} Spectral Representations

\vspace{3mm}

{\bf 2.3.} Dispersion Relations

\vspace{3mm}

{\bf 2.4.} Evolution Equations

\vspace{3mm}

{\bf 2.5.} Wave Functions

\vspace{3mm}

{\bf 2.6.} Scattering Matrix

\vspace{3mm}

{\bf 2.7.} Perturbation Theory

\vspace{3mm}

{\bf 2.8.} Excited States

\vspace{3mm}

{\bf 2.9.} Screened Potential

\vspace{3mm}

{\bf 2.10.} Heterophase States

\newpage

{\parindent=0pt

{\bf 3.} {\large{\bf Correlated Iteration}}}

\vspace{5mm}

{\bf 3.1.} Singular Potentials

\vspace{3mm}

{\bf 3.2.} Iterative Algorithm

\vspace{3mm}

{\bf 3.3.} Smoothing Function

\vspace{3mm}

{\bf 3.4.} Response Functions

\vspace{3mm}

{\bf 3.5.} Correlation Function

\vspace{2cm}

{\parindent=0pt

{\large{\bf Appendix}}

\vspace{2cm}

{\large{\bf References}}}

\newpage

\section{Basic Notions}

This part contains those main definitions which statistical mechanics is 
based on and which are necessary for understanding the following parts.

\subsection{Pure States}

Consider a system of $N$ objects which we shall traditionally call 
particles. Put into correspondence to each particle a variable $x\in\Bbb{V}$ 
whose particular nature is, for the time being, of no importance. This can 
be a multivariable incorporating the real--space coordinates, spin, isospin
and other degrees of freedom, such as colour and flavour.

For dealing with $N$ particles, we need $N$ such multivariables. Enumerating 
the variables (but not the particles) by the index $i=1,2,\ldots,N$, we 
have the set of $N$ variables $x_i\in\Bbb{V}_i$. For brevity, we shall use
the notation
\begin{equation}
\label{1.1}
x^n:=\{ x_i|\; i=1,2,\ldots,n\leq N\}
\end{equation}
for the set defined on the manifold
\begin{equation}
\label{1.2}
\Bbb{V}^n:=\Bbb{V}_1\times \Bbb{V}_2\times \ldots \times \Bbb{V}_n \; .
\end{equation}

In addition to variables (1.1) we need one more variable
$$ t \in \Bbb{R}^+ := [0,\infty) $$
called {\it time}.

Let functions $\Psi(x^N,t)$, given on the domain $\Bbb{V}^N\times\Bbb{R}^+$,
pertain to a Hilbert space ${\cal H}_\Psi$ with the scalar product
\begin{equation}
\label{1.3}
(\Psi,\Psi'):=\int\Psi^*(x^N,t)\Psi'(x^N,t)dx^N ,
\end{equation}
in which $\Psi^*$ means complex conjugate of $\Psi$, integration goes over 
$\Bbb{V}^N$ and the differential measure on (1.2) is denoted by
$$ dx^n:=dx_1dx_2\ldots dx_n \; .$$
As usual, the integration implies an actual integration over continuous 
manifolds and summation over discrete ones.

If the elements of the Hilbert space ${\cal H}_\Psi$ are normalized,
\begin{equation}
\label{1.4}
(\Psi,\Psi) = 1 ,
\end{equation}
then the functions $\Psi$ are called {\it wave functions} or {\it quantum
states}; respectively, the space ${\cal H}_\Psi$ is the {\it space of
quantum states}. The quantity
\begin{equation}
\label{1.5}
|\Psi(x^N,t)|^2 :=\Psi^*(x^N,t)\Psi(x^N,t)
\end{equation}
is interpreted as the {\it distribution of particle density}.

The {\it particles} are called {\it identical} or {\it indistinguishable} 
if their corresponding variables pertain to the same manifold, 
$\Bbb{V}_i=\Bbb{V}$, and there is no difference which variable is ascribed 
to which particle, so that the particle--density distribution (1.5) does
not change under the permutation of any two variables:
\begin{equation}
\label{1.6}
|P(x_i,x_j)\Psi(x^N,t)|^2 =|\Psi(x^N,t)|^2 ,
\end{equation}
where $P(x_i,x_j)$ is the operator permuting the variables 
$x_i,x_j\in\Bbb{V}$. Emphasize that it only makes sense of permuting
variables but not particles.

The simplest sufficient condition for (1.6) is the {\it symmetry principle}
\begin{equation}
\label{1.7}
P(x_i,x_j)\Psi(x^N,t) =\pm \Psi(x^N,t) ,
\end{equation}
stating that the quantum states associated with identical particles are 
either symmetric or antisymmetric with respect to the permutation of
variables. Symmetric quantum states describe {\it Bose particles} or {\it
bosons} while antisymmetric quantum states, {\it Fermi particles} or {\it 
fermions}.

A set $\{ A\}$ of self--adjoint linear operators $A(x^N)$ is given on the 
space of quantum states ${\cal H}_\Psi$, such that the mathematical
expectations
\begin{equation}
\label{1.8}
\langle A\rangle := (\Psi,A\Psi)
\end{equation}
define observable quantities. A linear algebra of these operators forms the
{\it algebra of observables} $\{ A\}$. The set $\{\langle A\rangle\}$ of
mathematical expectations (1.8) is called {\it observable state}.

One of the operators from the algebra of observables plays a special role. 
This is a self--adjoint operator $H$ whose mathematical expectation gives 
the energy of the system. The operator $H$, called {\it Hamiltonian},
defines the {\it evolution equation}
\begin{equation}
\label{1.9}
i\hbar\frac{\partial}{\partial t}\Psi(x^N,t) = H(x^N)\Psi(x^N,t)
\end{equation}
for quantum states. This is the {\it Schr\"odinger equation}, $\hbar$ 
being the Planck constant.

Using wave functions one may construct another type of objects, called
density matrices [1-3]. Introduce the {\it pure density matrix}
\begin{equation}
\label{1.10}
D_N(t) =\left [ D_N(x^N,y^N,t)\right ] ,
\end{equation}
whose elements
\begin{equation}
\label{1.11}
D_N(x^N,y^N,t) :=\Psi(x^N,t)\Psi^*(y^N,t)
\end{equation}
are products of wave functions satisfying the symmetry principle (1.7). 
Consequently,
\begin{equation}
\label{1.12}
P(x_i,x_j)D_N(x^N,y^N,t) =\pm D_N(x^N,y^N,t) .
\end{equation}
The trace of (1.10), in compliance with (1.4), is
\begin{equation}
\label{1.13}
Tr\; D_n(t) := (\Psi,\Psi) = 1.
\end{equation}

Also, introduce the operator matrices
\begin{equation}
\label{1.14}
A_N = [ A_N(x^N,y^N) ]
\end{equation}
with elements
\begin{equation}
\label{1.15}
A_N(x^N,y^N) := A(x^N)\delta (x^N - y^N) ,
\end{equation}
where
$$ \delta(x^n- y^n) :=\delta ( x_1 - y_1)\delta(x_2 -y_2)\ldots
\delta(x_n - y_n) $$
is a product of the Dirac $\delta$--functions for continuos variables and
of the Kroneker $\delta$--symbols for discrete. Using (1.10) and (1.14), 
the mathematical expectation (1.8) is written as
\begin{equation}
\label{1.16}
\langle A\rangle = Tr\; A_ND_N(t) .
\end{equation}

The evolution equation for (1.10) follows from the Schr\"odinger equation
(1.9),
\begin{equation}
\label{1.17}
i\hbar\frac{d}{dt}D_N(t) = [ H_N,D_N(t) ] ,
\end{equation}
where $[\; ,\; ]$ means the commutator
$$ [A,B] := AB - BA \; . $$
Eq. (1.17) is the {\it Liouville equation}. According to (1.17), the 
evolution equation for (1.16) is
\begin{equation}
\label{1.18}
i\hbar\frac{d}{dt} \langle A\rangle = i\hbar\langle
\frac{\partial A}{\partial t}\rangle + \langle [ A,H ]\rangle .
\end{equation}

The pure density matrix (1.10) and operators of observables (1.14) permit
several representations. To pass from the $x$--representation given by 
(1.11) and (1.15) to another representation, one chooses a complete basis 
$\{\Psi_q(x^N)\}$ of orthonormal quantum states in ${\cal H}_\Psi$,
$$ (\Psi_q,\Psi_p) =\delta_{qp}\; , $$
where $q$ is a multiindex labelling quantum states. With an expansion
$$ \Psi(x^N,t) =\sum_q C_q(t)\Psi_q(x^N) , $$
Eq. (1.4) gives
$$ \sum_q |C_q(t)|^2 = 1 . $$

Instead of (1.10), one may present the pure density matrix as
\begin{equation}
\label{1.19}
D_N(t) = [ D_{qp}(t) ]
\end{equation}
with elements
\begin{equation}
\label{1.20}
D_{qp}(t) := C_q(t) C_p^*(t) ,
\end{equation}
which again have the form of a product, the function $C_q(t)$ playing the 
role of the wave function in the $q$--representation. In the mathematical 
expectation (1.16) one has to put now, instead of (1.14),
\begin{equation}
\label{1.21}
A_N = [ A_{qp} ]
\end{equation}
with elements
\begin{equation}
\label{1.22}
A_{qp} := (\Psi_q, A\Psi_p ) .
\end{equation}

In any representation, the pure density matrix consists of elements that 
are  factorizable as in (1.11) or (1.20). Because of this and of
normalization (1.13), the convolution property
\begin{equation}
\label{1.23}
\int D_N(x^N,z^N,t) D_N(z^N,y^N,t) dz^N = D_N(x^N,y^N,t)
\end{equation}
holds, which can be written in an operator form as
\begin{equation}
\label{1.24}
D_N^2(t) = D_N(t) .
\end{equation}
The latter defines the pure density matrix as an indempotent operator.

A {\it pure state} is an observable state $\{\langle A\rangle\}$ consisting
of mathematical expectations (1.16) defined through a pure density matrix.
The {\it indempotency property} (1.24) is the necessary and sufficient
condition of a pure state.

Reducing the number of variables in (1.11) by integrating
\begin{equation}
\label{1.25}
D_n(x^n,y^n,t) :=\int D_N(x^n,z^{N-n},y^n,z^{N-n},t)dz^{N-n} ,
\end{equation}
one defines the {\it reduced density matrices}, or submatrices,
\begin{equation}
\label{1.26}
D_n(t) := [ D_n(x^n,y^n,t)] \qquad (n=1,2,\ldots, N) ,
\end{equation}
implying that $D_0:=1$. It is not important which of the variables are
integrated out in (1.25) since (1.11) is symmetric under the double 
permutation
$$ P(x_i,x_j)P(y_i,y_j)D_N(x^N,y^N,t) = D_N(x^N,y^N,t) . $$
The normalization for (1.26),
\begin{equation}
\label{1.27}
Tr\; D_n(t) := \int D_n(x^n,x^n,t) dx^n = 1 ,
\end{equation}
follows from (1.13). The elements (1.25) do not satisfy the convolution 
property (1.23), except for the trivial cases $n=0$ or $n=N$. Thus, the 
reduced density matrices (1.26) do not represent indempotent operators,
$$ D_n^2(t) \neq D_n(t) \qquad (n=1,2,\ldots, N-1) . $$

The reduced density matrices are useful for dealing with operators
depending only on a part of the variables $x^N$. For example, for an
operator $A(x^N)$ defining the operator matrix
\begin{equation}
\label{1.28}
A_n = [ A_n(x^n,y^n) ]
\end{equation}
with elements
\begin{equation}
\label{1.29}
A_n(x^n,y^n) := A(x^n)\delta( x^n - y^n ) ,
\end{equation}
one can write the expectation (1.16) as
\begin{equation}
\label{1.30}
\langle A\rangle = Tr\; A_nD_n(t) .
\end{equation}
In this way, to calculate (1.30) for an operator (1.28) whose elements
(1.29) involve $n<N$ variables, one needs to know only the corresponding
reduced density matrix (1.26). Reduced density matrices have become a very
convenient tool for the many--body problem [4-14].

\subsection{Mixed States}

In reality, any considered system of $N$ particles is always a subsystem
of a bigger system. Real observable systems can be never completely
isolated [15-19]. Therefore, in addition to the variables $x^N$ ascribed
to the subsystem of interest, there exists a set $X$ of other variables
associated with the rest of a bigger system, which can be called the 
{\it background}. The whole system consisting of a subsystem and a
background can be named the {\it universe}.

Assume that the observable state of the universe is a pure state
characterized by a pure density matrix
\begin{equation}
\label{1.31}
D(t) = [ D(x^N,X,y^N,Y,t) ] ,
\end{equation}
whose elements
\begin{equation}
\label{1.32}
D(x^N,X,y^N,Y,t) :=\Psi(x^N,X,t)\Psi^*(y^N,Y,t)
\end{equation}
are products of the {\it universe wave functions}.

The algebra of observables $\{ A\}$ related to the subsystem of interest
consists of operators $A(x^N)$ depending on the subsystem variables $x^N$.
The mathematical expectations can be defined as in (1.16) but with the
reduced density matrix $D_N(t)$, whose elements are
\begin{equation}
\label{1.33}
D_N(x^N,y^N,t) := \int\Psi(x^N,X,t)\Psi^*(y^N,X,t) dX\; .
\end{equation}

To pass to another representation, one has to choose an orthonormal basis
$\{\Psi_q(x^N)\}$ and to make an expansion
$$ \Psi(x^N,X,t) =\sum_q C_q(X,t)\Psi_q(x^N) . $$
Then, (1.33) may be written as
\begin{equation}
\label{1.34}
D_N(x^N,y^N,t) =\sum_{qp}D_{qp}(t)\Psi_q(x^N)\Psi_p^*(y^N) ,
\end{equation}
where
$$ D_{qp}(t) =\int C_q(X,t) C_p^*(X,t) dX $$
are the elements of the reduced density matrix $D_N(t)$ in the
$q$--representation.

There is a sense in considering a system of $N$ particles as a subsystem
of the universe solely if any of these particles has ever interacted with
its surrounding background. This is just what always happens in reality.
It would be an extremely unrealistic fantasy to imagine that a considered
subsystem has never interacted with anything else. The fact that a
subsystem cannot be completely separated from its background means that
the universe wave function $\Psi(x^N,X,t)$ cannot be factorized into a
product of functions each depending separately on $x^N$ and $X$. This is
called the {\it principle of nonseparability} [13,20-22].

If the universe wave function is not factorizable in the above sense, then
the elements of the reduced density matrix $D_N(t)$, given by (1.33), are
not factorizable into a product of functions depending separately on $x^N$
and $y^N$, as in (1.10). Therefore (1.33) cannot define a pure density
matrix.

A density matrix consisting of nonfactorizable elements will be called the
{\it mixed density matrix}. The observable state $\{\langle A\rangle\}$,
formed by mathematical expectations $\langle A\rangle$ which are defined
through a mixed density matrix, is named the {\it mixed state}.

The necessity of characterizing any system of particles by a mixed state
is not based only on the principle of nonseparability, as is described
above, but can even be strengthen. Actually, the notion of an observable
state itself presupposes the possibility of measuring observable
quantities. But the process of measurement is impossible without
interactions between the considered system and a measuring apparatus.
Consequently, the latter is to be treated as a part of the background from
which the system cannot be isolated.

Moreover, the concept of isolatedness of a subsystem from the rest of the
universe is contradictory in its own, since to realize the isolation and
to check that the subsystem is kept isolated, one again needs to use some
technical and measuring devices [23,24].

Suppose that to check the isolatedness of a subsystem we use such a gentle
procedure which involves only the {\it minimally disturbing measurement}
[23,24], defined as follows: A measurement, finding the subsystem in a
quantum state $\Psi_q(x^N)$, leads solely to the appearance of a phase
factor $e^{i\xi_q}$,
\begin{equation}
\label{1.35}
\Psi_q(x^N) \rightarrow e^{i\xi_q}\Psi_q(x^N) ,
\end{equation}
where $\xi_q$ is a random real number, $\xi_q\in [0,2\pi]$. The measuring
process (1.35) does not change the distribution of particle density
$|\Psi_q(x^N)|^2$ in the quantum state $q$, as well as the matrix elements
$(\Psi_q,A\Psi_q)$. However, the total wave function
\begin{equation}
\label{1.36}
\Psi(x^N,\xi,t) =\sum_q C_q(t)e^{i\xi_q}\Psi_q(x^N)
\end{equation}
acquires the dependence on a set $\xi:=\{\xi_q\}$ of random variables,
which plays here the role of the set $X$ of background variables. To
preserve the normalization
\begin{equation}
\label{1.37}
\int|\Psi(x^N,\xi,t)|^2dx^Nd\xi = 1 ,
\end{equation}
the differential measure $d\xi$ is to be defined as
\begin{equation}
\label{1.38}
d\xi :=\prod_q\frac{d\xi_q}{2\pi}\; , \qquad \xi_q\in [0,2\pi] .
\end{equation}
The elements of the reduced density matrix, according to (1.33), are given
by
\begin{equation}
\label{1.39}
D_N(x^N,y^N,t) :=\int\Psi(x^N,\xi,t)\Psi^*(y^N,\xi,t) d\xi\; .
\end{equation}
The mathematical expectation $\langle A\rangle$ of an operator $A(x^N)$ is
$Tr\; A_ND_N(t)$, where the elements of $A_N$ and $D_N(t)$ are defined
either by (1.15) and (1.39) or by (1.22) and
\begin{equation}
\label{1.40}
D_{qp}(t) = |C_q(t)|^2\delta_{qp} ,
\end{equation}
respectively. The latter follows from (1.36), (1.38) and from the integral
$$ \int e^{-i(\xi_q-\xi_p)}d\xi = \delta_{qp} . $$
Neither (1.39) nor (1.40) are factorizable. Hence, the subsystem is in a
mixed state even for such an ideal case of the minimally disturbing 
measurements.

The above discussion forces us to accept the {\it principle of
nonisolatedness} which tells that no system can be completely isolated and
has to be described by a mixed density matrix. This general statement,
however, does not forbid the possibility that for some particular cases
a pure density matrix could be a reasonable approximation.

We shall call a {\it system quasi--isolated} if it is characterized by a
mixed density matrix and its time evolution is governed by a Hamiltonian
$H(x^N)$ depending on the system variables $x^N$, but not on the
background variables $X$,
\begin{equation}
\label{1.41}
i\hbar\frac{\partial}{\partial t}\Psi(x^N,X,t) = H(x^N)\Psi(x^N,X,t) .
\end{equation}
The mixed density matrix of a quasi--isolated system satisfies the same
Liouville equation (1.17) as a pure density matrix.

The formal solution of the Liouville equation may be written as
\begin{equation}
\label{1.42}
D_N(t) = U_N(t)D_N(0)U_N^+(t) ,
\end{equation}
where $U_N(t)$ is the {\it evolution operator} which is a unitary operator
obeying the Schr\"odinger equation
\begin{equation}
\label{1.43}
i\hbar\frac{d}{dt}U_N(t) = H_NU_N(t)
\end{equation}
and having the properties
$$ U_N(t)U_N(t') = U_N(t+t') , $$
\begin{equation}
\label{1.44}
U_N^+(t) = U_N^{-1}(t) , \qquad U_N(0) = 1 .
\end{equation}
In a particular case, when $H_N$ does not depend explicitly on time, the
evolution operator, according to (1.43), is
\begin{equation}
\label{1.45}
U_N(t) =\exp\left ( -\frac{i}{\hbar} H_N(t)\right ) \qquad
\left (\frac{\partial H_N}{\partial t} = 0\right ) .
\end{equation}

With the help of the evolution operator the time evolution for the
operators of observables is given as
\begin{equation}
\label{1.46}
A_N(t) := U_N^+(t) A_NU_N(t) .
\end{equation}
The form $A_N(t)$ is named the {\it Heisenberg representation} for the
operators, as compared to that of $A_N$ which is called the {\it
Schr\"odinger representation}. From (1.46) and (1.43) the evolution
equation
\begin{equation}
\label{1.47}
i\hbar\frac{d}{dt} A_N(t) = i\hbar\frac{\partial}{\partial t} A_N(t) +
[A_N(t), H_N(t)]
\end{equation}
follows, in which
\begin{equation}
\label{1.48}
\frac{\partial}{\partial t}A_N(t) := 
U_N^+(t)\frac{\partial A_N}{\partial t} U_N(t) .
\end{equation}
Eq.(1.47) is called the {\it Heisenberg equation}. Its important
particular case is when the operator $A_N$ does not depend explicitly on
time, then (1.47) and (1.48) yield
\begin{equation}
\label{1.49}
i\hbar\frac{d}{dt} A_N(t) = [ A_N(t),H_N(t) ] \qquad
\left ( \frac{\partial A_N}{\partial t} = 0\right ) .
\end{equation}

Using the properties of the evolution operator and the equality $Tr\; AB=
Tr\; BA$, one can write the expectation value (1.16) in one of two forms
\begin{equation}
\label{1.50}
\langle A\rangle = Tr\; D_N(t)A_N = Tr\; D_N(0)A_N(t) .
\end{equation}
Respectively, the evolution equation for (1.50) may be presented in two
ways
$$ i\hbar\frac{d}{dt}\langle A\rangle = 
i\hbar\langle \frac{\partial A_N}{\partial t}\rangle + 
Tr\; [H_N,D_N(t)] A_N = $$
$$ = i\hbar\langle \frac{\partial A_N}{\partial t}\rangle +
Tr\; D_N(0) [ A_N(t), H_N(t) ] . $$
The identity of these two forms results from the properties of trace and
from the relation (1.46) between the Schr\"odinger and Heisenberg
representations.

For mixed states, the reduced density matrices can be introduced as for
pure states in (1.25) and (1.26), though now there is no principal
difference between $D_N(t)$ and $D_n(t)$ with $n<N$, since for a mixed
state $D_N(t)$ itself is a reduced density matrix. Dealing with mixed
states it is reasonable to consider all density matrices $D_n(t)$ as
reduced, with the index $n$ ranging from zero, when $D_0(t):=1$, up to
infinity.

The reduced density matrices for mixed states have the symmetry property
$$ P(x_i,x_j)D_n(x^n,y^n,t) =\pm D_n(x^n,y^n,t)$$
as (1.12) for pure states.

\subsection{Field Operators}

A powerful method of treating quasi--isolated systems is the field
approach based on the notion of field operators. This approach is also
known as the method of {\it second quantization}. There are several ways
of describing the latter. Here we expound a variant which is, to our mind,
the most general and best mathematically grounded [25].

Let ${\cal H}_n$ be a Hilbert space of square--integrable functions
$h_n(x^n)$ with a domain $\Bbb{V}^n$ and the scalar product
$$ (h_n,g_n) :=\int h^*_n(x^n)g_n(x^n)dx^n , $$
where the integration goes over $\Bbb{V}^n$. Construct a Hilbert space
${\cal H}$ as a direct sum
\begin{equation}
\label{1.51}
{\cal H} := \oplus_{n=0}^\infty{\cal H}_n \qquad ({\cal H}_0 :=\Bbb{C}).
\end{equation}
A vector $h\in{\cal H}$ is presented as a column
\begin{eqnarray}
\label{1.52}
h:= [h_n(x^n) ] =\left [ \begin{array}{c}
h_0 \\
h_1(x_1) \\
h_2(x_1,x_2) \\
. \\
. \\
. \\
h_n(x_1,x_2,\ldots,x_n) \\
. \\
. \\
. \\
\end{array}\right ] \; .
\end{eqnarray}

The latter is a {\it quantum state}, or {\it microstate}, and ${\cal H}$
is a space of states or {\it phase space}. The scalar product in (1.51)
is given by
$$ (h,g) := h^+g =\sum_{n=0}^\infty (h_n,g_n) . $$

Define matrices of the form
\begin{equation}
\label{1.53}
\hat A := [ A_{mn}(x^m,y^n) ] ,
\end{equation}
whose action on the vector (1.52) is a vector
$$ \hat A h = \left [ \sum_{n=0}^\infty\int A_{mn}(x^m,y^n) h_n(y^n)dy^n
\right ] , $$
and the multiplication of two such matrices is given by the rule
$$ \hat A\hat B = \left [\sum_{k=0}^\infty\int A_{mk}(x^m,z^k)
B_{kn}(z^k,y^n)dz^k\right ] . $$

Introduce the {\it annihilation operator}
\begin{equation}
\label{1.54}
\hat\psi(x) =\left [ \delta_{m\; n-1}\sqrt{n}\delta( x^{n-1} - y^{n-1})
\delta( x - y_n)\right ]
\end{equation}
and the {\it creation operator}
\begin{equation}
\label{1.55}
\hat\psi^\dagger(x) =\left [ \delta_{m\; n+1}\sqrt{n+1}\delta( x^n -
y^n ) \delta( x_{n+1} - x)\right ] ,
\end{equation}
having the form of the matrix (1.53) with the corresponding multiplication
rules. As follows from above, the operators (1.54) and (1.55) are adjoint
to each other; these are the {\it field operators}. Since the matrices
(1.54) and (1.55) contain the Dirac $\delta$--functions, the field
operators are operator distributions. The correct definition of
distribution presupposes the existence of a space of square--integrable
functions $f(x)$, so that
\begin{equation}
\label{1.56}
\hat\psi_f =\int f(x)\hat\psi(x)dx , \qquad \hat\psi_f^\dagger =
\int f(x)\hat\psi^\dagger(x)dx
\end{equation}
becomes operators in ${\cal H}$, whose domain of definition, as well as
their range, contain the set of all finite vectors.

Action of the field operators on the vector (1.52) gives
$$ \hat\psi(x) h =\left [\sqrt{n+1} h_{n+1}(x^n,x)\right ] , $$
$$ \hat\psi^\dagger(x) h =\left [\sqrt{n} h_{n-1}(x^{n-1})\delta(x_n -x)
\right ] , $$
so that $\hat\psi_fh$ and $\hat\psi_f^\dagger h$ are really finite
vectors.

The explicit definition of the field operators  (1.54) and (1.55) imposes
no simple commutation relations on these matrices. To understand this, it
is sufficient to examine their products
$$ \hat\psi(x)\hat\psi(x') = \left [\delta_{m\; n-2}\sqrt{n(n-1)}
\delta( x^{n-2} - y^{n-2})\delta ( x - y_{n-1})\delta (x' - y_n)
\right ] , $$
$$ \hat\psi^\dagger(x)\hat\psi^\dagger(x') = \left [
\delta_{m\;n+2}\sqrt{(n+1)(n+2)}
\delta( x^n - y^n)\delta ( x_{n+1} - x')\delta (x_{n+2} - x) \right ] , $$
$$ \hat\psi(x)\hat\psi^\dagger(x') = \left [\delta_{mn}(n+1)
\delta( x^n - y^n)\delta ( x - x' ) \right ] , $$
$$ \hat\psi^\dagger(x')\hat\psi(x) = \left [\delta_{mn}n
\delta( x^{n-1} - y^{n-1})\delta ( x_n - x' )\delta ( y_n - x)\right ] , $$

The algebraic properties of operators depend on three factors: the
explicit form of an operation, the properties of the space on which the
operators are defined, and the existence of some additional conditions.
This is well illustrated by the example of field operators. Their explicit
definition as matrices (1.54) and (1.55) does not permit to extract any
simple commutation relations. Even if we narrow the domain space to a
subspace enjoying some simple symmetry properties, this does not necessary
lead to simple commutation rules for operators. The latter can be
demonstrated by the actions
$$ \hat\psi(x)\hat\psi(x')h = 
\left [\sqrt{(n+1)(n+2)} h_{n+2}(y^n,x,x') \right ] , $$
$$ \hat\psi^\dagger(x)\hat\psi^\dagger(x')h = 
\left [\sqrt{n(n-1)} h_{n-2}(x^{n-2})\delta( x_{n-1} - x')
\delta (x_n - x) \right ] , $$
$$ \hat\psi(x)\hat\psi^\dagger(x')h = 
\left [ (n+1) h_n(x^n)\delta( x  - x') \right ] , $$
$$ \hat\psi^\dagger(x')\hat\psi(x) h = 
\left [ n h_n(x^{n-1},x)\delta( x_n  - x') \right ] \; . $$
Suppose that we are considering a subspace of (1.51) whose vectors consist
of elements having some fixed symmetry, for instance, the functions
$h_n(x^n)$ are symmetric or antisymmetric. But, as is seen from above, the
action of field operators does not preserve this symmetry, except for the
action of the annihilation operators, $\hat\psi(x)\hat\psi(x')$. This
happens because the domain of the field operators, when this domain is a
subspace of (1.51), generally, does not coincide with their range. Recall
that by the range of field operators we mean the range of operators
(1.56).

It is clear, that to obtain simple commutation relations, one needs to
invoke some additional conditions. Before doing this, let us explain why
(1.54) and (1.55) are called annihilation and creation operators,
respectively.

Choose as a basis of (1.51) that consisting of vectors
\begin{eqnarray}
\label{1.57}
|n\rangle := [\delta_{mn}h_n(x^n) ] =\left [ \begin{array}{c}
0 \\
0 \\
. \\
. \\
. \\
0 \\
h_n(x^n) \\
0 \\
0 \\
. \\
. \\
. \\
\end{array}\right ] .
\end{eqnarray}
A vector adjoint to (1.57) is a row
\begin{equation}
\label{1.58}
\langle n| :=\left [ 0\; 0\; \ldots h_n^*(x^n)\; 0 \; 0\; \ldots\right ] .
\end{equation}
The notation used in (1.57) and (1.58) is called the {\it Dirac notation}.

The scalar product of elements from the basis $\{ |n\rangle\}$ is
$$ \langle m|n\rangle =\delta_{mn}(h_n,h_n ) . $$
So, it is a complete orthogonal basis. It is easily normalizable by
putting
$$ \langle n|n\rangle = (h_n,h_n) = 1 . $$
Then, (1.57) is called the $n$--{\it particle state}.

The $0$--particle state, that is
\begin{eqnarray}
\label{1.59}
|0\rangle := [\delta_{m0}h_0 ] =\left [ \begin{array}{c}
h_0 \\
0 \\
0 \\
. \\
. \\
. \\
\end{array}\right ] ,
\end{eqnarray}
with $|h_0|=1$, is called the {\it vacuum vector} or {\it vacuum state}.

The action of the annihilation operator (1.54) on the vacuum vector (1.59)
is
\begin{equation}
\label{1.60}
\hat\psi(x) |0\rangle = 0 ,
\end{equation}
and its action on other vectors from the basis $\{ |n\rangle\}$, with
$n=1,2,\ldots$, gives
\begin{equation}
\label{1.61}
\hat\psi(x) |n\rangle = \sqrt{n}\left [\delta_{m\; n-1} h_n(x^{n-1},x)
\right ] .
\end{equation}
The action of the creation operator (1.55) on an $n$--particle vector
(1.57) yields
\begin{equation}
\label{1.62}
\hat\psi^\dagger(x) |n\rangle = \sqrt{n+1}\left [\delta_{m\; n+1}
h_n(x^n)\delta ( x_{n+1} - x ) \right ] .
\end{equation}

For the operators (1.56) we have from (1.60) and (1.61)
$$ \hat\psi_f|0\rangle = 0 , $$
$$ \hat\psi_f|n\rangle = \sqrt{n} [\delta_{m\; n-1} h_{n-1}'(x^{n-1})] , $$
with 
$$ h_{n-1}'(x^{n-1}) = \int h_n(x^{n-1},x)f(x) dx , $$
which shows that $\hat\psi_f$ transforms an $n$--particle vector to an
$(n-1)$--particle vector. In other words, $\hat\psi_f$ annihilates a
particle. From (1.62) we get
$$ \hat\psi_f^\dagger(x) |n\rangle = \sqrt{n+1}\left [\delta_{m\; n+1}
h_{n+1}'(x^{n+1}) \right ]  $$
with
$$ h_{n+1}'(x^{n+1}) = h_n(x^n) f(x_{n+1}) , $$
which shows that $\hat\psi_f^\dagger$ transforms an $n$--particle vector
to an $(n+1)$--particle vector, that is, creates a particle.

In the same way one can check that the double action of the field
operators, given by
$$ \hat\psi(x)\hat\psi^\dagger(x')|n\rangle = (n+1) \left [
\delta_{mn} h_n(x^n)\delta( x - x')\right ] , $$
$$ \hat\psi^\dagger(x')\hat\psi(x) |n\rangle = n \left [
\delta_{mn} h_n(x^{n-1},x)\delta( x_n - x')\right ] , $$
means that starting from an $n$--particle vector and first creating a
particle and then annihilating it, or first annihilating a particle and
then creating it, one always returns to an $n$--particle vector, though
not necessarily the same.

By acting $n$ times on the vacuum state (1.59), with $h_0=1$, by the
creation operator,
$$ \prod_{i=1}^n \hat\psi^\dagger(x_i') |0\rangle = \sqrt{n!}
\left [ \delta_{mn}\prod_{i=1}^n\delta( x_i - x_i')\right ] , $$
one can construct any $n$--particle vector (1.57) as
\begin{equation}
\label{1.63}
|n\rangle =\frac{1}{\sqrt{n!}}
\int h_n(x^n)\prod_{i=1}^n\hat\psi^\dagger(x_i)|0\rangle dx^n
\end{equation}
and any vector of space (1.51) as
$$ h =\sum_{n=0}^\infty C_n|n\rangle , $$
where $C_n$ are complex numbers.

Till now, we have not required from the vectors of the space (1.51) any
symmetry properties. But, according to the symmetry principle discussed in
Section 1.1, the quantum states of identical particles are to be either
symmetric or antisymmetric with respect to permutation of their variables.
Symmetric quantum states correspond to bosons, antisymmetric to fermions.
Thus, we have to construct the subspace of (1.51) possessing the desired
symmetry properties.

The set $x^n$ of the variables $x_1,x_2,\ldots,x_n$ permits one to make
$n!$ permutations. Enumerate these permutations by the index
$i=0,1,2,\ldots,(n-1)!$ and denote the corresponding permutation operators
by $P_i(x^n)$, so that
$$ P_0(x^n) = 1 , \qquad P_i^2(x^n) = 1 . $$

Using arbitrary functions $h_n(x^n)$, we may construct symmetric and
antisymmetric functions by the receipt
\begin{equation}
\label{1.64}
f_n^\pm(x^n) =\frac{1}{\sqrt{n!}}\sum_{i=0}^{(n-1)!} (\pm 1)^i
P_i(x^n)h_n(x^n) .
\end{equation}
By construction, from (1.64) it follows
\begin{equation}
\label{1.65}
P_i(x^n) f_n^\pm(x^n) = \pm f_n^\pm(x^n) \qquad (i\geq 1) .
\end{equation}
Denote a Hilbert space of such functions as
$$ {\cal H}_n^\pm = \{ f_n^\pm(x^n)\} . $$
Analogously to (1.51), define the Hilbert space
\begin{equation}
\label{1.66}
{\cal H}_\pm : =\oplus_{n=0}^\infty{\cal H}_n^\pm
\end{equation}
of vector columns
\begin{equation}
\label{1.67}
f_\pm = [ f_n^\pm(x^n) ]\in {\cal H}_\pm \; .
\end{equation}
The symmetric space ${\cal H}_+$ represents Bose states while the
antisymmetric space ${\cal H}_-$, Fermi states. A space having the
structure (1.66), possessing a vacuum vector, is a {\it Fock space}.

The symmetric and antisymmetric spaces in (1.66) are subspaces of (1.51),
and can be considered as projections
\begin{equation}
\label{1.68}
{\cal H}_\pm = P_\pm {\cal H}
\end{equation}
of ${\cal H}$ by means of projectors $P_+$ and $P_-$.

Introduce the field operators
$$ \psi_\pm(x) := P_\pm\psi(x)P_\pm , $$
\begin{equation}
\label{1.69}
\psi^\dagger_\pm(x) := P_\pm\psi^\dagger(x)P_\pm , 
\end{equation}
such that $\psi_{f+}$ and $\psi_{f+}^\dagger$, defined as in (1.56),
always act inside the symmetric space ${\cal H}_+$, while $\psi_{f-}$ and
$\psi^\dagger_{f-}$, inside the antisymmetric space ${\cal H}_-$. By this
definition
\begin{equation}
\label{1.70}
\psi_+(x)\psi_-(x') =\psi_+(x)\psi^\dagger_-(x') = 0 .
\end{equation}
In what follows, for simplicity of notation, we shall write $\psi(x)$,
representing either $\psi_+(x)$ or $\psi_-(x)$ from (1.69), and
$\psi^\dagger(x)$ being either $\psi^\dagger_+(x)$ or $\psi^\dagger_-(x)$.
Respectively, we will write $f$ instead of $f_\pm$ from (1.67). To
distinguish a commutator from an anticommutator, we denote them as
$$ [\hat A,\hat B]_\mp :=\hat A\hat B\mp \hat B\hat A . $$

Using the formulas of this Section for the field operator (1.54) and
(1.55), for the operators (1.69) we obtain the {\it commutation relations}
$$ [ \psi(x),\psi(x') ]_\mp f = 0 , $$
$$ [ \psi^\dagger(x),\psi^\dagger(x') ]_\mp f = 0 , $$
\begin{equation}
\label{1.71}
[ \psi(x),\psi^\dagger(x') ]_\mp f = \delta(x - x') f , 
\end{equation}
the upper sign being related to Bose-- and the lower, to Fermi--cases.
One also says that the commutators in (1.71) define {\it Bose statistics}
while the anticommutator, {\it Fermi statistics}. Since the relations
(1.71) are valid for all corresponding vectors from (1.66), a vector $f$
in (1.71) can be omitted, for example
$$ [ \psi(x),\psi^\dagger(x') ]_\mp  = \delta(x - x')\hat 1 , $$
where the unity matrix is
$$ \hat 1 := [\delta_{mn}\delta( x^n - y^n ) ] . $$
Usually, for brevity, one omits the unity matrix as well.

Recall again that $\psi(x)$ and $\psi^\dagger(x)$ are operator
distributions. Actual field operators are given, as in (1.56) by
\begin{equation}
\label{1.72}
\psi_f =\int f(x)\psi(x) dx , \qquad \psi_f^\dagger = 
\int f(x)\psi^\dagger(x) dx .
\end{equation}
The commutation relations for (1.72) are
$$ [\psi_f,\psi_g]_\mp = [\psi^\dagger_f,\psi^\dagger_g ]_\mp = 0 , $$
$$ [ \psi_f,\psi_g^\dagger ]_\mp = (f,g) $$
for any square--integrable functions $f(x)$ and $g(x)$.

The time evolution of the field operators is governed by a Hamiltonian
$\hat H$ which is a matrix of the form (1.53). The {\it evolution
operator} $U(t)$ is defined, analogously to (1.43) and (1.44), as a
unitary operator satisfying the Schr\"odinger equation
\begin{equation}
\label{1.73}
i\hbar\frac{d}{dt} U(t) =\hat H U(t)
\end{equation}
and possessing the properties
$$ U(t)U(t') = U(t + t') , $$
\begin{equation}
\label{1.74}
U^+(t) = U^{-1}(t) = U(-t) ,
\end{equation}
$$ U(0) = 1 . $$
An annihilation field operator at time $t$ is given, as in (1.46), by the
relation
\begin{equation}
\label{1.75}
\psi(x,t) := U^+(t)\psi(x,0)U(t) ,
\end{equation}
in which
\begin{equation}
\label{1.76}
\psi(x,0) :=\psi (x) .
\end{equation}
In the same way, the creation field operator $\psi^\dagger(x,t)$ is
defined.

The evolution equation for $\psi(x,t)$ follows from (1.75),
\begin{equation}
\label{1.77}
i\hbar\frac{\partial}{\partial t}\psi(x,t) = [ \psi(x,t) , \hat H(t) ]
\end{equation}
where $[,]$ means a commutator and
$$ \hat H(t) := U^+(t)\hat HU(t) . $$
The equation (1.77), together with its Hermitian conjugate giving the
evolution equation for $\psi^\dagger(x,t)$, are the {\it Heisenberg
equations} for the field operators.

\subsection{Local Observables}

The operators corresponding to observable quantities are constructed with
the help of the field operators. In doing this, one has to keep in mind
that the manifold $\Bbb{V}$ can be unbounded. Therefore, first one has to
consider submanifolds of $\Bbb{V}$ which are open bounded manifolds. Let
$\bf{\Lambda}\subset\Bbb{V}$ be such a submanifold. The general structure
of an {\it operator of a local observable} is
\begin{equation}
\label{1.78}
\hat A_\Lambda := \sum_{m,n=0}^\infty\frac{1}{\sqrt{m!n!}}
\int_{\bf{\Lambda}}\psi^\dagger(x_m)\ldots\psi^\dagger(x_1)
A(x^m,y^n)\psi(y_1)\ldots \psi(y_n)dx^mdy^n ,
\end{equation}
where $A(x^m,y^n)$ can be a generalized function or an operator defined on
functions with the domain ${\bf\Lambda}^m\times{\bf\Lambda}^n$.
Integration in (1.78) for all variables goes over the same submanifold
$\bf{\Lambda}$.

For example, the {\it local number--of--particles operator} is
\begin{equation}
\label{1.79}
\hat N_\Lambda =\int_{\bf{\Lambda}}\psi^\dagger(x)\psi(x)dx .
\end{equation}
Using the {\it density--of--particle operator}
\begin{equation}
\label{1.80}
\hat n(x) := \psi^\dagger(x)\psi(x) ,
\end{equation}
(1.79) can be written as
$$ \hat N_\Lambda =\int_{\bf{\Lambda}}\hat n(x)dx . $$
The action of (1.80) on an $n$--particle vector (1.57) gives
$$ \hat n(x)|n\rangle = 
n\left [\delta_{mn} f_n(x^{n-1},x)\delta( x_n - x)\right ] , $$
from where
$$ \hat N_\Lambda|n\rangle = n|n\rangle . $$
The $n$--particle vectors are eigenvectors of $\hat N_\Lambda$, the
numbers $n=0,1,2,\ldots$, being eigenvalues. This explains why (1.79) is
called the number--of--particle operator.

The time evolution of operators (1.78) follows from (1.75) giving
\begin{equation}
\label{1.81}
\hat A_\Lambda(t) = U^+(t)\hat A_\Lambda  U(t) .
\end{equation}
The evolution equation for (1.81), in compliance with (1.73), is the
Heisenberg equation
\begin{equation}
\label{1.82}
i\hbar\frac{d}{dt}\hat A_\Lambda(t) = 
i\hbar\frac{\partial}{\partial t}\hat A_\Lambda(t) + [\halt(t),\hat H(t)] ,
\end{equation}
where
\begin{equation}
\label{1.83}
\frac{\prt}{\prt t}\halt(t) := U^+(t)\frac{\prt\halt}{\prt t} U(t) ,
\end{equation}
which is equivalent to (1.47) and (1.48).

Introduce the {\it statistical operator}, or {\it density operator},
$\hat\rho(t)$, which is a positive bounded self--adjoint operator,
normalized to unity,
\begin{equation}
\label{1.84}
Tr\;\hat\rho(t) = 1 ,
\end{equation}
and satisfying the Liouville equation
\begin{equation}
\label{1.85}
i\hbar\frac{d}{dt}\hat\rho(t) = [\hat H(t),\hat\rho(t) ] .
\end{equation}
The time evolution, following from (1.85), is
\begin{equation}
\label{1.86}
\hat\rho(t) = U(t)\hat\rho U^+(t) ; \qquad \hat\rho :=\hat\rho(0) .
\end{equation}

The mathematical expectations for the operators of local observables are
called the {\it statistical averages} and are defined as
\begin{equation}
\label{1.87}
\lgl\halt\rgl := Tr\;\hat\rho(t)\halt = Tr\;\hat\rho\halt(t) ,
\end{equation}
which is analogous to (1.50). The trace in (1.87), as well as in (1.84), 
is taken over any basis from either ${\cal H}_+$ or ${\cal H}_-$ according
to the statistics of particles. Since the average (1.87) depends on time,
we may write
$$ \lgl\halt\rgl = \lgl\halt(t)\rgl . $$

Define the {\it statistical density matrices} through their elements
\begin{equation}
\label{1.88}
\rho_{mn}(x^m,y^n,t) = Tr\;\psi(x_1)\ldots\psi(x_m)
\hat\rho(t)\psi^\dagger(y_n)\ldots\psi^\dagger(y_1) .
\end{equation}
Then the average (1.87) takes the form
\begin{equation}
\label{1.89}
\lgl\halt\rgl = \sum_{m,n=0}^\infty
\frac{1}{\sqrt{m!n!}}\int_{\bf{\Lambda}} A(x^m,y^n)\rho_{mn}(y^n,x^m,t)
dx^m dy^n .
\end{equation}

The widespread situation is when
$$ A(x^m,y^n) =\delta_{mn} A(x^n,y^n) . $$
Then we need only the elements (1.88) that are diagonal with respect to
the indices $m,n$, that is
\begin{equation}
\label{1.90}
\rho_n(x^n,y^n,t) := \rho_{nn}(x^n,y^n,t ) .
\end{equation}
In this case, (1.89) reads
\begin{equation}
\label{1.91}
\lgl\halt\rgl =\sum_{n=0}^\infty \frac{1}{n!}
\int_{\bf{\Lambda}} A(x^n,y^n)\rho_n(y^n,x^n,t) dx^n,dy^n .
\end{equation}

A linear space
\begin{equation}
\label{1.92}
{\cal A}_\Lambda := \{\halt|\; \bf{\Lambda}\subset\Bbb{V}\}
\end{equation}
of the self--adjoint operators of local observables is an {\it algebra of
local observables} [26,27].

A set
\begin{equation}
\label{1.93}
\lgl{\cal A}_\Lambda\rgl := \{\lgl\halt\rgl\} ,
\end{equation}
formed by the statistical averages (1.89), is called the {\it statistical
state}.

If the manifold $\Bbb{V}$ is unbounded, we can define an ordered sequence
$\{\bf{\Lambda}_1\subset\bf{\Lambda}_2\subset\ldots\subset\bf{V}\}$ of
open submanifolds $\bf{\Lambda}_k$, such that
$$ \lim_{k\ra\infty}\bf{\Lambda}_k = \Bbb{V} , $$
and define a set of algebras, $\{{\cal A}_{\Lambda_1}\subset{\cal
A}_{\Lambda_2}\subset\ldots{\cal A}_{\Lambda_k}\subset\ldots\}$. The
inductive limit
\begin{equation}
\label{1.94}
{\cal A}_V :=\lim_{k\ra\infty}{\cal A}_{\Lambda_k}
\end{equation}
is called the {\it quasilocal algebra}, or {\it global algebra of
observables}.

\subsection{Order Indices}

The statistical density matrices
\begin{equation}
\label{1.95}
\rho_n(t) : = \left [ \rho_n(x^n,y^n,t)\right ] \qquad
(\rho_0 := 1 )
\end{equation}
with elements (1.90) are matrices in the $x-y$ space. Their action on a
vector
$$ f_n := [f(x^n)] , $$
being a column with respect to $x$, is given by the rule
$$ \rho_n(t)f_n := \left [\int \rho_n(x^n,y^n,t)f_n(y^n)dy^n\right ] . $$
The elements (1.90), owing to the evolution properties (1.75) and (1.86),
can be written in two forms:
$$ \rho_n(x^n,y^n,t) := Tr\;\psi(x_1)\ldots\psi(x_n)\hat\rho(t)
\psi^\dgr(y_n)\ldots\psi^\dgr(y_1) = $$
\begin{equation}
\label{1.96}
= Tr\; \psi(x_1,t)\ldots\psi(x_n,t)\hat\rho\psi^\dgr(y_n,t)\ldots
\psi^\dgr(y_1,t) .
\end{equation}
According to the commutation relations (1.71), the symmetry property
$$ P(x_i,x_j)\rho_n(x^n,y^n,t) =\pm\rho_n(x^n,y^n,t) $$
holds.

The elements (1.96) cannot, in general, be factorized into products of
functions depending separately on $x^n$ and $y^n$. This means that, by
construction, the statistical state is a mixed state. For latter, as is
discussed in Section 1.2, all density matrices with $n=0,1,2,\ldots$ can
be called reduced. This is especially natural when working in the
field--operator representation where the number of particles is not fixed.
Then one calls $\rho_n(t)$ simply an $n$--matrix.

However, we may fix the {\it average number of particles}
\begin{equation}
\label{1.97}
N :=\lgl\hat N_V\rgl = \int\rho_1(x,x,t) dx ,
\end{equation}
where the integration is over $\Bbb{V}$, and
\begin{equation}
\label{1.98}
\rho_1(x,x,t) =\lgl\hat n(x)\rgl = Tr\;\hat\rho(t)\hat n(x) .
\end{equation}
Then, it would be reasonable to distinguish among all $n$--matrices those
for which $n<N$ and to apply the name reduced only to them.

Let us fix the average number of particles (1.97) and consider a subspace
of the Fock space for which
\begin{equation}
\label{1.99}
\lgl\hat N_V\hat A_V\rgl = N\lgl\hat A_V\rgl .
\end{equation}
Then for the $2$--matrix with elements
\begin{equation}
\label{1.100}
\rho_2(x_1,x_2,y_1,y_2,t) = 
\lgl\psi^\dgr(y_2)\psi^\dgr(y_1)\psi(x_1)\psi(x_2)\rgl ,
\end{equation}
using the commutation relations (1.71), we can write
$$ \rho_2(x_1,x_2,y_1,y_2,t)
=\lgl\psi^\dgr(y_2)\psi(x_2)\psi^\dgr(y_1)\psi(x_1)\rgl -
\delta(x_2 - y_1)\lgl\psi^\dgr(y_2)\psi(x_1)\rgl . $$
From here, taking account of (1.99), we get
\begin{equation}
\label{1.101}
\int\rho_2(x_1,x,x_2,x,t)dx = (N-1)\rho_1(x_1,x_2,t) .
\end{equation}
Integrating (1.101) once more, and using (1.97), we find
$$ \int\rho_2(x_1,x_2,x_1,x_2,t) dx_1dx_2 = N(N-1) . $$
Generally, the following recursion relation takes place:
\begin{equation}
\label{1.102}
\int\rho_n(x^{n-1},x,y^{n-1},x,t)dx = 
(N - n + 1)\rho_{n-1}(x^{n-1},y^{n-1},t) .
\end{equation}
The normalization condition for an $n$--matrix is
\begin{equation}
\label{1.103}
\int\rho_n(x^n,x^n,t)dx^n =\frac{N!}{(N-n)!}\; .
\end{equation}

Note that the normalization (1.103) for $\rho_n(t)$ differs from the
normalization (1.27) for $D_n(t)$.

Consider a system with a fixed average number of particles $N$ and
occupying in the real space a volume $V$. One says that the {\it
thermodynamic limit} is performed if
\begin{equation}
\label{1.104}
N\ra\infty\; , \qquad V\ra\infty\; , \qquad \frac{N}{V}\ra const\; ,
\end{equation}
which is denoted as $\lim_{N\ra\infty}$ or $\lim_{V\ra\infty}$.

An ensemble of particles is called a {\it thermodynamic system}, if the
statistical state (1.93) under the thermodynamic limit (1.104) behaves so
that
\begin{equation}
\label{1.105}
\lim_{N\ra\infty}\left |\frac{1}{N}\lgl{\cal A}_V\rgl\right | < \infty .
\end{equation}
For example, the limit
\begin{equation}
\label{1.106}
\rho:=\lim_{V\ra\infty}\frac{1}{V}\lgl\hat N_V\rgl
\end{equation}
means the {\it average density of particles}, and the limit
\begin{equation}
\label{1.107}
\lim_{N\ra\infty}\frac{1}{N}E_N := \lim_{V\ra\infty}\frac{1}{N}
\lgl\hat H_V\rgl
\end{equation}
gives the {\it internal energy per particle}.

To understand the behaviour of density matrices under the thermodynamic
limit, define the norm
\begin{equation}
\label{1.08}
||\rho_n(t)|| :=\sup_{||f_n||\leq 1}\left ( \rho_n(t)f_n,\rho_n(t)f_n
\right )^{1/2} ,
\end{equation}
where $\rho_n(t)$ is the density matrix (1.95) and
$$ || f_n|| := ( f_n,f_n )^{1/2} . $$
The statistical operator $\hat\rho(t)$ is, by definition, a positive
self--adjoint operator. Then, the density matrix (1.95) with elements
(1.96) is also positive and self--adjoint. Therefore, the norm (1.108) can
also be presented as
$$ || \rho_n(t)|| =\sup_{||f_n||\leq 1}\left | (f_n,\rho_n(t)f_n)
\right | = $$
\begin{equation}
\label{1.109}
= \sup_{||f_n||\leq 1}\int f_n^*(x^n)\rho_n(x^n,y^n,t) f_n(y^n)dx^ndy^n .
\end{equation}

Define the {\it order indices} [28-31] for density matrices by the limit
\begin{equation}
\label{1.110}
\al_n(t) :=\lim_{N\ra\infty}\frac{\ln||\rho_n(t)||}{\ln N} ,
\end{equation}
where the thermodynamic limit (1.104) is assumed and $n=0,1,2,\ldots$.
Each density matrix $\rho_n(t)$ is positive and obeys the normalization
condition (1.103), because of which, using the Stirling formula for
$N\ra\infty$, we have
\begin{equation}
\label{1.111}
0\leq \al_n \leq n .
\end{equation}

If the eigenvalue problem
$$ \rho_n(t)\vp_{nq}(t) =\gamma_{nq}(t)\vp_{nq}(t) $$
is solvable for an $n$--matrix, then
$$ ||\rho_n(t)|| = \sup_q\gamma_{nq}(t) :=\gamma_n(t) , $$
and the order indices (1.110) can be expressed through the {\it largest
eigenvalues} of the corresponding $n$--matrices:
\begin{equation}
\label{1.112}
\al_n(t) =\lim_{N\ra\infty}\frac{\ln\gamma_n(t)}{\ln N}\; .
\end{equation}

This shows the importance of the largest eigenvalues of density matrices
in the thermodynamic limit.

\subsection{Information Functionals}

To define the density matrices with elements (1.96), we need to know the
statistical operator $\hat\rho(t)$. The latter obeys the Liouville
equation (1.85). This operator equation not always defines $\hat\rho(t)$
uniquely but some additional conditions may be required. Such conditions
can be formulated by invoking the notions of information theory [32].

In what follows, implying that the operators of local observables are
associated with a manifold $\Bbb{V}$, for simplicity, we shall not ascribe
the corresponding index at $\hat A_V$ writing instead just $\hat A$. Also,
remind that if $\ha$ is an operator from the algebra ${\cal A}$ of local
observables, and $F[\ha(t)]$ is a function of $\ha(t)$, then the evolution
of $F[\ha(t)]$ is given by the law
$$ F[\ha(t)] = F\left [ U^+(t)\ha U(t)\right ] = U^+(t)F[\ha]U(t) , $$
where $\ha:=\ha(0)$. This law can be considered either as a definition or
it can be proved if the function $F[\ha]$ is defined as a Taylor expansion
in powers of $\ha$. Similarly, if $F[\hr(t)]$ is a function of the
statistical operator, then, in compliance with (1.86),
$$ F[\hr(t)] = F\left [ U(t)\hr U^+(t)\right ] = U(t)F[\hr] U^+(t) , $$
where $\hr=\hr(0)$.

Introduce the {\it information operator}
\begin{equation}
\label{1.113}
\hat I(t) := \ln\hr(t) .
\end{equation}
The time evolution for (1.113), owing to the evolution laws discussed
above, is
\begin{equation}
\label{1.114}
\hat I(t) = U(t)\hat I(0) U^+(t) ,
\end{equation}
where
$$ \hat I(0) :=\ln\hr(0) =\ln\hr . $$
Consequently, the information operator (1.113) satisfies the Liouville
equation
\begin{equation}
\label{1.115}
i\hbar\frac{d}{dt} \hat I(t) =\left [\hat H(t),\hat I(t)\right ] .
\end{equation}
The {\it Shannon information} is defined as a functional
\begin{equation}
\label{1.116}
I_S(t) := Tr\; \hr(t)\hat I(t) = Tr\;\hr(t)\ln\hr(t) .
\end{equation}
Its relation with
\begin{equation}
\label{1.117}
I_S := I_S(0) = Tr\;\hr\ln\hr 
\end{equation}
easily follows from the equality
$$ Tr\;\hr(t)\hat I(t) = Tr\;\hr(0)\hat I(0) $$
resulting from (1.86) and (1.114). Therefore, the Shannon information
(1.116), actually, does not depend on time,
\begin{equation}
\label{1.118}
I_S(t) = I_S(0) .
\end{equation}
The equality (1.118) is the essence of the {\it Liouville theorem}.

The {\it information entropy} is given by
\begin{equation}
\label{1.119}
S:= - I_S = - Tr\;\hr\ln\hr .
\end{equation}
Since the statistical operator $\hr$ is a positive bounded operator
normalized to unity by (1.84), the information entropy (1.119) is 
positive  and, because of (1.118), it has no time evolution.

It is important to emphasize that the Shannon information (1.116) is not
an average of the information operator (1.113). An {\it average
information} can be defined as
\begin{equation}
\label{1.120}
I_A(t) := Tr\;\hr\hat I(t) = Tr\;\hr\ln\hr(t) .
\end{equation}
Its time evolution follows from the Liouville equation (1.115) giving
\begin{equation}
\label{1.121}
i\hbar\frac{d}{dt} I_A(t) = Tr\;\hr\left [\hat H(t),\hat I(t)\right ] ,
\end{equation}
while the Shannon information, in comformity with (1.118), is constant in
time. One may expect that the average information (1.120) decreases with
time because of the following.

\vspace{3mm}

{\bf Proposition 1.1.} ({\it Nonincrease of average information}).

\vspace{2mm}

Let the average information be given by the functional (1.120), and let an
inverse operator $\hr^{-1}(t)$ exist satisfying
$$ \hr(t)\hr^{-1}(t) = 1 , $$
then for $t\geq 0$
\begin{equation}
\label{1.122}
I_A(t)\leq I_A(0) .
\end{equation}

\vspace{3mm}

{\bf Proof:} Consider the difference
$$ I_A(0) - I_A(t) = Tr\hr\left [\ln\hr - \ln\hr(t)\right ] . $$
Due to the existence of the inverse operator $\hr^{-1}(t)$, we can write
$$ \ln\hr - \ln\hr(t) = \ln\hr\hr^{-1}(t) . $$
Since, by definition, $\hr(t)$ is a positive bounded operator in a Hilbert
space, its inverse is a positive operator. The product of two positive
operators is a positive operator, so $\hr\hr^{-1}(t)$ is a positive
operator. For a positive operator $\hat X$ in a Hilbert space the
inequality
$$ \ln\hat X \geq 1 - \hat X^{-1} \qquad (\hat X > 0) $$
holds, if the inverse operator exists. Applying this inequality to the
positive operator $\hr\hr^{-1}(t)$, we have
$$ \ln\hr\hr^{-1}(t) \geq 1 -\hr^{-1}\hr(t) . $$
Consequently,
$$ I_A(0) - I_A(t) \geq Tr\;\hr\left [ 1 - \hr^{-1}\hr(t)\right ] . $$
As far as the statistical operator $\hr(t)$ for any $t\geq 0$ is
normalized to unity by condition (1.84), we get
$$ Tr\;\hr\left [ 1 - \hr^{-1}\hr(t)\right ] = Tr\;\hr - Tr\hr(t) = 0 . $$
Thus, we find that $Tr\;\hr\ln\hr \geq Tr\hr\ln\hr(t)$ and
$$ I_A(0) - I_A(t) \geq 0 , $$
which prooves (1.122).

If, instead of the entropy (1.119), we introduce the {\it average entropy}
\begin{equation}
\label{1.123}
S_A(t) := -Tr\;\hr\ln\hr(t) ,
\end{equation}
then from the above proposition, it follows
\begin{equation}
\label{1.124}
S_A(t) \geq S_A(0) \qquad (t\geq 0) .
\end{equation}

The Shannon information (1.116) is an {\it unconditional information}. It
may happen that some additional knowledge about the properties of a
considered system is available. Then conditional information functionals
can be defined.

Assume that for a set $\{\ha_i(t)|\; i=0,1,2,\ldots,n\}$ of operators from
the algebra ${\cal A}$ of local observables some statistical averages
\begin{equation}
\label{1.125}
A_i(t) := \lgl\ha_i(t)\rgl = Tr\;\hr\ha_i(t) = Tr\;\hr(t)\ha_i
\end{equation}
are known. One obvious case is for the unity operator $\ha_0(t)=\hat 1$,
for which we know that
\begin{equation}
\label{1.126}
\lgl\hat 1\rgl = Tr\;\hr(t) = 1 .
\end{equation}

Introduce the {\it operator of conditional information}
\begin{equation}
\label{1.127}
\hat I_{con}(t) := \ln\hr(t) +\sum_{i=0}^n\lambda_i(t) \left [\ha_i -
A_i(t)\right ]
\end{equation}
and the {\it conditional information}
$$ I_{con}(t) := Tr\;\hr(t)\hat I_{con}(t) = $$
\begin{equation}
\label{1.128}
= I_S + \sum_{i=0}^n\lambda_i(t)\left [ Tr\;\hr(t)\ha_i - A_i(t)\right ] ,
\end{equation}
where $\lambda_i(t)$ are {\it Lagrange multipliers} satisfying the
variational condition
\begin{equation}
\label{1.129}
\frac{\delta I_{con}(t)}{\delta\lambda_i(t)} = 0 \; .
\end{equation}
The meaning of (1.129) is to guarantee the validity of (1.125). Note that,
contrary to the Shannon information (1.116) that does not depend on time,
according to (1.118), the conditional information (1.128), generally, is a
function of time.

If the behaviour of some observable quantities, characterized by a set
$\{\ha_i(x,t)|\; i=0,1,2,\ldots,n\}$ of operators $\ha_i(x,t)$ is assumed
to be known in each local vicinity of $x$ and $t$, this is described by
the averages 
\begin{equation}
\label{1.130}
A_i(x,t) :=\lgl\ha_i(x,t)\rgl = Tr\;\hr(t)\ha_i(x) .
\end{equation}
In such a case, one can define the {\it operator of local conditional
information}
\begin{equation}
\label{1.131}
\hat I_{loc}(t) := \ln\hr(t) +\sum_{i=0}^n\int\lambda_i(x,t)\left [
\ha_i(x) - A_i(x,t)\right ] dx
\end{equation}
and, respectively, the {\it local conditional information}
\begin{equation}
\label{1.132}
I_{loc}(t) := Tr\;\hr(t)\hat I_{loc}(t) ,
\end{equation}
in which the variational condition
\begin{equation}
\label{1.133}
\frac{\delta I_{loc}(t)}{\delta\lambda_i(x,t)} = 0
\end{equation}
defines the {\it Lagrange multipliers} $\lambda_i(x,t)$ guaranteeing the
validity of (1.130). The local conditional information (1.132) also
depends on time.

Finally, when the transition amplitudes
\begin{equation}
\label{1.134}
A_i(x,t,x',t') := \lgl\ha_i(x,t,x',t')\rgl
\end{equation}
are given as the averages of operators $\ha_i(x,t,x',t')$ with the
evolution property
$$ \ha_i(x,t,x',t') = U^+(t)\ha_i(x,0,x',t')U(t) , $$
then one can introduce the {\it operator of nonlocal conditional
information}
$$ \hat I_{non}(t) := \ln\hr(t) +\sum_{i=0}^n\int\lambda_i(x,t,x',t')
\times $$
\begin{equation}
\label{1.135}
\times  \left [\ha_i(x,0,x',t') - A_i(x,t,x',t')\right ] dxdx'dt' 
\end{equation}
and the {\it nonlocal conditional information}
\begin{equation}
\label{1.136}
I_{non}(t) := Tr\;\hr(t)\hat I_{non}(t)
\end{equation}
with the Lagrange multipliers defined as in (1.129) and (1.133).
Undoubtedly, (1.136) is a function of time.

When the Liouville equation (1.85) does not allow a unique definition of
the statistical operator, then one can invoke the {\it principle of
minimal information} stating that the statistical operator must have such
a form that provides the minimum of information. This implies the validity
of the extremum condition
\begin{equation}
\label{1.137}
\frac{\delta I(t)}{\delta\hr(t)} = 0 ,
\end{equation}
in which $I(t)$ is an information functional for either the conditional
information (1.128), local conditional information (1.132), or nonlocal
conditional information (1.136), depending on a case considered. Also, the
inequality
\begin{equation}
\label{1.138}
\frac{\delta^2I(t)}{\delta\hr^2(t)}\ge 0
\end{equation}
is to be true, meaning that the left--hand--side of (1.138) is a positive
operator. The corresponding Lagrange multipliers are given by the
variational conditions as in (1.129) or (1.133). Performing the variation
of information with respect to the statistical operator $\hr(t)$, one
assumes the existence of the inverse operator $\hr^{-1}(t)$, so that
$$ \frac{\delta}{\delta\hr(t)}\ln\hr(t) =\hr^{-1}(t) , $$
and one uses the equality
$$ \frac{\delta}{\delta\hr(t)}Tr\;\hr(t)\ha_i(t) =\ha_i(t) . $$

For any kind of a statistical operator, the normalization condition
(1.126) is supposed to hold. So, one of the Lagrange multipliers, namely
$\lambda_0(t)$, should have the same form for all statistical operators.
This Lagrange multiplier may be related to the {\it partition function}
$Z(t)$ through the equation
\begin{equation}
\label{1.139}
Z(t) =\exp\{ 1 + \lambda_0(t)\} \; .
\end{equation}

In the case of the conditional information (1.128), the principle (1.137)
yields
\begin{equation}
\label{1.140}
\hr(t) =\frac{1}{Z(t)}\exp\left\{ -\sum_{i=1}^n\lambda_i(t)\ha_i\right\}
\end{equation}
where the partition function
\begin{equation}
\label{1.141}
Z(t) = Tr\exp\left\{ - \sum_{i=1}^n\lambda_i(t)\ha_i\right\}
\end{equation}
plays the role of a normalization factor guaranteeing the validity of
(1.126). The inequality (1.138) holds true since it tells that
$$ \frac{\delta^2I(t)}{\delta\hr^2(t)} =\hr^{-1}(t) > 0 , $$
which is actually the case, as far as $\hr(t)$ is a positive operator.

The local conditional information (1.132), in compliance with the extremum
equation (1.137), leads to the {\it local statistical operator}
\begin{equation}
\label{1.142}
\hr_{loc}(t) =\frac{1}{Z_{loc}(t)}\exp\left\{ -
\sum_{i=1}^n\int\lambda_i(x,t)\ha_i(x)dx\right\}
\end{equation}
with the {\it local partition function}
\begin{equation}
\label{1.143}
Z_{loc}(t) = Tr\;\exp\left\{ - \sum_{i=1}^n\int\lambda_i(x,t)
\ha_i(x)dx\right\}\; .
\end{equation}

In the same way, for the nonlocal conditional information (1.136) one
finds the {\it nonlocal statistical operator}
\begin{equation}
\label{1.144}
\hr_{non}(t) =\frac{1}{Z_{non}(t)}\exp\left\{ - \sum_{i=1}^n
\int\lambda_i(x,t,x',t')\ha_i(x,0,x',t') dxdx'dt'\right\}
\end{equation}
with the {\it nonlocal partition function}
\begin{equation}
\label{1.145}
Z_{non}(t) = Tr\;\exp\left\{ - \sum_{i=1}^n
\int\lambda_i(x,t,x',t')\ha_i(x,0,x',t')dxdx'dt'\right\} .
\end{equation}

Although the found statistical operators $\hr(t),\;\hr_{loc}(t)$, and
$\hr_{non}(t)$ depend on time, however this dependence does not
necessarily coincide with the time evolution prescribed by the Liouville
equation (1.85). To correct this deficiency, one can do the following.
Introduce the {\it Liouville super--operator} $\hat{\cal L}$, acting on
the statistical operator $\hr(t)$ by the rule
\begin{equation}
\label{1.146}
\hat{\cal L}\hr(t) := -\frac{1}{\hbar}\left [ \hat H(t),\hr(t)\right ] ,
\end{equation}
and the operator
\begin{equation}
\label{1.147}
\hat O := \left (\frac{d}{dt} - i\hat{\cal L}\right )\hr(t) .
\end{equation}
Define the {\it operator of generalized conditional information}
\begin{equation}
\label{1.148}
\hat I_{gen}(t) := \hat I_{con}(t) +\lambda(t)\hat O
\end{equation}
and the {\it generalized conditional information}
\begin{equation}
\label{1.149}
I_{gen}(t) := I_{con}(t) +\lambda(t) Tr\;\hr(t)\hat O.
\end{equation}
Using the principle of minimal information, we obtain the {\it generalized
statistical operator}
\begin{equation}
\label{1.150}
\hr(t) =\frac{1}{Z_{gen}(t)}\exp\left\{ - \sum_{i=1}^n\lambda_i(t)
\ha_i - \lambda(t)\hat O\right\}
\end{equation}
with the {\it generalized partition function}
\begin{equation}
\label{1.151}
Z_{gen}(t) = Tr\;\exp\left\{ - \sum_{i=1}^n\lambda_i(t)\ha_i
-\lambda(t)\hat O\right\} .
\end{equation}
The Lagrange multiplier $\lambda(t)$ is to be defined from the variational
condition
$$ \frac{\delta I_{gen}(t)}{\delta\lambda(t)} = 0 , $$
which yields
$$ Tr\;\hr(t)\hat O = Tr\;\hr(t)\left (\frac{d}{dt} - i\hat{\cal L}
\right )\hr(t) = 0 . $$
The latter equation, together with (1.146) and (1.147), gives
\begin{equation}
\label{1.152}
Tr\;\hr(t)\left\{ i\hbar\frac{d}{dt}\hr(t) -\left [\hat H(t),\hr(t)
\right ]\right\} = 0 ,
\end{equation}
that is the averaged Liouville equation (1.85). In the same way, one may
correct the local statistical operator (1.142) and nonlocal statistical
operator (1.144).

\subsection{Equilibrium States}

The statistical state $\lgl{\cal A}\rgl$, which is the set
$\{\lgl\ha(t)\rgl\}$ of expectation values $\lgl\ha(t)\rgl$ of the
operators $\ha(t)$ representing observables quantities, generally, depends
on time. It may happen, that this dependence is absent, so that
\begin{equation}
\label{1.153}
\lgl{\cal A}(t) \rgl = \lgl{\cal A}(0)\rgl \; .
\end{equation}
Then (1.153) presents an {\it equilibrium state}, or {\it time--invariant
state}.

Since (1.153) must hold for all operators from the algebra of observables
${\cal A}$, the necessary and sufficient condition for (1.153) is the time
invariance of the {\it equilibrium statistical operator},
\begin{equation}
\label{1.154}
\hr(t) =\hr(0) :=\hr .
\end{equation}
This means that
\begin{equation}
\label{1.155}
\frac{d}{dt}\hr(t) = 0 ,
\end{equation}
and the Liouville equation (1.85) yields
\begin{equation}
\label{1.156}
\left [\hh(t),\hr(t)\right ] = 0 .
\end{equation}
Eq.(1.156) is valid if and only if the statistical operator in (1.154) is
a function of $\hh(t)$ and of operators $\hc_i(t)$, with $i=1,2,\ldots,n$,
for which
\begin{equation}
\label{1.157}
\left [\hc_i(t),\hh(t)\right ] = 0 , \qquad
\frac{\prt}{\prt t}\hc_i(t) = 0 .
\end{equation}
The operators satisfying (1.157) are called the {\it integrals of motion},
because, according to (1.82) and (1.157),
$$ \frac{d}{dt}\hc_i(t) = 0 , \qquad
\hc_i(t) =\hc_i(0) := \hc_i\; . $$
From (1.154) and (1.156) it is evident that the Hamiltonian
\begin{equation}
\label{1.158}
\hh(t)=\hh(0) :=\hh :=\hc_1
\end{equation}
is one of such integrals of motion. The expectation value
\begin{equation}
\label{1.159}
E :=\lgl\hh\rgl
\end{equation}
is called the {\it internal energy} of the system.

The knowledge that the equilibrium statistical operator (1.154) is a
function of integrals of motion is not sufficient for defining its
concrete form. The latter can be found with the help of the principle of
minimal information as described in the previous Section. To this end, one
constructs the conditional information (1.128) with the operators $\ha_i$
identified with integrals of motion,
$$ \ha_i =\hc_i \qquad (i=1,2,\ldots,n); \quad \ha_0 =\hat 1 \; . $$
As a result, one comes to (1.140) and (1.141) but without time dependence,
\begin{equation}
\label{1.160}
\hr =\frac{1}{Z}\exp\left ( -\sum_{i=1}^n\lambda_i\hc_i\right ) ,
\end{equation}
where
\begin{equation}
\label{1.161}
Z = Tr\;\exp\left ( -\sum_{i=1}^n\lambda_i\hc_i\right ) .
\end{equation}
The statistical operator of the form (1.160) is named the {\it Gibbs
statistical operator} and the statistical state $\lgl{\cal A}\rgl$,
composed of averages $\lgl\ha\rgl$ defined as $Tr\;\hr\ha$ with this
statistical operator, is a {\it Gibbs state} [33-35].

There can be different Gibbs states depending on the given integrals of
motion. For example, if the sole integral of motion is a Hamiltonian
$\hh$, then the conditional information (1.128) becomes
$$ I_{con} = Tr\;\hr\ln\hr +\lambda_0\left ( Tr\;\hr - 1\right ) +
\lbd_1\left ( Tr\;\hr\hh - E\right ) , $$
where
$$ \lbd_0 :=\ln Z - 1 , \qquad \lbd_1 := \bt \; . $$
The principal of minimal information gives
\begin{equation}
\label{1.162}
\hr =\frac{1}{Z}e^{-\bt\hh}
\end{equation}
with 
\begin{equation}
\label{1.163}
Z = Tr\;e^{-\bt\hh}\; .
\end{equation}
The parameter $\bt$ is called the {\it inverse temperature}, and it is
related to the {\it absolute temperature} measured either in energy units,
$\Th$, or in Kelvin degrees, $T$, by the equations
$$ \bt\Th = 1 , \qquad \Th = k_B T , $$
in which $k_B$ is the Boltzmann constant.

The statistical operator (1.162) is named the {\it Gibbs canonical
operator}. One says that a physical system which it describes is
represented by the {\it Gibbs canonical ensemble}.

A quantity $F$ satisfying the equation
\begin{equation}
\label{1.164}
Tr\;\exp\left ( -\sum_{i=1}^n\lbd_i\hc_i\right ) =\exp\left ( -\bt F
\right )
\end{equation}
is termed the {\it thermodynamic potential}. Another form of (1.164) is
the definition
\begin{equation}
\label{1.165}
F:= -\Th\ln Tr\;\exp\left ( - \sum_{i=1}^n\lbd_i\hc_i\right ) .
\end{equation}

The thermodynamic potential for the Gibbs canonical ensemble, according to
(1.163) and (1.165), is
\begin{equation}
\label{1.166}
F := -\Th\ln Z = -\Th\ln Tr\; e^{-\bt\hh} .
\end{equation}
For a system of $N$ particles inside a real--space volume $V$, the
thermodynamic potential (1.166) which is a function
\begin{equation}
\label{1.167}
F = F(\Th,V,N)
\end{equation}
of temperature, volume, and number of particles, is called the free
energy.

Writing for (1.167) the differential form
\begin{equation}
\label{1.168}
dF = - S_c d\Th - p_c dV + \mu_c dN ,
\end{equation}
we define the {\it canonical entropy}
\begin{equation}
\label{1.169}
S_c := -\left (\frac{\prt F}{\prt\Th}\right )_{V,N}\; ,
\end{equation}
{\it canonical pressure}
\begin{equation}
\label{1.170}
p_c := -\left (\frac{\prt F}{\prt V}\right )_{\Th,N} ,
\end{equation}
and the {\it canonical chemical potential}
\begin{equation}
\label{1.171}
\mu_c :=\left (\frac{\prt F}{\prt N}\right )_{\Th,V}\; .
\end{equation}

Consider now the case of two integrals of motion, one of which is the
Hamiltonian $\hh$ and the second is the number--of--particles operator
\begin{equation}
\label{1.172}
\hn =\hn(t); \qquad \left [ \hn,\hh\right ] = 0 .
\end{equation}
The average number of particles is
\begin{equation}
\label{1.173}
N :=\lgl\hn\rgl = Tr\;\hr\hn\; .
\end{equation}
As the conditional information (1.128) we have
$$ I_{con} = Tr\;\hr\ln\hr +\lbd_0\left ( Tr\; \hr - 1\right ) +
\lbd_1\left ( Tr\;\hr\hh - E \right ) + \lbd_2\left ( Tr\;\hr\hn 
- N\right ) , $$
in which
$$ \lbd_0 :=\ln Z - 1 , \qquad \lbd_1 :=\bt , \qquad 
\lbd_2 :=- \bt\mu\; . $$
The principle of minimal information yields
\begin{equation}
\label{1.174}
\hr =\frac{1}{Z} e^{-\bt(\hh -\mu\hn)}
\end{equation}
with the partition function
\begin{equation}
\label{1.175}
Z = Tr\; e^{-\bt(\hh -\mu\hn)}\; .
\end{equation}
The parameter $\mu$ introduced above is the {\it chemical potential}.

The statistical operator (1.174) is called the {\it grand canonical
operator}, and one says that a physical system described by it is
represented by the {\it grand canonical ensemble}.

For the thermodynamic potential (1.165), we get
\begin{equation}
\label{1.176}
\Om := -\Th\ln Tr\; e^{-\bt(\hh -\mu\hn)} \; ,
\end{equation}
which is termed the {\it grand potential}. The latter is a function
\begin{equation}
\label{1.177}
\Om = \Om (\Th,V,\mu)
\end{equation}
of temperature, volume, and chemical potential.

The differential form
\begin{equation}
\label{1.178}
d\Om = - S_g d\Th - p_g dV - N_g d\mu
\end{equation}
for (1.177) defines the {\it grand entropy}
\begin{equation}
\label{1.179}
S_g := -\left (\frac{\prt\Om}{\prt\Th}\right )_{V,\mu} \; ,
\end{equation}
the {\it grand pressure}
\begin{equation}
\label{1.180}
p_g := -\left (\frac{\prt\Om}{\prt V}\right )_{\Th,\mu} \; ,
\end{equation}
and the {\it grand number of particles}
\begin{equation}
\label{1.181}
N_g := -\left (\frac{\prt\Om}{\prt\mu}\right )_{\Th,V} \; .
\end{equation}
Similarly, one may construct other statistical operators and the
corresponding thermodynamic potentials.

The quantities obtained by differentiating thermodynamic potentials are
named {\it thermodynamic characteristics}, or {\it thermodynamic
functions}, and their variables, {\it thermodynamic parameters}, or {\it
thermodynamic variables}.

Different thermodynamic characteristics defined for different ensembles
are, of course, different functions not necessary having simple relations
with each other. This is because, generally, different ensembles represent
different physical systems. An ensemble correctly representing a given
physical system is a {\it representative ensemble}.

It may happen that the same system of particles is equally representable by
several ensembles, which means that the analogous thermodynamic
characteristics of different ensembles can be transformed to each other by
a corresponding change of thermodynamic variables.

Representative ensembles equally describing the thermodynamic properties
of a physical system are {\it thermodynamically equivalent ensembles}.

For example, if the canonical and grand canonical ensembles are
thermodynamically equivalent, then the canonical entropy (1.169) and the
grand entropy (1.179) are related with each other by the equations
$$ S_c(\Th,V,N_g(\Th,V,\mu)) = S_g(\Th,V,\mu) , $$
$$ S_g(\Th,V,\mu_c(\Th,V,N)) = S_c(\Th,V,N) , $$
the canonical pressure (1.170) and the grand pressure (1.180) satisfy the
relations
$$ p_c(\Th,V,N_g(\Th,V,\mu)) = p_g(\Th,V,\mu) , $$
$$ p_g(\Th,V,\mu_c(\Th,V,N)) = p_c(\Th,V,N) , $$
and the relations between the canonical chemical potential (1.171) and the
chemical potential $\mu$, and between the grand number of particles
(1.181) and the number of particles $N$ are
$$ \mu_c(\Th,V,N_g(\Th,V,\mu)) =\mu \; , $$
$$ N_g(\Th,V,\mu_c(\Th,V,N)) = N \; . $$
In such a case the differential forms (1.168) and (1.178) are related with
each other by a Legendre transformation which gives
$$ \Om(\Th,V,\mu) = F(\Th,V,N_g(\Th,V,\mu)) - \mu N_g(\Th,V,\mu) , $$
$$ F(\Th,V,N) = \Om(\Th,V,\mu_c(\Th,V,N)) + \mu_c(\Th,V,N) N . $$
Keeping this change of variables in mind, it is customary to write the
latter relation as
\begin{equation}
\label{1.182}
\Om = F -\mu N \; .
\end{equation}
Traditionally, one also omits the indices at the thermodynamic
characteristics writing
$$ S = -\left (\frac{\prt F}{\prt\Th}\right )_{V,N} = 
- \left (\frac{\prt\Om}{\prt\Th}\right )_{V,\mu} , $$
\begin{equation}
\label{1.183}
p = -\left (\frac{\prt F}{\prt V}\right )_{\Th,N} = 
- \left (\frac{\prt\Om}{\prt V}\right )_{\Th,\mu} , 
\end{equation}
$$ \mu = \left (\frac{\prt F}{\prt N}\right )_{\Th,V} \; , \qquad
N =- \left (\frac{\prt\Om}{\prt\mu}\right )_{\Th,V} \; , $$
and employing in each particular case the corresponding dependence on
thermodynamic variables.

For real physical systems the thermodynamic equivalence of ensembles
usually occurs only in the thermodynamic limit (1.104). Then the above
relations involving the change of thermodynamic variables are to be
considered as asymptotic equalities.

The thermodynamic equivalence of ensembles does not necessarily involve
the equivalence of Gibbs states. This means the following. Suppose the
canonical ensemble is thermodynamically equivalent to the grand canonical
ensemble resulting in (1.182) and (1.183). However, the set of averages
for the operators from the algebra of observables, calculated with the
Gibbs canonical operator (1.162) does not necessarily coincide with the
set calculated with the grand canonical operator (1.174).

Let us denote by $\lgl{\cal A}\rgl_c$ and $\lgl{\cal A}\rgl_g$ the Gibbs
states defined through the Gibbs canonical operator (1.162) and the grand
canonical operator (1.174), respectively. We shall write
\begin{equation}
\label{1.184}
\lgl{\cal A}\rgl_c = \lgl{\cal A}\rgl_g\; , 
\end{equation}
if all corresponding averages coincide with each other. Call 
$\lgl{\cal A}\rgl_c$ the {\it Gibbs canonical state} and
$\lgl{\cal A}\rgl_g$ the {\it grand canonical state}. Then the equivalence
of the statistical states (1.184) is termed the {\it statistical
equivalence of ensembles}. The thermodynamic equivalence follows from the
latter, but not reverse.

The definition of entropy given for the Gibbs canonical state in (1.169)
or for the grand canonical state in (1.179) can be generalized for an
arbitrary Gibbs state with a thermodynamic potential (1.165) as
\begin{equation}
\label{1.185}
S := -\frac{\prt F}{\prt\Th} \; ,
\end{equation}
where the partial derivative with respect to temperature is taken under all
other thermodynamic variables kept fixed. So defined entropy (1.185) is
the {\it thermodynamic entropy}. Its particular examples are the canonical
entropy (1.169) and the grand entropy (1.170). We may remember also that
we have had the definition of the information entropy (1.119). What are
conditions under which the thermodynamic and information entropies
coincide?

\vspace{3mm}

{\bf Proposition 1.2.} ({\it Criterion for entropy equivalence}).

\vspace{2mm}

The thermodynamic entropy (1.185) for a Gibbs state with the Gibbs
statistical operator (1.160) coincides with the information entropy
(1.119), that is
\begin{equation}
\label{1.186}
-\frac{\prt F}{\prt\Th} = - Tr\;\hr\ln\hr\; ,
\end{equation}
if and only if the Lagrange multipliers satisfy the equality
\begin{equation}
\label{1.187}
\sum_{i=1}^n\left ( \frac{\prt\lbd_i}{\prt\bt} - \frac{\lbd_i}{\bt}
\right ) \lgl\hc_i\rgl = 0 \; .
\end{equation}

\vspace{2mm}

{\bf Proof:} From the definition of the thermodynamic entropy (1.185),
with the thermodynamic potential (1.165), we have
\begin{equation}
\label{1.188}
S =\bt\left ( \sum_{i=1}^n\frac{\prt\lbd_i}{\prt\bt}\lgl\hc_i\rgl
- F\right ) \; .
\end{equation}

Now, let us consider the information entropy (1.119). For the statistical
operator (1.160), we get
\begin{equation}
\label{1.189}
\ln\hr = - \sum_{i=1}^n\lbd_i\hc_i - \ln Z \; .
\end{equation}
With the thermodynamic potential $F$ defined in (1.166), this results in
\begin{equation}
\label{1.190}
-Tr\;\hr\ln\hr =\bt\left ( \sum_{i=1}^n\frac{\lbd_i}{\bt}\lgl
\hc_i\rgl - F \right ) .
\end{equation}
The comparison of expressions (1.188) and (1.190) shows that they coincide
if and only if the condition (1.187) holds, which concludes the proof.

\vspace{3mm}

{\bf Corollary.} The canonical entropy (1.169) and the grand entropy
(1.179) coincide with the information entropy (1.119) expressed in the
corresponding variables.

\vspace{2mm}

In the case of the Gibbs canonical state,
$$ \lbd_1 =\bt , \qquad \frac{\prt\lbd_1}{\prt\bt} = 1 , $$
and the criterion (1.187) is valid. The thermodynamic entropy has the form
$$ S = \bt\left ( E - F\right ) $$
coinciding with the information entropy.

For the grand canonical state the Lagrange multipliers are
$$ \lbd_1 =\bt, \qquad \lbd_2 = -\bt\mu $$
and the criterion (1.187) holds since
$$ \frac{\prt\lbd_1}{\prt\bt} = 1 , \qquad
\frac{\prt\lbd_2}{\prt\bt} = -\mu\; . $$
The thermodynamic entropy (1.188) equals the information entropy (1.119)
being
$$ S =\bt\left ( E - \mu N -\Om\right ) . $$

Note that in these two cases we have the equality
\begin{equation}
\label{1.191}
\frac{\prt\lbd_i}{\prt\bt} =\frac{\lbd_i}{\bt}\; ,
\end{equation}
which is a sufficient condition for (1.187)

Similarly to the considered cases, we may define other statistical states
described by the Gibbs statistical operator (1.160) with other integrals
of motion. For example, if the real--space volume of the system is not
fixed, we can introduce the {\it volume operator} $\hat V$, which is a
projector from the real space to a chosen volume, such that
\begin{equation}
\label{1.192}
\lgl\hat V\rgl = V\; .
\end{equation}
Then, constructing the corresponding conditional information and
minimizing it, one comes to the {\it Gibbs isobaric state} with the {\it
Gibbs isobaric operator}
\begin{equation}
\label{1.193}
\hr =\frac{1}{Z} e^{-\bt(\hh - p\hat V)} \; ,
\end{equation}
in which
$$ Z = Tr\; e^{-\bt(\hh - p\hat V)} \; . $$

The thermodynamic potential related to the Gibbs isobaric state,
\begin{equation}
\label{1.194}
G := -\Th\ln Tr\; e^{-\bt(\hh - p\hat V)} \; ,
\end{equation}
is called the {\it Gibbs potential}. It is a function
\begin{equation}
\label{1.195}
G = G(\Th,p,N)
\end{equation}
of temperature, pressure and number of particles, with the differential
form
\begin{equation}
\label{1.196}
dG = - S d\Th + V dp + \mu dN\; .
\end{equation}
The later defines the {\it isobaric entropy}
\begin{equation}
\label{1.197}
S := -\left (\frac{\prt G}{\prt\Th}\right )_{p,N}\; ,
\end{equation}
the {\it average volume}
\begin{equation}
\label{1.198}
V := \left (\frac{\prt G}{\prt p}\right )_{\Th,N}\; ,
\end{equation}
and the {\it isobaric chemical potential}
\begin{equation}
\label{1.199}
\mu :=\left (\frac{\prt G}{\prt N}\right )_{\Th,p} \; .
\end{equation}
In formulas (1.196) to (1.199), following tradition, the index specifying
the affiliation of the thermodynamic characteristics to an ensemble, in
this case to the {\it isobaric ensemble}, for brevity, is omitted. The
relation of the Gibbs potential (1.194) with the free energy and the grand
potential is
\begin{equation}
\label{1.200}
G = F + pV = \Om + \mu N + p V\; .
\end{equation}

The general form of the thermodynamic potential (1.165) has remarkable
convexity properties with respect to Lagrange multipliers. The first
derivative of (1.165) is
$$ \frac{\prt F}{\prt\lbd_i} = \Th\lgl \hc_i\rgl \; . $$
Differentiating this once more, we find
$$ \frac{\prt^2F}{\prt \lbd_i^2} = -\Th\left ( \lgl\hc_i^2\rgl -
\lgl\hc_i\rgl^2\right ) \; . $$
For any self--adjoint operator $\ha$, the quantity
$$ \lgl\ha^2\rgl - \lgl\ha\rgl^2 =
\lgl\left ( \ha -\lgl\ha\rgl\right )^2\rgl $$
is nonnegative. If integrals of motion pertain to the algebra of
observables, they are self--adjoint. In such a case,
\begin{equation}
\label{1.201}
\frac{\prt^2F}{\prt\lbd_i^2}\leq 0 \qquad (i\geq 1) ,
\end{equation}
that is, the thermodynamic potential is a concave function with respect to
each of its Lagrange multipliers. Inequalities (1.201) are named the {\it
conditions of thermodynamic stability}.

Defining the {\it specific heat}
\begin{equation}
\label{1.202}
C_V := \frac{1}{N}\left (\frac{\prt E}{\prt\Th}\right )_{V,N}
= \frac{\Th}{N}\left (\frac{\prt S}{\prt\Th}\right )_{V,N}
= -\frac{\Th}{N}\left (\frac{\prt^2 F}{\prt\Th^2}\right )_{V,N}
\end{equation}
and the {\it isothermal compressibility}
\begin{equation}
\label{1.203}
K_T := -\frac{1}{V}\left (\frac{\prt V}{\prt p}\right )_{\Th,N}
= \frac{1}{\rho}\left (\frac{\prt\rho}{\prt p}\right )_{\Th,N}
= -\frac{1}{V}\left (\frac{\prt^2 G}{\prt p^2}\right )_{\Th,N}\; ,
\end{equation}
we see that they are expressed through the second derivatives of
thermodynamic potentials, whence are to be sign--definite. Really, for the
specific heat (1.202) we get
\begin{equation}
\label{1.204}
C_V =\frac{\bt^2}{N}\left (\lgl\hh^2\rgl - \lgl\hh\rgl^2\right ) \; .
\end{equation}
The isothermal compressibility (1.203) can also be found from the
expressions
$$ K_T^{-1} := - V \left (\frac{\prt p}{\prt V}\right )_{\Th,N}
= \rho\left (\frac{\prt p}{\prt\rho }\right )_{\Th,N}
= V\left (\frac{\prt^2 F}{\prt V^2}\right )_{\Th,N}\; . $$
Finally, one obtains
\begin{equation}
\label{1.205}
K_T =\frac{\bt V}{N^2}\left (\lgl\hn^2\rgl - \lgl\hn\rgl^2\right ) \; .
\end{equation}
The quantities (1.204) and (1.205) are positively defined,
\begin{equation}
\label{1.206}
C_V \geq 0 , \qquad K_T \geq 0 \; .
\end{equation}
These are the most often used stability conditions. When at least one of
the inequalities in (1.206) becomes an equality, this defines the {\it
stability boundary}.

\subsection{Quasiequilibrium States}

A {\it quasiequilibrium state}, or {\it locally equilibrium state}, is a
statistical state $\lgl{\cal A}\rgl$ formed by the operator averages with
a local statistical operator [36,37] (see also [19]).

To concretize the construction of the latter, introduce the {\it
Hamiltonian density} $\hat H(x,t)$ such that
\begin{equation}
\label{1.207}
\hh(t) =\int\hat H(x,t) dx ,
\end{equation}
and recall that the number--of--particles operator can be written as
\begin{equation}
\label{1.208}
\hn(t) =\int\hat n(x,t) dx\; ,
\end{equation}
where the density--of--particle operator
\begin{equation}
\label{1.209}
\hat n(x,t) =\psi^\dgr(x,t)\psi(x,t)
\end{equation}
is defined analogously to (1.80).

As the local averages (1.130), choose the {\it energy density}
\begin{equation}
\label{1.210}
E(x,t) := \lgl\hh(x,t)\rgl
\end{equation}
and the {\it particle density}
\begin{equation}
\label{1.211}
\rho(x,t) := \lgl\hat n(x,t)\rgl \; .
\end{equation}
The integration of (1.210),
$$ E(t) =\int E(x,t) dx\; , $$
gives the internal energy (1.159). The integral of (1.211) yields the
average number of particles $$ N(t) =\int\rho(x,t) dx \; . $$
Note that, according to (1.98), the particle density (1.211) is a diagonal
element of the reduced density matrix
$$ \rho_1(x,x,t) =\rho(x,t) \; . $$

With the given local averages (1.210) and (1.211), the local conditional
information (1.132) becomes
$$ I_{loc}(t) = Tr\;\hr(t)\ln\hr(t) +\lbd_0(t)\left [ Tr\;\hr(t) - 1
\right ] + $$
$$ + \int\lbd_1(x,t)\left [ Tr\;\hr(t)\hh(x,t) - E(x,t)\right ] dx + $$
\begin{equation}
\label{1.212}
+ \int\lbd_2(x,t)\left [ Tr\;\hr(t)\hat n(x,t) -\rho(x,t)\right ] dx ,
\end{equation}
where
$$ \lbd_0(t) := \ln Z_{loc}(t) - 1 , \qquad \lbd_1(x,t) := \bt(x,t) , 
\qquad \lbd_2(x,t) := -\bt(x,t)\mu(x,t) \; . $$

The function $\bt(x,t)$ is the {\it local inverse temperature} connected
with the {\it local absolute temperature} $\Th(x,t)$ by the relation
$$ \bt(x,t)\Th(x,t) = 1 \; , $$
and $\mu(x,t)$ is the {\it local chemical potential}.

The principle of minimal information for (1.212) yields the local
statistical operator
\begin{equation}
\label{1.213}
\hr_{loc}(t) =\frac{1}{Z_{loc}(t)} \exp\left \{ - \int\bt(x,t)\left [
\hh(x,t) -\mu(x,t)\hat n(x,t)\right ] dx\right \} ,
\end{equation}
with the local partition function
\begin{equation}
\label{1.214}
Z_{loc}(t) = Tr\exp\left \{ - \int\bt(x,t) \left [ \hh(x,t) -
\mu(x,t)\hat n(x,t)\right ] dx\right \} \; .
\end{equation}

Introduce the {\it quasiequilibrium potential} $Q(t)$ requiring that
$$ Z_{loc}(t) = e^{-Q(t)}\; , $$
which gives the definition
$$ Q(t) := -\ln Z_{loc}(t) = $$
\begin{equation}
\label{1.215}
= -\ln Tr\;\exp\left \{ -\int\bt(x,t)\left [
\hh(x,t) - \mu(x,t)\hat n(x,t)\right ] dx\right \} \; .
\end{equation}

One may also define the {\it local grand potential} $\Om(x,t)$ by the
equation
\begin{equation}
\label{1.216}
Z_{loc}(t) :=\exp\left \{ -\int\bt(x,t)\Om(x,t) dx\right \} \; .
\end{equation}
From (1.215) and (1.216) it follows that
\begin{equation}
\label{1.217}
Q(t) =\int\bt(x,t)\Om(x,t) dx \; .
\end{equation}

We need to find how the local grand potential changes when varying the
local absolute temperature $\Th(x,t)$ and the local chemical potential
$\mu(x,t)$. To this end, we, first, obtain from (1.215) the variational
derivatives
$$ \frac{\delta Q(t)}{\delta\bt(x,t)} = E(x,t) -\mu(x,t)\rho(x,t) , \qquad
\frac{\delta Q(t)}{\delta\mu(x,t)} = -\bt(x,t)\rho(x,t) . $$
Then, we find the same derivatives from (1.217),
$$ \frac{\delta Q(t)}{\delta\bt(x,t)} = \Om(x,t) +\bt(x,t)
\frac{\delta \Om(x,t)}{\delta\bt(x,t)} \; , \qquad
\frac{\delta Q(t)}{\delta\mu(x,t)} = \bt(x,t)
\frac{\delta \Om(x,t)}{\delta\mu(x,t)} \; . $$
Comparing the derivatives with respect to $\bt(x,t)$, we get
$$ \frac{\delta\Om(x,t)}{\delta\bt(x,t)} = \Th(x,t)\left [
E(x,t) -\mu(x,t)\rho(x,t) - \Om(x,t)\right ] \; . $$
Using the relation
$$ \frac{\delta\Om(x,t)}{\delta\bt(x,t)} = - \Th^2(x,t)
\frac{\delta\Om(x,t)}{\delta\Th(x,t)} \; , $$
we have
$$ \frac{\delta\Om(x,t)}{\delta\Th(x,t)} = \bt(x,t)\left [ \Om(x,t)
+ \mu(x,t)\rho(x,t) - E(x,t)\right ] \; . $$

Analogously to the equilibrium case (1.179), we may define the {\it local
thermodynamic entropy}
\begin{equation}
\label{1.218}
S(x,t) := -\frac{\delta\Om(x,t)}{\delta\Th(x,t)}\; .
\end{equation}
Substituting to the right--hand side of (1.218) the variational derivative
found above, we obtain
\begin{equation}
\label{1.219}
S(x,t) = \bt(x,t)\left [ E(x,t) - \mu(x,t)\rho(x,t) - \Om(x,t)\right ] \; .
\end{equation}

Comparing the derivatives of $Q(t)$ with respect to $\mu(x,t)$, we find
\begin{equation}
\label{1.220}
\rho(x,t) = -\frac{\delta\Om(x,t)}{\delta\mu(x,t)}\; .
\end{equation}

In compliance with (1.218) and (1.220), the variational equation
\begin{equation}
\label{1.221}
\delta\Om(x,t) = - S(x,t)\delta\Th(x,t) -\rho(x,t)\delta\mu(x,t)
\end{equation}
holds, named the {\it Gibbs--Duhem relation}.

The {\it local pressure} is defined as
\begin{equation}
\label{1.222}
p(x,t) := -\Om(x,t) \; .
\end{equation}
The variation of (1.219) gives two other useful relations
$$  \bt(x,t) =\frac{\delta S(x,t)}{\delta E(x,t)} , \qquad
\mu(x,t) = -\Th(x,t)\frac{\delta S(x,t)}{\delta\rho(x,t)}\; , $$
which show how the local inverse temperature and the local chemical
potential are connected with the variation of the local thermodynamic
entropy.

In addition to the latter, we can introduce the {\it quasiequilibrium
information entropy}
\begin{equation}
\label{1.223}
S(t) := -  Tr\;\hr_{loc}(t)\ln\hr_{loc}(t) \; .
\end{equation}
With the local statistical operator (1.213), this gives 
$$ S(t) = \int\bt(x,t)\left [ E(x,t) -\mu(x,t)\rho(x,t)\right ] dx -
Q(t) \; , $$
where $Q(t)$ is defined in (1.215). Taking into account (1.217) and
(1.219), we obtain the relation
\begin{equation}
\label{1.224}
S(t) = \int S(x,t) dx
\end{equation}
between the quasiequilibrium information entropy (1.223) and the local
thermodynamic entropy (1.218).

The local statistical operator (1.213), generally, does not satisfy the
Liouville equation (1.85), thus, its time evolution cannot be described by
a simple unitary transformation, like (1.86). Therefore, the
quasiequilibrium information entropy (1.223), contrary to the information
entropy (1.119), depends on time. This permits one to use the {\it
quasiequilibrium ensemble}, with the local statistical operator (1.213),
as a representation of {\it open}, or {\it non--isolated systems} whose
time evolution cannot be described by the Liouville equation.

When applying the local statistical operator to quasi--isolated systems
whose time evolution obeys the Liouville equation, one can correct the
time behaviour of this operator by introducing one more Lagrange
multiplier, as is described in Section 1.6. One may also employ the local
statistical operator as a starting approximation in constructing
perturbation theory for the Liouville equation. Then, the higher--order
approximations of perturbation theory will automatically correct the
starting approximation adjusting its time behaviour to the Liouville law.
Finally, the local statistical operator $\hr_{loc}(0)$ can be used as an
initial condition for the Liouville equation.

\subsection{Evolution Equations}

The statistical state $\lgl{\cal A}(t)\rgl$ which is not time--invariant
is a {\it nonequilibrium state}. The statistical operator for which
(1.154) is not valid is a {\it nonequilibrium statistical operator}. The
time evolution of the latter is given by the Liouville equation (1.85).

However, what one finally needs is not a statistical operator itself, but
the averages of operators associated with observable quantities. For an
arbitrary operator $\ha$, the statistical averages, as defined in (1.87),
satisfy the property
\begin{equation}
\label{1.225}
Tr\;\hr(t)\ha(0) = Tr\;\hr(0)\ha(t)\; ,
\end{equation}
where $\ha(t)$ is the Heisenberg representation of an operator $\ha$ with
the time evolution (1.81). Therefore, the equation of motion for the
average (1.225) can be derived by using either the Liouville equation for
the statistical operator $\hr(t)$ or the Heisenberg equation for the
operator $\ha(t)$.

Keeping in mind the general case, when the number of particles in the
system is not fixed, one defines the {\it generalized Hamiltonian}
\begin{equation}
\label{1.226}
H(t) := \hh(t) - \mu\hn(t) \; ,
\end{equation}
which can also be called the {\it grand Hamiltonian}, since this form is
preferable for dealing with the grand ensemble. The time evolution is
assumed to be governed by (1.226).

In order to simplify formulas, it is customary to pass to the system of
units in which the repeatedly appearing Planck constant is set to unity,
$\hbar = 1$.

In this notation, the Liouville equation for the statistical operator
becomes
\begin{equation}
\label{1.227}
i\frac{d}{dt}\hr(t) = \left [ H(t),\hr(t)\right ]\; ,
\end{equation}
and the Heisenberg equation for an operator $\ha(t)$ is
\begin{equation}
\label{1.228}
i\frac{d}{dt}\ha(t) = i\frac{\prt}{\prt t}\ha(t) +
\left [\ha(t),H(t)\right ]\; ,
\end{equation}
where the partial time derivative is defined in (1.83). The evolution
equations governed by a Hamiltonian define the {\it Hamiltonian dynamics}.

The time evolution of a statistical average $\lgl\ha(t)\rgl$, given by one
of the forms in (1.225), can be presented by two equations. Using the
left--hand side of (1.225) and (1.227), one has
\begin{equation}
\label{1.229}
i\frac{d}{dt}\lgl\ha(t)\rgl = Tr\; \left [ H(t),\hr(t)\right ]
\ha(0) \; .
\end{equation}
From the right--hand side of (1.225), together with (1.228), one obtains
\begin{equation}
\label{1.230}
i\frac{d}{dt}\lgl\ha(t)\rgl  = \lgl i\frac{\prt}{\prt t}\ha(t)\rgl +
\lgl \left [\ha(t),H(t)\right ]\rgl \; .
\end{equation}
These two equations are completely equivalent, and the choice of one of
them is a matter of taste and convenience [38-42]. According to our own
experience, the evolution equation (1.230) is usually easier to treat.

To give examples of evolution equations, we need to concretize the grand
Hamiltonian (1.226). For a nonrelativistic system of particles with a
two--body {\it interaction potential} $V(x,x')$, the generally accepted
Hamiltonian is
$$ H(t) := \int\psi^\dgr(x,t)\left [ K(\stackrel{\ra}{\nabla}) +
U(x,t) -\mu\right ] \psi(x,t) dx + $$
\begin{equation}
\label{1.231}
+ \frac{1}{2} \int\psi^\dgr(x,t)\psi^\dgr(x',t)V(x,x')\psi(x',t)\psi(x,t)
dxdx' \; ,
\end{equation}
where $K(\stackrel{\ra}{\nabla})$ is a {\it kinetic--energy operator} with
$\stackrel{\ra}{\nabla}:=\prt/\prt\stackrel{\ra}{r}$, and 
$\stackrel{\ra}{r}:=\{ r_\al|\; \al=1,2,3\}$ being the real--space vector;
the multivariable $x=\{\stackrel{\ra}{r},\ldots,\}$ is assumed to include
$\stackrel{\ra}{r}$; $U(x,t)$ is an {\it external potential}; and $\mu$,
chemical potential. The field operators satisfy the commutation relations
\begin{equation}
\label{1.232}
\left [\psi(x,t)\psi^\dgr(x',t)\right ]_\mp = \delta( x - x') , \qquad
\left [\psi(x,t)\psi(x',t)\right ]_\mp = 0 , 
\end{equation}
which follows from (1.71) and the Heisenberg evolution (1.75), the upper
sign corresponding to Bose--, and lower, to Fermi statistics. Emphasize
that (1.232) holds only for the field operators at coinciding times.

From the Heisenberg equation for an annihilation operator
\begin{equation}
\label{1.233}
i\frac{\prt}{\prt t}\psi(x,t) = \left [ \psi(x,t), H(t)\right ] \; ,
\end{equation}
employing (1.231) and (1.232), one gets
$$ i\frac{\prt}{\prt t}\psi(x,t) = \left [ K(\stackrel{\ra}{\nabla}) 
+ U(x,t) - \mu \right ]\psi(x,t) + $$
\begin{equation}
\label{1.234}
+\int\Phi(x,x')\psi^\dgr(x',t)\psi(x',t)\psi(x,t) dx'\; ,
\end{equation}
where there appears the {\it symmetrized interaction potential}
\begin{equation}
\label{1.235}
\Phi(x,x') :=\frac{1}{2}\left [ V(x,x') + V(x',x)\right ]\; ,
\end{equation}
having the property
$$ \Phi(x,x') = \Phi(x',x)\; . $$
Note that the same equation (1.234) may be obtained from (1.231) by the
variational procedure
\begin{equation}
\label{1.236}
i\frac{\prt}{\prt t}\psi(x,t) = 
\frac{\delta H}{\delta\psi^\dgr(x,t)}\; .
\end{equation}
The time evolution of the creation operator $\psi^\dgr(x,t)$ is given by
the Hermitian conjugate of (1.233)--(1.236).

To write the equation of motion for the density--of--particles operator
(1.209), take the kinetic--energy operator as
\begin{equation}
\label{1.237}
K(\srn) := -\frac{\nabla^2}{2m}\; ,
\end{equation}
where $m$ is a {\it particle mass}. Introduce the {\it density--of--current 
operator}
\begin{equation}
\label{1.238}
\hat{\sjj}(x,t) := -\frac{i}{2m}\left\{ \psi^\dgr(x,t)\srn\psi(x,t) -
\left [\srn\psi^\dgr(x,t)\right ]\psi(x,t)\right\} \; .
\end{equation}
The evolution equation for the density--of--particles operator can be
reduced to the form
\begin{equation}
\label{1.239}
\frac{\prt}{\prt t}\hat n(x,t) + {\rm div}\hat{\sjj}(x,t) = 0\; ,
\end{equation}
called the {\it continuity equation}. Here ${\rm div}\sjj\equiv\srn\sjj$.

Using again (1.234), together with its Hermitian conjugate, we may derive
the equations for the components $\hat j_\al(x,t)$ of the
density--of--current operator 
$\hat{\sjj}(x,t) =\{\hat j_\al(x,t)|\;\al=1,2,3\}$ given in (1.238). For
simplicity, consider the case of no external potentials, $U(x,t)=0$. Then
we find
$$ \frac{\prt}{\prt t}\hat j_\al(x,t) +\frac{1}{2m^2}\sum_\bt
\frac{\prt}{\prt r_\bt}\left\{ \left [ \frac{\prt}{\prt r_\al}
\psi^\dgr(x,t)\right ]\frac{\prt}{\prt r_\bt}\psi(x,t) +\right. $$
$$ +\left. \left [\frac{\prt}{\prt r_\bt}\psi^\dgr(x,t)\right ]
\frac{\prt}{\prt r_\al}\psi(x,t) - \frac{1}{2}
\frac{\prt^2\hat n(x,t)}{\prt r_\al\prt r_\bt} \right\} = $$
\begin{equation}
\label{1.240}
= -\frac{1}{m}\int\left [ \frac{\prt}{\prt r_\al}\Phi(x,x')\right ]
\psi^\dgr(x,t)\psi^\dgr(x',t)\psi(x',t)\psi(x,t)dx'\; .
\end{equation}

The {\it density--of--energy operator} is defined as
$$ \hh(x,t) :=\frac{1}{2m}\srn\psi^\dgr(x,t)\cdot \srn\psi(x,t) + $$
\begin{equation}
\label{1.241}
+\frac{1}{2}\int V(x,x')\psi^\dgr(x,t)\psi^\dgr(x',t)\psi(x',t)\psi(x,t) 
dx' \; .
\end{equation}
This is a particular case of the Hamiltonian density (1.207). The
evolution equation for (1.241) is
$$ \frac{\prt}{\prt t}\hh(x,t) + {\rm div}\left \{ 
\frac{i}{(2m)^2}\left [ \nabla^2\psi^\dgr(x,t)\cdot\srn\psi(x,t) -
\srn\psi^\dgr(x,t)\cdot \nabla^2\psi(x,t)\right ] + \right. $$
$$\left. + \frac{1}{2}\int 
\Phi(x,x')\psi^\dgr(x,t)\hat{\sjj}(x,t)\psi(x',t) dx' \right\} = $$
\begin{equation}
\label{1.242}
= -\frac{1}{2}\int\srn\Phi(x,x')\cdot \left [\psi^\dgr(x',t)\hat{\sjj}(x,t)
\psi(x',t) + \psi^\dgr(x,t)\hat{\sjj}(x',t)\psi(x,t)\right ] dx' \; .
\end{equation}
The evolution equations (1.239), (1.240), and (1.242) are called the {\it
local conservation laws}. Accomplishing statistical averaging of these
laws, one derives the equations of motion for observables quantities as in
(1.230).

Another way of describing statistical states is based on the fact that
these states are completely defined by the reduced density matrices with
elements (1.96). For example, the average number of particles (1.97)
involves only $1$--matrix. The internal energy (1.159), with the
Hamiltonian $\hh(t)$ given by (1.226) and (1.231), that is
\begin{equation}
\label{1.243}
E(t) =\lgl\hh(t)\rgl = \lgl H(t) +\mu\hn\rgl\; ,
\end{equation}
can be expressed through $1$-- and $2$--matrices,
$$ E(t) =\int K(x_1,x_2,t)\rho_1(x_2,x_1,t)dx_1dx_2 + $$
\begin{equation}
\label{1.244}
+\frac{1}{2}\int V(x_1,x_2)\rho_2(x_2,x_1,x_2,x_1,t) dx_1dx_2\; ,
\end{equation}
where the notation for the {\it one--particle kernel}
$$ K(x_1,x_2,t) := \delta( x_1 - x_2)K_1(x_2,t) $$
and the {\it one--particle reduced Hamiltonian}
\begin{equation}
\label{1.245}
K_1(x,t) := K(\srn) + U(x,t)
\end{equation}
are introduced. When the number of particles, $N$, is fixed, then, using
relation (1.101), one can cast (1.244) into the form
\begin{equation}
\label{1.246}
E(t) =\int\delta(x_1 - y_1) \delta( x_2 - y_2)\frac{K_2(x_1,x_2,t)}{2(N-1)} 
\rho_2(x_1,x_2,y_1,y_2,t)dx_1dx_2dy_1dy_2\; ,
\end{equation}
with the {\it two--particle reduced Hamiltonian}
\begin{equation}
\label{1.247}
K_2(x_1,x_2,t) := K_1(x_1,t) + K_1(x_2,t) +( N - 1)V(x_1,x_2)\; .
\end{equation}
To find the internal energy (1.246), one needs to know only $2$--matrix.

Generally, all averages corresponding to observable quantities can be
expressed through several, usually a few, low--order density matrices.
Therefore, one has to find out the evolution equations for the latter.
These equations can be derived by using (1.234). Thus, for $1$--matrix we
have
$$ i\frac{\prt}{\prt t}\rho_1(x_1,x_2,t) =\left [ K_1(x_1,t) - K_1(x_2,t)
\right ] \rho_1(x_1,x_2,t) + $$
\begin{equation}
\label{1.248}
 +\int\left [ \Phi(x_1,x_3) - \Phi(x_2,x_3)\right ]
\rho_2(x_3,x_1,x_3,x_2,t) dx_3 \; .
\end{equation}
For $2$--matrix, we obtain
$$ i\frac{\prt}{\prt t}\rho_2(x_1,x_2,y_1,y_2,t) =\left [ H(x_1,x_2,t) -
H(y_1,y_2,t) \right ] \rho_2(x_1,x_2,y_1,y_2,t) + $$
\begin{equation}
\label{1.249}
 +\int\left [ \Phi(x_1,x_2,x_3) - \Phi(y_1,y_2,x_3)\right ]
\rho_3(x_3,x_1,x_2,x_3,y_1,y_2,t) dx_3 \; ,
\end{equation}
where the {\it two--particle Hamiltonian}
\begin{equation}
\label{1.250}
H(x_1,x_2,t) := K_1(x_1,t) + K_1(x_2,t) +\Phi(x_1,x_2)\; ,
\end{equation}
and the {\it three--particle interaction}
$$ \Phi(x_1,x_2,x_3) := \Phi(x_1,x_3) +\Phi(x_2,x_3) $$
are introduced.

If the kinetic--energy operator has the form of (1.237), then, using the
property
$$ \nabla_1^2 -\nabla_2^2 =\left (\srn_1 +\srn_2\right )
\left(\srn_1 - \srn_2\right ) , $$
the $1$--matrix equation (1.248) can be written as
$$ i\frac{\prt}{\prt t}\rho_1(x_1,x_2,t) = 
-i\left (\srn_1 +\srn_2 \right )\sjj(x_1,x_2,t) + 
\left [ U(x_1,t) - U(x_2,t) \right ]  \rho_1(x_1,x_2,t) + $$
\begin{equation}
\label{1.251}
 +\int\left [ \Phi(x_1,x_3) - \Phi(x_2,x_3)\right ]
\rho_2(x_3,x_1,x_3,x_2,t) dx_3 \; ,
\end{equation}
where the {\it density--of--current matrix} is defined by
\begin{equation}
\label{1.252}
\sjj(x_1,x_2,t) := \frac{\srn_1 -\srn_2}{2mi}\rho_1(x_1,x_2,t) \; .
\end{equation}
The relation between (1.252) and the density--of--current operator (1.238)
reads
$$ \lgl\hat{\sjj}(x,t)\rgl =\lim_{x'\ra x} \sjj(x,x',t)\; . $$
The evolution equation for the matrix in (1.252) is
$$ i\frac{\prt}{\prt t}\sjj(x_1,x_2,t) =
\left [ K_1(x_1,t) - K_1(x_2,t) \right ] \sjj(x_1,x_2,t) - $$
$$ -\frac{i}{2m}\left [\srn_1 U(x_1,t) +\srn_2 U(x_2,t)\right ]
\rho_1(x_1,x_2,t) - $$
$$ -\frac{i}{2m}\int \left [\srn_1\Phi(x_1,x_3) +\srn_2\Phi(x_2,x_3)
\right ] \rho_2(x_3,x_1,x_3,x_2,t)dx_3  - $$
\begin{equation}
\label{1.253}
-\frac{i}{2m}\int \left [\Phi(x_1,x_3) - \Phi(x_2,x_3) \right ]
\left (\srn_1 - \srn_2\right )\rho_2(x_3,x_1,x_3,x_2,t)dx_3 \; ,
\end{equation}
where the $2$--matrix satisfies (1.249).

The evolution equations for matrices are to be supplemented with initial
and boundary conditions. However, the system of these equations is not
closed, since the equation for $\rho_1$ or $\sjj$ involves $\rho_2$, the
equation for $\rho_2$ involves $\rho_3$. Generally, an equation for
$\rho_n$ always contains $\rho_{n+1}$, of course, if the interaction
potential is not zero. In this way, reduced density matrices satisfy an
infinite system of integrodifferential equations, which is called the {\it
hierarchical chain of equations}. To obtain a closed set of equations one
has to break this chain by approximately expressing one of $\rho_n$
through the lower--order density matrices. Then one gets a closed system
of $n-1$ equations for $\rho_1,\rho_2,\ldots,\rho_{n-1}$. This kind of
approximation is called the {\it decoupling procedure}. It is important
that any approximation used would preserve the general properties of
density matrices following from their definition (1.96). This is the
conjugation property
\begin{equation}
\label{1.254}
\rho_n^*(x^n,y^n,t) =\rho_n(y^n,x^n,t) ;
\end{equation}
the symmetry property
\begin{equation}
\label{1.255}
P(x_i,x_j)\rho_n(x^n,y^n,t) =\pm \rho_n(y^n,x^n,t) \; ,
\end{equation}
where $x_i,x_j\in x^n$; the non--negativeness of the diagonal elements
\begin{equation}
\label{1.256}
\rho_n(x^n,x^n) \geq 0;
\end{equation}
and the normalization condition (1.103). For the density--of--current
matrix, given by (1.252), in compliance with (1.254), one has
$$ \sjj^*(x_1,x_2,t) =\sjj(x_2,x_1,t)\; . $$

Note that in some cases the symmetry property (1.255) can be broken
giving, nevertheless, a good approximation [43-46]. For example, the
widely used {\it Hartree approximation}
\begin{equation}
\label{1.257}
\rho_2(x_1,x_2,y_1,y_2,t) \cong \rho_1(x_1,y_1,t)\rho_1(x_2,y_2,t)\; ,
\end{equation}
obviously does not obey (1.255). The property (1.255) holds for the
{\it Hartree--Fock approximation}
\begin{equation}
\label{1.258}
\rho_2(x_1,x_2,y_1,y_2,t) \cong \rho_1(x_1,y_1,t)\rho_1(x_2,y_2,t)
\pm \rho_1(x_1,y_2,t)\rho_1(x_2,y_1,t)\; .
\end{equation}

Substituting a decoupling, like (1.257) or (1.258), into (1.248), one
obtains a closed equation for $1$--matrix,
$$ i\frac{\prt}{\prt t}\rho_1(x_1,x_2,t) =
\left [ K_1(x_1,t) - K_1(x_2,t) \right ] \rho_1(x_1,x_2,t) + $$
$$ +\int\left [ \Phi(x_1,x_3) - \Phi(x_2,x_3)\right ]\left [
\rho_1(x_1,x_2,t)\rho_1(x_3,x_3,t) + \right. $$
\begin{equation}
\label{1.259}
\left. + \ep \rho_1(x_1,x_3,t)\rho_1(x_3,x_2,t)\right ] dx_3 \; ,
\end{equation}
in which $\ep=0$ corresponds to the Hartree approximation and $\ep=\pm 1$,
to the Hartree--Fock approximation. Initial and boundary conditions are
supposed to be given for equation (1.259) whose solution has also to
satisfy the conditions
$$ \rho_1^*(x_1,x_2,t) =\rho_1(x_2,x_1,t)\; , \qquad
\rho_1(x,x,t) \geq 0 \; , $$
\begin{equation}
\label{1.260}
\frac{1}{V}\int\rho_1(x,x,t) dx =\rho\; , 
\end{equation}
where $\rho$ is an average density of particles.

Actually, equation (1.259) is a set of two equations, since $1$--matrix is
a complex function
\begin{equation}
\label{1.261}
\rho_1(x_1,x_2,t) := a(x_1,x_2,t) + ib(x_1,x_2,t);
\end{equation}
here $a$ and $b$ are, respectively, the real and imaginary parts of
$\rho_1$. Owing to (1.260), we have
$$ a(x_2,x_1,t) = a(x_1,x_2,t) , \qquad b(x_2,x_1,t) = -b(x_1,x_2,t) , $$
$$ a(x,x,t) \geq 0 , \qquad b(x,x,t) = 0 , \qquad
\frac{1}{V}\int a(x,x,t) dx =\rho\; . $$
Then, from (1.259) and (1.261) we get
$$ \frac{\prt}{\prt t} a(x_1,x_2,t) =
\left [ K_1(x_1,t) - K_1(x_2,t) \right ] b(x_1,x_2,t) + $$
$$ +\int\left [ \Phi(x_1,x_3) - \Phi(x_2,x_3)\right ]
\left\{ a(x_3,x_3,t)b(x_1,x_2,t) + \right. $$
$$ \left. + \ep\left [ a(x_1,x_3,t)b(x_3,x_2,t) +
a(x_3,x_2,t)b(x_1,x_3,t)\right ]\right\} dx_3 \; , $$
$$ \frac{\prt}{\prt t} b(x_1,x_2,t) =
\left [ K_1(x_2,t) - K_1(x_1,t) \right ] a(x_1,x_2,t) + $$
$$ +\int\left [ \Phi(x_2,x_3) - \Phi(x_1,x_3)\right ]
\left\{ a(x_1,x_2,t)a(x_3,x_3,t) + \right.$$
$$ \left. + \ep\left [ a(x_1,x_3,t)a(x_3,x_2,t) +
b(x_1,x_3,t)b(x_3,x_2,t)\right ]\right\} dx_3 \; . $$
These equations describe the dynamics of all, diagonal as well as
off-diagonal, elements of $1$--matrix.

In addition to the Hartree approximation (1.257) and Hartree--Fock
approximation (1.258), another widely used decoupling is the {\it
Hartree--Fock--Bogolubov approximation} [39]
$$ \rho_2(x_1,x_2,y_1,y_2,t)\cong \rho_1(x_1,y_1,t)\rho_1(x_2,y_2,t) 
\pm $$
\begin{equation}
\label{1.262}
\pm \rho_1(x_1,y_2,t)\rho_1(x_2,y_1,t) +\lgl\psi(x_1)\psi(x_2)\rgl
\lgl\psi^\dgr(y_2)\psi^\dgr(y_1)\rgl
\end{equation}
containing the {\it anomalous averages} $\lgl\psi\psi\rgl$ and
$\lgl\psi^\dgr\psi^\dgr\rgl$. In this case, to find a closed set of
evolution equations for $1$--matrix, one needs to write down the evolution
equations for the anomalous averages. Group--theoretical analysis of the
Hartree--Fock--Bogolubov approximation reveals a rich variety of admissible
physical states [47].

\subsection{Coherent States}

The notion of coherent states appears in many applications, therefore it
is worth presenting here the general mathematical definition and the main
properties of these states [48-51].

{\it Coherent states} are microstates that are normalized eigenvectors of
the annihilation operator. Taking the latter in the form (1.54), we have,
by definition,
\begin{equation}
\label{1.263}
\hat\psi(x) h_\eta := \eta(x)h_\eta\; ,
\end{equation}
where $\eta(x)$ is a {\it coherent field}, and the coherent state
\begin{equation}
\label{1.264}
h_\eta = [\eta_n(x^n) ] \in {\cal H}
\end{equation}
pertains to the Hilbert space ${\cal H}$ given by (1.51).

The action of the annihilation operator (1.54) on the vector (1.264)
yields
$$ \hat\psi(x) h_\eta =\left [ \sqrt{n+1}\eta_{n+1}(x^n,x)\right ] \; . $$
Substituting this into (1.263), we have
$$ \sqrt{n+1}\eta_{n+1}(x^n,x) =\eta(x)\eta_n(x^n) \; . $$
The solution to this recursion relation reads
\begin{equation}
\label{1.265}
\eta_n(x^n) =\frac{\eta_0}{\sqrt{n!}}\prod_{k=1}^n\eta(x_k)\; ,
\end{equation}
where $\eta_0\in\Bbb{C}$ and $\eta(x)$ is an arbitrary complex function of
a Hilbert space with the scalar product
$$ (\eta,\xi) := \int\eta^*(x)\xi(x)dx \; . $$

The scalar product of coherent states is
$$ h_\eta^+ h_\xi =\sum_{n=0}^\infty (\eta_n,\xi_n) =
\sum_{n=0}^\infty\frac{\eta_0^*\xi_0}{n!}(\eta,\xi)^n \; , $$
which can be reduced to
$$ h^+_\eta h_\xi =\eta_0^*\xi_0 e^{(\eta,\xi)}\; . $$
From the normalization condition
$$ h_\eta^+ h_\eta = 1 $$
we find
\begin{equation}
\label{1.266}
|\eta_0| =\exp\left\{ -\frac{1}{2}(\eta,\eta)\right\} \; . 
\end{equation}
Thus, coherent states are defined by (1.264), (1.265), and (1.266).

For the scalar product of coherent states, within a constant phase factor,
we have
\begin{equation}
\label{1.267}
h_\eta^+ h_\xi =\exp\left\{ -\frac{1}{2}(\eta,\eta) + (\eta,\xi) -
\frac{1}{2}(\xi,\xi)\right \} \; ,
\end{equation}
which shows that different states are not orthogonal to each other, since
$$ 0 < | h_\eta^+ h_\xi| \leq 1\; . $$
Introducing the corresponding functional integral with a differential
measure ${\cal D}\eta$, one can get the resolution of unity
\begin{equation}
\label{1.268}
\int h_\eta h_\eta^+ {\cal D}\eta = \hat 1\; .
\end{equation}
The formal expression (1.268) is to be interpreted in the sense of weak
equality, that is, as equality of arbitrary matrix elements of the
indicated expression.

Coherent states form a basis but it is overcomplete because of the absence
of orthogonality. Also, coherent states are eigenvectors of only the
annihilation operator (1.54), but not of the creation operator (1.55),
whose action on (1.264) gives
$$ \hat\psi^\dgr h_\eta = \left [ \sqrt{n}\eta_{n-1}(x^{n-1})
\delta( x_n -x )\right ] \; . $$
However, there is a relation
\begin{equation}
\label{1.269}
h_\eta^+\hat\psi^\dgr(x) = h_\eta^+\eta^*(x)
\end{equation}
following from the Hermitian conjugation of (1.263).

Considering the double action
$$ \hat\psi(x)\hat\psi(x') h_\eta = \eta(x)\eta(x') h_\eta \; , $$
we see that, on the class of coherent states, annihilation operators
commute, $$ \left [ \hat\psi(x),\hat\psi(x')\right ] h_\eta = 0\; . $$
But there are no simple commutators involving creation operators, which is
clear from the actions
$$ \hat\psi^\dgr(x)\hat\psi^\dgr(x') h_\eta =\left [ \sqrt{n(n-1)}
\eta_{n-2}(x^{n-2})\delta(x_{n-1} - x')\delta(x_n - x)\right ]\;  , $$
$$ \hat\psi(x)\hat\psi^\dgr(x') h_\eta =
\left [ (n+1) \eta_n(x^n)\delta(x  - x') \right ]\; , $$
$$ \hat\psi^\dgr(x')\hat\psi(x) h_\eta =\left [ n
\eta_n(x^{n-1},x)\delta(x_n - x')\right ] \; . $$

Notice that function (1.265) is symmetric with respect to permutation of
any two variables $x_i,x_j\in x^n$,
$$ P(x_i,x_j)\eta_n(x^n) =\eta_n(x^n) \; . $$
Consequently, $\eta_n(x^n)\in{\cal H}_n^+$ and $h_\eta\in {\cal H}_+$,
where ${\cal H}_n^+$ and ${\cal H}_+$ are defined in Section 1.3. In the
same way as in (1.69), we may introduce the Bose field operators $\psi_+$
and $\psi_+^\dgr$, which have, on the class of coherent states, simple
commutation relations given in (1.71).

To introduce coherent states for the Fermi field operators, according to
(1.69), requires that the function (1.265) would be antisymmetric. This is
achievable if the coherent field $\eta(x)$ is not a complex function, as
in the case of Bose statistics, but an element of the Grassmann algebra,
satisfying the relation
$$ \eta_F(x)\eta_F(x') +\eta_F(x')\eta_F(x) = 0\; , $$
in which the index $F$ refers to the Fermi case. If $\eta(x)$ is a {\it
Grassmann variable}, then the function (1.265) is antisymmetric with
respect to permutation of variables $x_i,x_j\in x^n$. Hence,
$\eta_n(x^n)\in{\cal H}_n^-$, and $h_\eta\in{\cal H}_-$. Therefore, one
can define the Fermi field operators $\psi_-$ and $\psi_+$, as in (1.69),
which have, on the class of coherent states, the standard anticommutation
relations, as in (1.71).

In this way, coherent states can be introduces as the normalized
eigenvalues of Bose, as well as Fermi operators,
\begin{equation}
\label{1.270}
\psi(x) h_\eta := \eta(x) h_\eta\; ,
\end{equation}
where $\psi(x)$ is either $\psi_+$ or $\psi_-(x)$. The difference is that
for Bose statistics $\eta(x)$ is a complex function, while for Fermi
statistics $\eta(x)$ is an element of the Grassmann algebra. These two
cases may be combined in one condition for the coherent fields
\begin{equation}
\label{1.271}
\left [\eta(x),\eta(x')\right ]_\mp = 0 \; ,
\end{equation}
in which the upper sign, as usual, is related to Bose statistics, and the
lower, to Fermi statistics.

Coherent states (1.264) can be obtained from vacuum by means of field
operators. By analogy with (1.57) and (1.63), we may define {\it
n--particle coherent states}
\begin{equation}
\label{1.272}
\left [ \delta_{mn}\eta_n(x^n)\right ] :=
\frac{1}{\sqrt{n!}}\int \eta_n(x^n)\prod_{k=1}^n
\psi^\dgr(x_k)|0\rgl dx^n\; ,
\end{equation}
where $\eta(x)$ entering into (1.265), depending on the type of statistics
considered, is either a complex function or a Grassmann variable, as is
discussed above, and $\psi^\dgr(x)$ is, respectively, either a Bose
creation operator, or Fermi creation operator. With the form (1.265), we
can present (1.272) as
$$ \left [\delta_{mn}\eta_n(x^n)\right ] =
\frac{\eta_0}{n!}\left (\int\psi^\dgr(x)\eta(x)dx\right )^n | 0\rgl\; . $$
Then, for the coherent state (1.264) we get
$$ h_\eta =\sum_{n=0}^\infty\frac{\eta_0}{n!} \left (\int\psi^\dgr(x)
\eta(x) dx\right )^n| 0\rgl\; . $$
As is easy to notice, the latter expression is a series expansion of an
exponential
$$ h_\eta =
\eta_0\exp\left\{ \int\psi^\dgr(x)\eta(x) dx\right\} |0\rgl \; . $$
Taking onto account the normalization (1.266), we come to 
\begin{equation}
\label{1.273}
h_\eta =\exp\left\{ -\frac{1}{2}(\eta,\eta) +\int\psi^\dgr(x)\eta(x)dx
\right\} |0\rgl\; .
\end{equation}
This is a general formula for constructing coherent states from vacuum.

To find the time evolution of coherent states, let us generalize the
definition (1.263) to the form
\begin{equation}
\label{1.274}
\psi(x,t) h_\eta =\eta(x,t) h_\eta \; ,
\end{equation}
in which $\psi(x,t)$ is given by (1.75). Whence, the {\it time--dependent
coherent state} is
\begin{equation}
\label{1.275}
h_\eta(t) := U(t) h_\eta\; ,
\end{equation}
for which
$$ \psi(x,0) h_\eta(t) =\eta(x,t) h_\eta(t)\; . $$
The Hermitian conjugate to (1.274) is
$$ h^+_\eta \psi^\dgr(x,t) = h^+_\eta \eta^*(x,t) \; . $$
The evolution equation for (1.275) follows from the equation
$$ i\frac{d}{dt} U(t) = HU(t) $$
for the unitary operator, which yields
\begin{equation}
\label{1.276}
i\frac{d}{dt} h_\eta(t) = H h_\eta(t) \; . 
\end{equation}
The scalar product of time--dependent coherent states
$$ h_\eta^+(t) h_\xi(t) = 
h_\eta^+ h_\xi = \eta^*_0\xi_0 e^{(\eta,\xi)} $$ 
does not depend on time.

Coherent states permit us to write simple expressions for the matrix
elements of the operators from the algebra of observables. Thus, for the
number--of--particle operator (1.208), using
$$ h^+_\eta\psi^\dgr(x,t)\psi(x',t) h_\xi =
\eta_0^*\xi_0 e^{(\eta,\xi)}\eta^*(x,t)\xi(x',t) , $$
we get
$$ h^+_\eta\hn(t) h_\xi =\eta_0^*\xi_0e^{(\eta,\xi)}
\int\eta^*(x,t)\xi(x,t) dx \; . $$
The diagonal elements of this is
\begin{equation}
\label{1.277}
h^+_\eta\hn(t) h_\eta =\int\left |\eta^*(x,t)\right |^2 dx\; ,
\end{equation}
where the normalization (1.266) is invoked. For the Hamiltonian (1.226),
the diagonal element reads
$$ h_\eta^+ H(t) h_\eta = \int\eta^*(x,t)\left [ K(\srn) + U(x,t) 
-\mu\right ]\eta(x,t) dx + $$
\begin{equation}
\label{1.278}
+\frac{1}{2}\int |\eta(x,t)|^2 V(x,x')|\eta(x',t)|^2 dx dx'\; .
\end{equation}

The evolution equation for a coherent field $\eta(x,t)$ can be derived
from equation (1.234) for the field operator and the use of the equality
\begin{equation}
\label{1.279}
h_\eta^+ i\frac{\prt}{\prt t}\psi(x,t) h_\eta =
i\frac{\prt}{\prt t}\eta(x,t)\; .
\end{equation}
In this way we obtain the {\it coherent--field equation}
\begin{equation}
\label{1.280}
\left\{ K(\srn) + U(x,t) - \mu +\int\Phi(x,x')|\eta(x',t)|^2 dx'
\right\}\eta(x,t) = i \frac{\prt}{\prt t}\eta(x,t)\; .
\end{equation}
The latter has the structure of a Schr\"odinger equation with an {\it
effective coherent potential}
$$ U_{eff}(x,t) := U(x,t) - \mu +
\int\Phi(x,x') |\eta(x',t)|^2 dx' \; . $$
Since the part of (1.280), containing interparticle interactions is
nonlinear with respect to the coherent field $\eta(x,t)$, the equation
(1.280) is a {\it nonlinear Schr\"odinger equation}. This equation defines
the coherent field up to a phase factor.

It is important to notice that nontrivial coherent states exist not for
any interaction potential, but only for those that are integrable, so that
$$ \left | \int \Phi(x,x')dx' \right | < \infty\; . $$
If it is not so, then the corresponding term in (1.280) diverges and the
coherent--field equation has only a trivial solution $\eta(x,t)\equiv 0$.

\end{sloppypar}


\section{Correlated Iteration}

The Green function technique expounded in the previous Chapter can be 
applied to many statistical systems. However, there exists an important 
class of systems which exhibits such a strong correlation between 
particles that a direct application of the methods discussed in the 
previous Chapter becomes impossible. For such systems there should be 
developed a special approach taking into account strong interparticle
correlations from the very beginning.

In this Chapter, we shall present such an approach to treating strongly 
correlated systems, based on Refs. [81-84].

\subsection{Singular Potentials}

The iteration procedures for developing a perturbation series which we have 
discussed hitherto are convenient for systems of particles interacting through
non--singular or weakly singular potentials. We make these terms more 
precise by classifying potentials according to their behaviour as 
$r_{12}\ra 0$, where $r_{12}\equiv |{\bf r}_{12}|$, and ${\bf r}_{12}={\bf 
r}_2-{\bf r}_1$. We shall say that $\Phi(x_1,x_2)$ is a {\it non--singular
potential} if
\be
\label{1}
\lim_{r_{12}\ra 0}|\Phi(x_1,x_2)| < \infty .
\ee
A {\it singular potential} is one for which
\be
\label{2}
\lim_{r_{12}\ra 0}|\Phi(x_1,x_2)| = \infty .
\ee
A {\it weakly singular potential} is singular but is integrable in the sense
\be
\label{3}
\left | \int_{\cal R}\Phi(x_1,x_2)d{\bf r}_{12}\right | < \infty \; ,
\ee
where ${\cal R}$ is a finite region including $r_{12}=0$. A {\it strongly 
singular potential} is singular and is not integrable so that
\be
\label{4}
\left | \int_{\cal R}\Phi(x_1,x_2)d{\bf r}_{12}\right | = \infty .
\ee

For example, if the behaviour of a potential at short distances is 
governed by the power law
$$ \Phi(x_1,x_2)\sim\frac{1}{r_{12}^n} \qquad (r_{12}\ra 0) \; , $$
and ${\cal R}$ is a small finite ball centered at the origin, then the 
potential, $\Phi$ is nonsingular if $n=0$, weakly singular if $0<n\leq 2$,
and strongly singular if $n>2$. Thus the Coulomb--like potentials with 
$n=1$, are weakly singular. Whereas the Lennard--Jones potential with 
$n=12$, which is sometimes used to model interaction of molecules, is 
strongly singular.

For systems interacting via a strongly singular potential, any iteration 
procedure starting from the Hartree--Fock approximation breaks down since
in the very first step the Hartree potential is divergent. This is because 
the input to the Hartree approximation assumes independent particles and 
does not take into account short--range correlations which could 
effectively shield the singularity. These short--range correlations might 
be retained by a decoupling procedure in which the three--particle 
propagator is approximated by a linear combination of two--particle 
propagators.  However, decoupling at any stage effectively precludes 
subsequent approximations.

What we really need is a perturbation procedure which satisfies the 
following conditions: 1) it should start from an approximation  
containing no divergencies and must therefore take into account strong 
short--range correlations, 2) it should provide an algorithm which 
permits recursive calculation of higher--order corrections. We now 
propose a procedure which we believe meets these desiderata and which we 
have named the {\it correlated iteration theory}.

\subsection{Iterative Algorithm}

Let us consider a situation with a strongly singular interaction 
potential $\Phi(12)$ and let $G_2^0$ be given by (2.75). Suppose we can
find  a function, $s(12)$, such that
\be
\label{5}
G_2^{app}(1221) := -s(12)\rho(1)\rho(2) , \qquad t_1 > t_2 \; ,
\ee
is an approximation for the $2$--particle propagator. We shall call 
$s(12)$ a {\it smoothing function}. It follows from this equation that 
\be
\label{6}
s(12)=s(21) .
\ee
We shall require that the smoothed potential 
\be
\label{7}
\bar\Phi(12) :=s(12)\Phi(12)
\ee
be nonsingular or weakly singular and therefore integrable in the sense 
of (\ref{3}).

We now introduce a function, $D(123)$, satisfying the relation
\be
\label{8}
G_2(1223) :=s(12)\int D(124)G(43)d(4) ,
\ee
which we call the {\it doubling function}. This function will be used to 
transform a $1$--particle into a $2$--particle propagator, while the 
smoothing function takes inter--particle correlations into account. An 
initial approximation, $D_0$, can be obtained by finding a solution of 
the integral equation obtained by combining (\ref{5}) with (\ref{8}), giving
\be
\label{9}
G_2^0(1223) =\int D_0(124)G(43)d(4) .
\ee
It is not difficult to verify that
\be
\label{10}
D_0(123) :=\delta(13)G(22)\pm G(12)\delta(23)
\ee
is a solution of (\ref{9}). Using (\ref{8}), the exact self--energy  
becomes
\be
\label{11}
\Sigma(12) =\pm i\int\bar\Phi(13)D(132)d(3)
\ee
which, in view of (\ref{10}), has as first approximation
\be
\label{12}
\Sigma_1(12) = i\bar\Phi(12)G(12) +\delta(12)\int\bar\Phi(13)\rho(3)d(3) .
\ee
This differs from the Hartree--Fock form in that the bare interaction 
potential $\Phi$ has been replaced by the smoothed potential $\bar\Phi$. 
Therefore, no divergencies appear in (\ref{12}) which may be called the 
{\it first correlated approximation} or the {\it correlated Hartree--Fock
approximation}.

In order to define a recursive procedure for obtaining higher--order 
approximations we shall need a set of equations which determine the 
doubling function and the self--energy. Establishing these equations is the
next, and most complicated step in our theory. By a variation of the 
equation of motion (2.69) with respect to $\mu$, we have
$$ \frac{\delta G(12)}{\delta\mu(3)} = -\int G(14)
\frac{\delta G^{-1}(45)}{\delta\mu(3)}G(52)d(45) . $$
From the definition of the inverse propagator it follows that
$$ \frac{\delta G^{-1}(12)}{\delta\mu(3)} =\delta(12)\delta(13) -
\frac{\delta\Sigma(12)}{\delta\mu(3)} , $$
where
$$ \frac{\delta\Sigma(12)}{\delta\mu(3)} =\int
\frac{\delta\Sigma(12)}{\delta G(45)}
\frac{\delta G(45)}{\delta\mu(3)} d(45) . $$
Therefore,
$$ \frac{\delta G(12)}{\delta\mu(3)} = - G(13)G(32) + $$
$$ + \int G(14)G(52)\frac{\delta\Sigma(45)}{\delta G(67)}
\frac{\delta G(67)}{\delta\mu(3)} d(4567) . $$
Substituting in this the expression
$$ \frac{\delta G(12)}{\delta\mu(3)} =\pm\left [ G(12)G(33) -G_2(1332)
\right ] , $$
which results from the variational representation (2.151), we obtain the 
equation 
\be
\label{13}
G_2(1223) = G_2^0(1223) +
\ee
$$ + \int G(14)\frac{\delta\Sigma(45)}{\delta G(67)}\left [
G_2(6227) -G(67)G(22)\right ] G(53) d(4567) $$
for the $2$--propagator. By using the relation (\ref{8}) which defines 
the doubling function, we get
\be
\label{14}
s(12)D(123) = D_0(123) +
\ee
$$ +\int G(14)\frac{\delta\Sigma(43)}{\delta G(56)}\left [
s(52)D(527)G(76) - G(22)G(56)\delta(67)\right ] d(4567) . $$

This suggests that we should define an integral operator ${\bf X}$ on 
$H^3$ as follows
\be
\label{15}
{\bf X}f(123) :=s(12)f(123) +
\ee
$$ +\int G(14)\frac{\delta\Sigma(43)}{\delta G(56)}\left [
G(22)G(56)\delta(67) - s(52)f(527)G(76)\right ] d(4567) . $$
Then (\ref{14}) may be rewritten as
\be
\label{16}
{\bf X}D(123) = D_0(123) .
\ee

Now define, as left inverse of ${\bf X}$, the operator ${\bf Y}$ such that
\be
\label{17}
{\bf Y} :={\bf X}^{-1} , \qquad {\bf YX}={\bf 1} ,
\ee
where ${\bf 1}$ is the identity operator. Acting with ${\bf Y}$ on 
(\ref{16}),
we obtain
\be
\label{18}
D(123) ={\bf Y}D_0(123) ,
\ee
which is the main equation for the doubling function.

To obtain an iteration procedure for solving (\ref{18}), we proceed as 
follows. Assume that our initial approximation, (\ref{5}), models the 
major properties of the system under consideration so that $D$ will be 
close to $D_0$. In this case, we shall assume that ${\bf X}$ is close to 
the identity. We therefore try to obtain ${\bf Y}$ by expressing it as
\be
\label{19}
{\bf Y} =[{\bf 1} -({\bf 1}-{\bf X})]^{-1} ,
\ee
which leads to the Neumann series expansion
\be
\label{20}
{\bf Y} =\sum_{k=0}^{\infty} ({\bf 1}-{\bf X})^k .
\ee
From (\ref{15}) we see that ${\bf X}$ is a functional of the self--energy 
$\Sigma$,
\be
\label{21}
{\bf X} ={\bf X}[\Sigma] .
\ee
Let
\be
\label{22}
{\bf X}_k ={\bf X}[\Sigma_k]  \qquad ({\bf X}_0:= 1) 
\ee
be the same functional acting on the $k$--approximation to the 
self--energy. Define the $k$--approximation to ${\bf Y}$ by
\be
\label{23}
{\bf Y}_k :=\sum_{p=0}^{k}({\bf 1}-{\bf X}_k)^p  \qquad 
({\bf Y}_0 := {\bf 1})\; .
\ee
If the sequence ${\{\bf Y}_k\}$ has a limit, then
\be
\label{24}
{\bf Y} =\lim_{k\ra\infty}{\bf Y}_k ,
\ee
provided that $\Sigma=\lim_{k\ra\infty}\Sigma_k$.

With equations (\ref{11}), (\ref{18}), and (\ref{20}) we have reached our 
immediate goal of being able to define an iteration process for the 
calculation of the doubling function and the self--energy. The recursion 
can be organized according to the scheme:
\be
\label{25}
D_k\ra\Sigma_{k+1}\ra {\bf Y}_{k+1}\ra D_{k+1} .
\ee
For $k=0$ we start from ({\ref{10}) leading to (\ref{12}). Then because 
of the variational derivative
$$ \frac{\delta\Sigma_1(12)}{\delta G(34)} = i\bar\Phi(14)\left [
\delta(13)\delta(24) \pm \delta(12)\delta(34)\right ] , $$
we find
\be
\label{26}
{\bf Y}_1 = 2 -{\bf X}_1 .
\ee
It follows that (\ref{18}) gives
$$ D_1(123) = [2-s(12)]D_0(123) \pm $$
\be
\label{27}
\pm i\int\left [ G(14)G(23)\pm G(13)G(24)\right ] G(42)\bar\Phi(43)d(4)\mp
\ee
$$ \mp i\int\left [ G(13)G(44)\pm G(14)G(43)\right ]G(22)\left [ 1 -
s(42)\right ]\bar\Phi(43) d(4) . $$
Combining this with (\ref{11}) yields the {\it second correlated 
approximation} for the self--energy:
$$ \Sigma_2(12) = 
i\bar\Phi(12)G(12) + \delta(12)\int\Phi_2(12)\rho(3)d(3) -
$$
\be
\label{28}
- \int\bar\Phi(13)G_2^0(1324)G(43)\bar\Phi(42)d(34) +
\ee
$$ +\int\Phi(134) G_2^0(1332) G(44)\bar\Phi(32)d(34) , $$
which involves the following two new functions, the {\it twice smoothed
potential}
$$ \Phi_2(12) :=\bar\Phi(12)[2-s(12)] = s(12)[2-s(12)]\Phi(12) , $$
and the {\it effective triple potential}
$$ \Phi(123) :=\bar\Phi(13)[1-s(23)] = s(13)[1-s(23)]\Phi(13) . $$
Provided that the smoothing function, $s(12)$, is known we may find 
higher--order approximations by following the scheme (\ref{25}).

\subsection{Smoothing Function}

According to (\ref{5}) the smoothing function is
\be
\label{29}
s(12) = - \frac{G_2^{app}(1221)}{\rho(1)\rho(2)} \qquad 
(t_{12} > 0 ) .
\ee
In view of the relations
$$ G(11) = \mp i\rho(1) , \qquad G(12)G(21) =\mp i\Pi_0(12) , $$
the denominator in (\ref{29}) is given by
$$ \rho(1)\rho(2)= -G_2^0(1221)  -i\Pi_0(12) . $$
Since it is clear that the diagonal two--particle propagator, 
$G_2(1221)$, plays a special role in our correlated approximation 
procedure, we analyze some of its general properties.

Consider the density--of--particle operator
\be
\label{30}
{\hat n}(1) :=\psi^\dgr(1)\psi(1) ,
\ee
which has expected value
\be
\label{31}
\rho(1) =\lgl{\hat n}(1)\rgl ,
\ee
equal to the density--of--particles. It is easy to check that when $t_1=t_2$,
$$ [{\hat n}(1),{\hat n}(2)] = 0 . $$
Thus we may define the chronological product
\begin{eqnarray}
\label{32}
\hat T{\hat n}(1){\hat n}(2) :=\left\{ \begin{array}{cc}
{\hat n}(1){\hat n}(2) , & t_1 > t_2
\\
{\hat n}(2){\hat n}(1) , & t_1 < t_2\end{array}\right. \; .
\end{eqnarray}
We may then, for the diagonal propagator $G_2$, obtain the expression
\be
\label{33}
G_2(1221) = -\lgl \hat T{\hat n}(1){\hat n}(2)\rgl .
\ee
Note the symmetry property
\be
\label{34}
G_2(1221) = G_2(2112) ,
\ee
which holds both for $t_1>t_2$ and $t_1<t_2$. This is in contrast to the 
{\it density--density correlation} function
\be
\label{35}
R(12) :=\lgl {\hat n}(1){\hat n}(2)\rgl
\ee
for which, in general, $R(12) = R(21)$ only when $t_1=t_2$.

The correlation function (\ref{35}) is related to the second--order density 
matrix by the relation
$$ \lim_{t_2\ra t_1}R(12) =\delta(x_1-x_2)\rho(1) 
+\rho_2(x_1,x_2,x_1,x_2,t_1) , $$
which implies that the propagator (\ref{33}) may be expressed in the form
\be
\label{36}
G_2(1221) = -\Theta(t_{12})R(12) - \Theta(t_{21})R(21) ,
\ee
where $t_{12}:=t_1-t_2$.

A quantity which is often considered is the {\it density--density Green's
function} defined by
\be
\label{37}
G_\rho(12) := -i\lgl\hat T {\hat n}(1){\hat n}(2)\rgl ,
\ee
which satisfies spectral representations and dispersion relations similar 
to those of the Bose propagators studied in Sections 2.2 and 2.3. The 
propagators (\ref{36}) and (\ref{37}) differ only by a factor $i$,
$$ G_\rho(12) = iG_2(1221) . $$

Another important quantity is the {\it pair correlation function},
\be
\label{38}
g(12) := \frac{\lgl {\hat n}(1){\hat n}(2)\rgl }{\lgl {\hat n}(1)\rgl
\lgl {\hat n}(2)\rgl } .
\ee
Since this is related to (\ref{35}) by the equation
\be
\label{39}
R(12) =\rho(1)\rho(2)g(12) ,
\ee
it follows that $g(12)$ and $R(12)$ have the same symmetry properties. In 
particular,
$$ g(12) =g(21), \qquad  (t_1=t_2) \; . $$
By (\ref{38}), the propagator (\ref{33}) can be expressed by the formula
$$ G_2(1221) = -\rho(1)\rho(2)\left [\Theta(t_{12})g(12) +\Theta(t_{21})g(21)
\right ] . $$

The equations of motion imply that $G_2$ differs from $G_2^0$ only in 
that it takes into account interactions between particles. Thus if 
$\Phi(12)$ approaches zero, then $G_2(1234)$ approaches $G_2^0(1234)$.
This
implies that the smoothing function (\ref{29}) must satisfy the 
asymptotic condition
\be
\label{40}
s(12)\ra 1 , \qquad | x_1 - x_2| \ra \infty\;  .
\ee
The Bethe--Salpeter equation (2.86) will sometimes enable us to find an 
approximate $G_2^{app}$ needed for the equation (\ref{29}) defining the 
smoothing function.

\subsection{Response Functions}

The functions discussed in this section enable us to characterize 
the changes in a system under the action of some external source or when 
the corresponding macroscopic quantity is varied. For example, the {\it 
density response function}
\be
\label{41}
\chi(12) :=-\frac{\delta\rho(1)}{\delta\mu(2)} \; ,
\ee
describes the change in the average density
$$ \delta\rho(1) = -\int\chi(12)\delta\mu(2) $$
when the effective chemical potential is varied.

It is also convenient to introduce the {\it generalized response function}
\be
\label{42}
\chi(123) :=\mp i\frac{\delta G(12)}{\delta\mu(3)}
\ee
of which (\ref{41}) is the particular case
\be
\label{43}
\chi(12)=\chi(112) .
\ee
From the variational representation (2.151) we have
\be
\label{44}
\chi(123) = i\left [ G_2(1332) - G(12)G(33)\right ] .
\ee
By substituting $G_2^0$ we find
$$ \chi_0(123) = i\left [ G_2^0(1332) - G(12)G(33)\right ] , $$
which defines in zero--approximation
\be
\label{45}
\chi_0(123) = \pm i G(13)G(32)
\ee
for the response function (\ref{42}).

Because of (\ref{43}) the density response function (\ref{41}) can be 
expressed by
\be
\label{46}
\chi(12) = i\left [ G_2(1221) +\rho(1)\rho(2)\right ] .
\ee
In view of (\ref{34}) this shows that the response function is also 
symmetric:
\be
\chi(12) =\chi(21) = -\frac{\delta\rho(2)}{\delta\mu(1)} \; .
\ee

An equation for the determination of the generalized response function 
can be obtained by using (\ref{44}) and (\ref{14}), giving
\be
\label{48}
\chi(123) =\chi_0(123) +\int G(14)G(52)\frac{\delta\Sigma(45)}{\delta G(67)}
\chi(673)d(4567) .
\ee
In order to make effective use of this equation we need an approximation 
for $\Sigma$. It is perhaps worth stressing again that in the case of 
strongly singular potentials we cannot employ the Hartree--Fock 
approximation for the self--energy. Some modified approximation such 
as that provided by the correlated iteration procedure is essential.

From the {\it correlated Hartree approximation}
\be
\label{49}
\Sigma_{CH}(12) :=\delta(12)\int\bar\Phi(13)\rho(3)d(3) ,
\ee
since
$$ \frac{\delta\Sigma_{CH}}{\delta G(34)} =\pm 
i\delta(12)\bar\Phi(23)\delta(34) , $$
noting (\ref{48}) we conclude that
$$ \chi(123) =\chi_0(123) +\int\chi_0(124)\bar\Phi(45)\chi(553)d(45) . $$
By specialization this shows that the equation for (\ref{43}) takes the form
\be
\label{50}
\chi(12) =\chi_0(12) +\int\chi_0(13)\bar\Phi(34)\chi(42)d(34) ,
\ee
where
$$ \chi_0(12) :=\pm i G(12)G(21) =\Pi_0(12) . $$
We shall say that the solution of (\ref{50}) is the response function in 
{\it correlated random--phase approximation}. In the standard 
random--phase approximation, the smoothed potential, $\bar\Phi$, would be
replaced by the bare potential, $\Phi$. If we employ the correlated 
Hartree--Fock approximation (\ref{12}) for $\Sigma$ in (\ref{48}) we are
led to the equation
\be
\label{51}
\chi(123) =\chi_0(123) \pm
\ee
$$ 
\pm i\int G(14)\bar\Phi(45)[G(42)\chi(553) \pm G(52)\chi(453)] d(45) .
$$

We note in Chapter 2 that the poles of the two--particle propagators 
define the spectrum of collective excitations. The relation (\ref{46}) 
shows that these propagators and the density response functions have the 
same poles. Since the equations for the response functions are usually 
simpler than those for the $2$--propagators, the response functions are 
commonly considered when the collective excitations spectrum is of interest. 
We describe two possible approaches to finding this spectrum.

\vspace{2mm}

{\bf 1.} Substitute an approximation for the self--energy, $\Sigma$, in 
(\ref{48}) which results in an equation for the {\it generalized} response 
function. If possible, specialize to obtain an equation for the density 
response function. Except for usual cases, this specialization is feasible 
only when the Hartree approximation is employed for the self--energy, 
but then as follows from (\ref{51}) this specialization will be impossible 
in the next approximation. Further, to proceed from the first approximation 
one needs to solve the resulting equation for the density response function. 
Because of the product in the integrand of the right--hand side of (\ref{50}) 
this can be quite difficult unless one uses merely a random--phase type of 
approximation. In this latter case we find a solution for $\chi(12)$ in the 
form
\be
\label{52}
\chi_{CRP} =\frac{\chi_0}{1-\chi_0\bar\Phi}\; ,
\ee
where the subscript, $CRP$, means {\it correlated random--phase} 
approximation. The spectrum of collective excitations is then 
approximated by the zeros of the equation
\be
\label{53}
1 -\chi_0\bar\Phi = 0 .
\ee
However, it is not always easy to solve (\ref{51}) for the density 
response function.

\vspace{1mm}

{\bf 2.} A second approach to finding the spectrum of the collective 
excitations begins by seeking an approximate iterative solution of (\ref{48}) 
in the form of a series in powers of $\bar\Phi$. Since the poles of $\chi$ 
are the zeros of $\chi^{-1}$, the spectrum is characterized by the equation 
$\chi^{-1}=0$ after $\chi$ is replaced by a polynomial in $\bar\Phi$ of 
degree equal to the order of approximation one is using for $\chi$. For 
example, taking
\be
\label{54}
\chi\simeq \chi_0 +\chi_0\bar\Phi\chi_0 
\ee
leads to
\be
\label{55}
\chi^{-1} \simeq \left ( 1 -\chi_0\bar\Phi\right )\chi_0^{-1} ,
\ee
when the infinite series for $\chi^{-1}$ is truncated at the first order 
in $\bar\Phi$. The zeros of this equation are the same as those of 
equation (\ref{53}). Notice that it might happen that (\ref{54}) has no 
poles even though (\ref{55}) has zeros! This could result because the 
truncation meant that these two equations are not strictly equivalent.

\vspace{2mm}

In order to carry this second approach through effectively we need to 
have at our disposal a "regular" perturbation procedure, -- that is a 
procedure which does not terminate abruptly due to the occurrence of 
singularities. We can define such a procedure by employing the same trick 
for inverting operators as we used in our correlated iteration theory for 
the doubling function and self--energy in Section 3.2.

We introduce an integral operator, ${\bf Z}$, on $H^3$ such that ${\bf 
Z}f$ is defined by
\be
\label{56}
{\bf Z}f(123) := \int G(14)G(52)\frac{\delta\Sigma(45)}{\delta G(67)}
f(673)d(4567) .
\ee
Then the basic equation, (\ref{48}), for the generalized response 
function can be written as
\be
\label{57}
({\bf I} -{\bf Z})\chi(123) =\chi_0(123) .
\ee
This is equivalent to
\be
\chi(123) = ({\bf I} -{\bf Z})^{-1}\chi_0(123) ,
\ee
in which the operator can be expanded in the series,
\be
\label{59}
({\bf I} -{\bf Z})^{-1} =\sum_{k=0}^\infty {\bf Z}^k .
\ee
It is clear from (\ref{56}) that ${\bf Z}$ is a functional of the 
self--energy. We symbolize this by
\be
\label{60}
{\bf Z} = {\bf Z}[\Sigma] ,
\ee
and denote the operator $k$--approximation to ${\bf Z}$ by
\be
\label{61}
{\bf Z}_k :={\bf Z}[\Sigma]_k, \qquad {\bf Z}_0 := {\bf I} ,
\ee
where $\Sigma_k$ denotes the $k$--approximation for the self--energy in 
the correlated iterative scheme. The $k$--approximation for the 
generalized response function is then $\chi_k$, defined by
\be
\label{62}
\chi_k(123) :=\sum_{p=0}^k {\bf Z}_k^p\chi_0(123) .
\ee
In the previous Section we showed how to obtain successive approximations 
to the self--energy by means of correlated iterations, so following the 
scheme
\be
\label{63}
\Sigma_k \ra {\bf Z}_k \ra \chi_k
\ee
we may obtain successive approximations to the generalized response 
function. According to (\ref{43}) the $k$--approximation to the density 
response function is obtained by the specialization
\be
\label{64}
\chi_k(12) =\chi_k(112) .
\ee

For example, with $k=1$ in (\ref{62}) we have
$$ \chi_1(123) =\chi_0(123) \pm $$ 
$$ 
\pm i\int G(14)\bar\Phi(45)[G(42)\chi_0(553)\pm G(52)\chi_0(453)]d(45) . 
$$
It follows that
\be
\label{65}
\chi_1(12) =\chi_0(12) +\int\chi_0(13)\bar\Phi(34)\chi_0(42)d(34) \pm 
\ee
$$ \pm \int\chi_0(123)\bar\Phi(34)\chi_0(214)d(34) . $$

The schemes (\ref{25}) and (\ref{63}) provide a recursive procedure for 
obtaining a sequence of successive approximations to the density 
response function in powers of $\bar\Phi$ which define $\chi$ 
asymptotically. The poles of the function so--defined, or the zeros of 
$\chi^{-1}$, characterize the spectrum of collective excitations of the 
system.

\subsection{Correlation Function}

From the above analysis it follows that the smoothing function and the
pair correlation function are intimately related with each other. Really,
comparing the definitions (3.5) and (3.38), we see that the smoothing
function can be identified with an approximation for the pair correlation
function. Therefore, to make the correlated iteration theory complete, we
need to present a way for finding approximate correlation functions.

The latter, according to (\ref{38}), can be defined as
\begin{equation}
\label{66}
g(12) = -\frac{G_2(1221)}{\rho(1)\rho(2)} \qquad ( t_{12} > 0 ) .
\end{equation}
If we substitute in (\ref{66})
$$ G_2^0(1221) = -\rho(1)\rho(2) \pm G(12) G(21)\; , $$
then we get
\begin{equation}
\label{67}
g_0(12) = 1 \mp \frac{G(12)G(21)}{\rho(1)\rho(2)}\; .
\end{equation}

A general equation for $g(12)$ can be obtained employing in (\ref{13})
definition (\ref{66}), which gives
$$
g(12) = g_0(12) + \int\frac{G(13)G(41)}{\rho(1)\rho(2)}\cdot
\frac{\delta\Sigma(34)}{\delta G(56)}\times 
$$
\begin{equation}
\label{68}
\times \left [ G(56)G(22) - g(52)\int D(527)G(76) d(7)\right ] d(3456) ,
\end{equation}
keeping in mind the condition $t_{12}>0$. Using in (\ref{68}) approximate
expressions for $\Sigma$ and $D$, we then come to a set of approximate
equations defining $g(12)$. For instance, putting $\Sigma=\Sigma_0=0$ in
(\ref{68}) immediately returns us to (\ref{67}). If we take for $\Sigma$
and $D$ the first approximations in (\ref{12}) and (\ref{37}),
respectively, then ({\ref{68}) transforms to
\begin{equation}
\label{69}
g(12) = g_0(12) -\int A(1234) g(34)\Phi(34) d(34)\; ,
\end{equation}
where we keep only the terms up to first order in $\Phi$ and
$$ A(1234) := i\frac{G(13)G(41)}{\rho(1)\rho(2)}\left \{
g(32)\left [ 2 - g(32)\right ] G_2^0(3224) - G(34) G(22)\right \} \pm  $$
$$ \pm i \frac{G(13)G(31)}{\rho(1)\rho(2)}\left \{ g(42)\left [ 2 -
g(42)\right ] G_2^0(4224) - G(44)G(22)\right \}\; . $$
Equation (\ref{69}) defines first--order approximation for $g(12)$.

Let us notice that (\ref{69}) cannot be iterated starting with $g_0(12)$ in
the integrand, since then integral would diverge because of singularity in
$\Phi(34)$. A simplification is possible only for $A(1234)$. If we put in
the latter $g=1$, then
$$ A(1234) = i\frac{G(13)G(24)}{\rho(1)\rho(2)} G_2^0(3421) \; . $$
Taking $g=1$ implies that the Hartree approximation is used for which
$$ G_2^0(1234) \cong G(14)G(23) , \qquad g_0(12) \cong 1\; . $$
With this in mind, we can make a further simplification getting
$$ A(1234) \cong i\frac{G(13)G(31)G(24)G(42)}{\rho(1)\rho(2)}\; . $$

If we substitute in (\ref{68}) the second--order approximations for
$\Sigma_2$ and $D_2$, then  we come to an equation defining the
second--order approximation for the correlation function $g(12)$. However,
this equation is rather cumbersome. Instead, we may improve the
first--order approximation for $g(12)$ in the following way. We can
consider (\ref{69}) as an asymptotic expansion in powers of $\Phi$. An
effective limit of an asymptotic series can be defined by using the
self--similar approximation theory [85--97]. Employing the algebraic
variant of this theory [98-100], we obtain
\begin{equation}
\label{70}
g(12) = g_0(12)\exp\left\{ -\int A(1234)\frac{g(34)}{g_0(12)}
\Phi(34) d(34)\right \} \; .
\end{equation}

This equation, as is explained above, because of divergences, cannot be
iterated starting with $g_0(12)$. One direct option would be to solve
(\ref{70}) numerically for a given interaction potential $\Phi$. Another
way can be by iterating the right--hand side of (\ref{70}) starting with
an approximate $g_{app}(12)$ taking into account interparticle
correlations. For example, we may start [82] with
$$ g_{app}(12) = |\Psi(12)|^2 \; , $$
where $\Psi(12)$ is a pair wave function normalized so that to satisfy the
condition
$$ g_{app}(12)\ra 1 , \qquad | x_1 - x_2| \ra \infty\; , $$
similar to (\ref{40}). In this way, the correlated iteration theory is
completed.

\newpage

{\Large{\bf Appendix}}

\vspace{5mm}

It was mentioned in Preface that statistical mechanics is the most general
physical theory. Here we concretize this statement.

From the mathematical point of view, one theory is more general than
another if the latter can be derived from the former by means of some
mathematical operations, but not conversely. A particular theory can be
considered as a limiting case of a more general theory. The corresponding
limit can be defined explicitly in an unambiguous mathematical sense. The
procedure of taking a limit is, of course, to be accompanied by the
related physical interpretation.

The classification of physical theories can be done by considering three
parameters: velocity of light, $c$, Plank constant, $\hbar$, and Boltzmann
constant, $k_B$. The choice of these parameters is caused by the fact that
each of them is related to a principal physical notion. The light
velocity, according to the Einstein postulate, is the same in all
reference systems. The Plank constant defines a minimal quant of action.
And the Boltzmann constant connects the entropy with a statistical
distribution.

A theory with the finite velocity of light, $c$, is called {\it 
relativistic}. The limit $c\ra\infty$ leads to a {\it nonrelativistic}
theory. A theory containing the Plank constant, $\hbar$, is termed {\it
quantum}. Taking the limit $\hbar\ra 0$, one passes to a {\it classical}
theory. A theory dealing with probability distributions and, thus,
containing the Boltzmann constant, $k_B$, is named {\it statistics}. The
limit $k_B\ra 0$ reduces the latter to a particular kind of theory for
which there is no a commonly accepted name. For such theories, we use
below the word "mechanics", although this term is more general and is
applicable not only to theories with $k_B=0$. So that the general name
"mechanics" is to be considered not separately but in the combination
with other terms characterizing a theory. The whole theoretical physics
can be subdivided into $8$ theories according as the principal parameters
from the triplet $\{ c,\hbar,k_B\}$ are finite or some limits are taken.
These theories are as follows:

\vspace{2mm}

(1) Relativistic Quantum Statistics: $\{ c,\hbar,k_B\}$;

\vspace{1mm}

(2) Nonrelativistic Quantum Statistics: $\{ c\ra\infty,\hbar,k_B\}$;

\vspace{1mm}

(3) Relativistic Classical Statistics: $\{ c,\hbar\ra 0,k_B\}$;

\vspace{1mm}

(4) Nonrelativistic Classical Statistics: $\{ c\ra\infty,\hbar\ra 0,k_B\}$;

\vspace{1mm}

(5) Relativistic Quantum Mechanics: $\{ c,\hbar,k_B\ra 0\}$;

\vspace{1mm}

(6) Nonrelativistic Quantum Mechanics: $\{ c\ra\infty,\hbar,k_B\ra 0\}$;

\vspace{1mm}

(7) Relativistic Classical Mechanics: $\{ c,\hbar\ra 0,k_B\ra 0\}$;

\vspace{1mm}

(8) Nonrelativistic Classical Mechanics: 
$\{ c\ra\infty,\hbar\ra 0,k_B\ra 0\}$.

\vspace{3mm}

As is clear from this classification, Relativistic Quantum Statistics is
the most general theory from which all other theories can be derived by
means of appropriate limits. If not to subdivide the theories onto
relativistic and nonrelativistic, then Quantum Statistics is the most
general physical theory incapsulating all other theories.